\documentclass{article}
\usepackage[utf8]{inputenc}
\usepackage{authblk}
\usepackage{setspace}
\usepackage[margin=1.25in]{geometry}
\usepackage{graphicx}
\graphicspath{ {./figures/} }
\usepackage{subcaption}
\usepackage{amsmath}
\usepackage{hyperref}
\usepackage{xcolor}
\usepackage{latexsym}
\usepackage{amssymb}
\usepackage{amsfonts}
\usepackage{amsthm}
\usepackage{bbm}
\usepackage{color}
\usepackage{cancel} 
\usepackage{ulem}
\usepackage{bm}
\usepackage{soul}
\usepackage{array}
\usepackage{multirow}
\usepackage[all]{xy}
\usepackage[bottom]{footmisc}

\hypersetup{
	bookmarksnumbered,
	unicode,			
	colorlinks,			
	citecolor=[rgb]{.9,0,.5},	
	urlcolor=[rgb]{0,0,1},	
	linkcolor=[rgb]{0,.7,0}	
	}

\usepackage[
    natbib=true,
    style=numeric-comp,
    sorting=none
]{biblatex}

\catcode`@=11
\def\secteqno{\@addtoreset{equation}{section}%
	\def\theequation{\thesection.\arabic{equation}}}
\catcode`@=12
\secteqno
\def\dd{\hbox{\,\Large$\triangleright$}}
\newcommand{\be}{\begin{equation}}
	\newcommand{\ee}{\end{equation}}
\newcommand{\bea}{\begin{eqnarray}}
	\newcommand{\eea}{\end{eqnarray}}
\newcommand{\bref}[1]{(\ref{#1})}
\newcommand{\nn}{\nonumber}

\def\dig#1{\setbox0=\hbox{$#1M$}
	\hskip.06\wd0 \vrule width.07\wd0 height.63\wd0 depth.01\wd0 
	\vrule width.37\wd0 height.63\wd0 depth-.56\wd0 \hskip-.4\wd0
	\vrule width.25\wd0 height.35\wd0 depth-.28\wd0 
	\vrule width.07\wd0 height.35\wd0 depth-.17\wd0 \hskip.14\wd0}
\def\digamma{{\mathpalette\dig{}}}
\def\={{\;=\;}}\def\+{{\;+\;}}

\def\dig#1{\setbox0=\hbox{$#1M$}
	\hskip.06\wd0 \vrule width.08\wd0 height.63\wd0 depth.01\wd0 
	\vrule width.37\wd0 height.63\wd0 depth-.55\wd0 \hskip-.4\wd0
	\vrule width.25\wd0 height.36\wd0 depth-.28\wd0 
	\vrule width.08\wd0 height.36\wd0 depth-.17\wd0 \hskip.14\wd0}
\def\digamma{{\mathpalette\dig{}}}
\def\bop#1{\setbox0=\hbox{$#1M$}\mkern1.5mu
	\vbox{\hrule height0pt depth.1\ht0
		\hbox{\vrule width.1\ht0 height.8\ht0 \kern.8\ht0
			\vrule width.1\ht0}\hrule height.1\ht0}\mkern1.5mu}

\def\dd{\hbox{\,\Large$\triangleright$}} 

\def\don#1#2{{\buildrel{\mkern2.5mu\raise-.1em\hbox{$\scriptstyle#1$}\mkern-2.5mu}\over{#2}}}	
\def\dron#1#2{{\buildrel{{\raise-.1em\hbox{$\scriptstyle#1$}}}\over{#2}}}		
\newcommand{\ctext}[1]{\raise0.2ex\hbox{\textcircled{\scriptsize{#1}}}}

\def\mapright#1{\smash{\mathop{\longrightarrow}\limits^{\textstyle#1}}}	

\title{${\cal A}$-theory\\ - A brane world-volume theory with manifest U-duality -}

\author[$\heartsuit$$\diamondsuit$]{Machiko Hatsuda}
\author[$\ddagger$]{Ond\v{r}ej Hul\'{\i}k}
\author[$\#$]{ William D. Linch }
\author[$\clubsuit$$\spadesuit$]{ Warren D. Siegel }
\author[$\clubsuit$]{\\Di Wang}
\author[$\clubsuit$]{ Yu-Ping Wang \footnote{Email: \href{yu-ping.wang@stonybrook.edu}{yu-ping.wang@stonybrook.edu} } }

\affil[$\clubsuit$]{\textit{ Department of Physics, SUNY Stony Brook University, Stony Brook, NY 11794, USA}}
\affil[$\spadesuit$]{\textit{C. N. Yang Institute for Theoretical Physics. Stony Brook, NY 11794, USA}}
\affil[$\heartsuit$]{\textit{Department of Radiological Technology, Faculty of Health Science, Juntendo University
Yushima, Bunkyou-ku, Tokyo 113-0034, Japan}}
\affil[$\diamondsuit$]{\textit{KEK Theory Center, High Energy Accelerator Research Organization
Tsukuba, Ibaraki 305-0801, Japan}}
\affil[$\ddagger$]{\textit{Theoretische Natuurkunde, Vrije Universiteit Brussel
Pleinlaan 2, B-1050 Brussels, Belgium}}
\affil[$\#$]{\textit{Mercersburg Academy, Mercersburg, PA 17236, USA} }

\date{\today}

\begin{document}

\newgeometry{top=0.1in,bottom=1in,right=1in,left=1in}
\hfill YITP-SB-2023-09
\vskip 0.1in
\hfill KEK-TH-2523
{\let\newpage\relax\maketitle}

\begin{abstract}
	In this paper, the ${\cal A}$-theory, an extension of F-theory, is described as a fully U-duality covariant brane theory.
	This theory has some distinguishing features not known from world-sheet models.
	In particular, seen as a sigma model, both world-volume and target space coordinates are specific representations of the same group (the U-duality group).
	The U-duality group in question is an exceptional group (a split form of the $E_d$ series).
The structure of this group allows it to encompass both the T-duality group of string theory as well as the general linear symmetry group of ${\cal M}$-theory. 	
	${\cal A}$-theory is defined by the current algebras in Hamiltonian formalism, or by world-volume actions in Lagrangian formalism.
	The spacetime coordinates are selfdual gauge fields on the world-volume, requiring the Gau\ss{} law constraints tying the world-volume to spacetime.
	Solving the Gau\ss{} law constraints/the Virasoro constraints gives the world-volume/spacetime sectioning from ${\cal A}$-theory to ${\cal T}$-theory/${\cal M}$-theory respectively. 
The	${\cal A}$-theory Lagrangian admits extended symmetry which has not been observed previously in the literature,
	where the background fields include both the spacetime and the world-volume gravitational fields.
	We also constructed the four-point amplitude of ${\cal A}$-theory in the low energy limit. The amplitude is written in a way that the U-duality symmetry
	is manifest, but after solving the section condition, it reduces to the usual four-graviton amplitude. 
	
	In the previous papers, we have referred to this model as F-theory, however, F-theory initiated by Vafa is now a big branch of string theory as the study of elliptic fibrations, so we refer to these constructions as generalized models of theory for all dimensions with all duality symmetries as ${\cal A}$-theory.
\end{abstract}

\restoregeometry

\tableofcontents

\newpage

\section{Introduction}

\subsection{Overview}

\label{section:1}

The five superstring theories and M-theory are related by string duality. In other words, string duality is the key to obtaining a unified picture of superstring theory.  Soon after Witten's articulation of the M-theory paradigm \cite{Witten:1995ex}, Vafa attempted to extend the geometrical realization of the strong coupling limit to the duality group of the type IIB string \cite{Vafa:1996xn}. In that paper, it was proposed that some yet-to-be-understood world-volume dynamics would spontaneously reduce the $2+2$-dimensional world-volume theory to the world-sheet of the IIB string while simultaneously reducing the $10+2$-dimensional target space down to the familiar 9+1-dimensional vacuum solutions. 

In an earlier line of research, one of us investigated the possibility of reformulating the superstring world-sheet theories to represent the T-duality groups that emerge in toroidal compactifications \cite{Siegel:1993th, Siegel:1993xq, Siegel:1993bj} (see also \cite{Tseytlin:1990va}). 
In this work, the doubled set of target space coordinates makes T-duality symmetry manifest.
The left and right moving currents in the doubled space separate the physical current (selfdual) from the unphysical current which relates the original coordinate and the dual coordinate as the selfduality constraint.
The Virasoro operators in the extended space, to which we will refer as $\mathcal S$, is bilinear in these currents.  When this constraint is imposed on the algebra of string fields, it implies that for each dimension, observables will either be independent of the position coordinate or of the winding coordinate. This is similar to the choice of polarization in quantum mechanics.
In more modern times, the (super)gravity limit of this T-dual string is known as double field theory (DFT), and in this context, the $\mathcal S =0$ constraint reduces to what is called ``the section conditions".

Continuing in anti-chronological order, Duff \cite{Duff:1985bv} had earlier suggested that the Cremmer-Julia symmetry (${\rm E}_n$ for $n$ dimensional torus) of toroidal compactifications of eleven-dimensional supergravity \cite{Cremmer:1979up} ought to be considered as a hidden {\it spacetime} symmetry already in the uncompactified theory. This idea can be incorporated to enhance the double field theory paradigm to exceptional field theory (EFT). In these theories, the classical limit of the U-duality group of Cremmer and Julia is made manifest by the addition of coordinates corresponding to the higher winding modes of the branes of ${\cal M}$-theory \cite{Berman:2012vc}. 

In \cite{Polacek:2014cva, Linch:2015fya, Linch:2015qva} two of us initiated a program to extend the world-volume approach of the T-dual string to Vafa's idea of representing S-duality group manifestly but extended to the U-duality group as envisioned by Duff. In Duff's interpretation of the U-duality group, the rank of this group will {\it grow} as we make more Poincar\'e symmetry manifest.

We started with the model of type II superstring with manifest U-duality symmetry, SL(5) case \cite{Linch:2015qva} 
and SO(5,5) case \cite{Linch:2015qva} and work our way up from there. The resulting models represented what we believed to be an extension of Vafa's original work  (albeit in toy model form) since the one dimensional case of this model can be seen as a full geometrization of the SL(2) S-duality group in type IIB string theory \cite{Berman:2015rcc, Chabrol:2019kis}.

In all the previous papers \cite{Linch:2015fya, Linch:2015qva, Linch:2016ipx, Linch:2015fca, Siegel:2018puf, Linch:2015lwa, Siegel:2019wrr, Hatsuda:2021wpb, Hatsuda:2021ezo, Siegel:2016dek, Siegel:2020gro, Ju:2016hla, Siegel:2020qef, Siegel:2019wrr, Linch:2017eru}, we refer these models as ``F-theory'', but we decided to rename our ``F-theory'' to  ${\cal A}$-theory as to avoid 
confusion. This is because (1) the later development of the original F-theory took on a different direction. It is used as a tool to understand string compactification on Calabi-Yau manifold and for model building. (See the review paper \cite{Weigand:2018rez} and the references within.) (2) There are many features in our ``F-theory'' that are not envisioned in Vafa's original paper, such as the geometrization of a higher rank exceptional group, the extension from $G$-symmetry to $A$-symmetry, and the fact that duality group acts on both the target space and world-volume coordinates. All of these will be elaborated on in this paper.

We will explain how the U-duality group acts not only on fields of the low energy limit but also the massive fields.
To this end both the world volume coordinates as well as the target space coordinates of the model are extended such that the U duality group is represented on these in a linear manner. Due to this extension extra constraints need to be introduced to eventually reduce the space down to it's original dimension.
In this paper, we will review the progress made to date on ${\cal A}$-theory in some detail, but it may help to first have an overview of the elements involved. 
To this end, suppose we have a classical type II superstring theory with a D-dimensional target space. We will call this the S-theory; it has global structure GL(D) and Lorentz group SO(D$-$1,1), so the vielbein of gravity represents the coset GL(D)/SO(D$-$1,1). 
The low energy limit of M-theory 
of this S-theory is $N=1$ supergravity in D+1 dimensions. It has global structure GL(D$+$1) and local SO(D,1), so the gravity coset is GL(D$+$1)/SO(D,1). In the target space supergravity theory, reduction from the M-supergravity to the S-supergravity proceeds by the usual circle compactification.

 ${\cal T}$-theory is its generalization with manifest T-symmetry O(D,D)/SO(D$-$1,1)$^2$.  Its massless sector is DFT, often without the use of its world-sheet origin.   $G/H$ for S,${\cal T}$,${\cal M}$ are all finite classical groups, for all D.  The spacetime supersymmetric algebras allow backgrounds in curved superspace.  	Similarly, all bosonic component fields represent $G$, while fermions represent only $H$ \cite{Polacek:2014cva, Polacek:2016nry, Hatsuda:2014qqa, Hatsuda:2014aza, Hatsuda:2015cia}.

 ${\cal A}$-theory encompasses all of these aforementioned theories. Its $G$ is the exceptional group E$_{\rm{D+1(D+1)}}$; and it's  $H$ follows from extended supergravities in various dimensions. 
 We want  gravitational fields to be coset elements of  $G/H$  rather than compactification scalars of the usually extended supergravities. 
 Our $H$ of the coset $G/H$  (a maximal compact subgroup of $G$) includes time as well as space. 
 No specific background or compactification is considered here.
We consider that our D-dimensional space of S-theory includes time and that duality symmetry is associated with D.
The critical dimension is achieved by adding a D$'$=10-D dimensional space.
 Dimensions of the groups may be infinite for higher D, due to the nature of exceptional groups.

The supersymmetric ${\cal A}$-theories described here include all string states.  The usual non-perturbative states are expected to show up already in the classical field theory level in compactification as well as in 1st-quantized theory around classical solution.  Thus STU-duality symmetries should be linearly realized in the classical field theory action 
	to some extent in the classical mechanics. Usual D-branes are described non-perturbatively with the Wess-Zumino terms,
but instead, ${\cal A}$-theory includes all brane states perturbatively.
Therefore ${\cal A}$-theory actions are world-volume actions.  Free brane actions require
 the spacetime coordinates to be non-vector representations of U-duality symmetry groups.

 When considering massless backgrounds, sectioning conditions are imposed:  
 For example, in ${\cal T}$-theory sectioning eliminates winding modes in the uncompactified limit\cite{Siegel:1993bj}; the condition $L_0=\bar{L}_0$ mixes massive winding modes with massive oscillator states.

These ${\cal A,T,M,}$S-theories are defined by world-volume current algebras in Hamiltonian
	formalism, or by world-volume actions in Lagrangian formalism. 
The right-hand sides of the current algebras include the $G$-symmetry group invariant metrics 
with two spacetime indices and a world-volume index.
For ${\cal T}$-theory case the  $G$-invariant metric is the O(D,D) invariant metric
where the string world-volume index is hidden.
For ${\cal A}$-theories the $G$-invariant metric carries both the spacetime indices and the world-volume index. In Lagrangian formalism,
the world-volume current is the selfdual field strength whereas the spacetime coordinate
is the gauge field. Then both the world-volume index and the spacetime index represent $G$-symmetry
but in different representations.
The gauge symmetry of the gauge field is generated by the Gau\ss{} law constraint $\mathcal U=0$. The Gau\ss{} law constraint is a product of the spacetime field strength and the world-volume derivative.

The reduction from ${\cal M}$-theory to S-theory is carried out by sectioning with 
the Gau\ss{} law constraint. Since the Gau\ss{} law constraint involves world-volume derivatives,
it gives sectioning of the world-volume coordinates reducing a brane to a string.
As discovered in \cite{Siegel:1993bj}, a reduction from ${\cal T}$-theory to S-theory proceeds by solving the new Virasoro condition $\mathcal S =0$. 
The 0-modes of the ${\cal S}=0$ give the strong and weak section conditions on fields of the 0-mode coordinates.
The nonzero modes of the ${\cal S}=0$ reduce the whole coordinates as the dimensional reduction.
 We can summarize all of this with the  diagram: 
 \bea
& {\cal M}\textrm {-theory}\stackrel{\mathcal U}\longrightarrow \textrm{S-theory} \stackrel{\mathcal S}\longleftarrow 
 {\cal T}\textrm {-theory} &~\nn\\
 \nn\\
 &\left\{
\begin{array}{ccl}
	{\cal S}&:&{\textrm{spacetime~sectioning~with~Virasoro~constraints}}\\
	{\cal U}&:&{\textrm{world-volume~sectioning~with~ Gau\ss{}~law~constraints}}
\end{array}\right.\nn &
 \eea
 
\par\noindent
The ${\cal A}$-theory program proposes to extend this structure by adding a fourth vertex ${\cal A}$ and arranging it in a commuting diamond-shaped diagram,
\begin{align}
	\xymatrix{
		   &{\cal A}\textrm {-theory}\ar[dl]_{\mathcal S} \ar[dr]^{\mathcal U} &\\
	   {\cal M}\textrm {-theory}\ar[dr]_{\mathcal U}&&{\cal T}\textrm {-theory}\ar[dl]^{\mathcal S}\\
		  &\textrm S\textrm {-theory}&
	  \nonumber	}
	  \end{align}

\noindent
The ${\cal A}$-theory is then constructed so that solving the strong sectioning constraint reduces ${\cal A} \longrightarrow {\cal M}$ and solving the Gau\ss{} law sectioning reduces $ {{\cal A} \longrightarrow  \cal T}$.

At the level of supergravity, this reduces to
\begin{align}
	\xymatrix{
	   &\textrm{EFT}\ar[dl]_{{\mathcal S}|_{\textrm{0-modes}}}
     &\\
	  \textrm{M-theory SG}&&\textrm{DFT}\ar[dl]^{{\mathcal S}|_{\textrm{0-modes}}}\\
	  &\textrm{type II SG}&\\
	}
  \nonumber  
  \end{align}
by construction. 
\subsection{Summary}

In this paper, we give a review of what is currently known about ${\cal A}$-theory. In section \ref{section:1}, we give
the motivations for why incorporating the various duality symmetry groups for the string or brane theories.   
 
In section \ref{section:2}, we systematically study these duality symmetry groups. $G$-symmetries are manifest 
duality symmetries of the current algebras. They also correspond to the conventional duality symmetries first discovered in string theory 
and supergravity theories.
One can extend the $G$-symmetry to $A$-symmetry in Lagrangian formalism.  After fixing the flat background the symmetries will be broken down 
to $L$-symmetry in Lagrangian formalism, and $H$-symmetry in Hamiltonian formalism. The symmetry groups
for various $A, G, L, H$-symmetries are listed in Table \ref{Table:2-1} and Table \ref{Table:2-2}.  The 
relationships between these groups becomes apparent through Dynkin diagrams as explained in section \ref{section:2-1}.
One notable feature of these symmetry groups is that they act on not only the target space coordinates but also the 
world-volume coordinates. The target space, world-volumes, field strengths, and many other physical quantities are representations
of the duality symmetry groups. These representations are listed in tables in section \ref{section:2-3} and \ref{section:2-4}.

In section \ref{section:3}, we treat the current algebras of ${\cal A}$-theory with full details. As an example, we give the construction 
of the current algebra of ${\cal T}$-theory in section \ref{section:3-1}; the explicit construction of current algebras for D$=$3
theories are given at \ref{section:3-2}.
The construction of current algebra in general dimensions is given in section \ref{section:3-3}. We discuss the various constraint 
that the current algebra gives at \ref{section:3-4}, and how the current algebras generate the Dorfman bracket in \ref{section:3-5}.
In order to calculate the curvature tensors the Lorentz generator must be included in the current algebra. The consistent extension requires that includes both the Lorentz generators and the supersymmetry
generators, The details of this extension are discussed in section \ref{section:3-6}.

As discussed in this paper, one can reduce ${\cal A}$-theory is reduced to ${\cal M}$ or ${\cal T}$-theories by solving the constraints. 
We explicitly solve these constraints by dimensions in section \ref{section:4}. D$=$3 case discussed in
\ref{section:4-1}, and D$=$4 case is discussed in \ref{section:4-2}.

In section \ref{section:5}, we introduce currents in curved backgrounds.
In section \ref{section:5-1}, we apply the background fields to the D$=$3 current algebras. The current algebra itself 
generates the gauge transformation of the background gauge field, and the orthogonality conditions pick out the field content in the vielbein. 
In section \ref{section:5-2}, we add the backgrounds of the full non-degenerate Poincar\'e current algebras.
By doing so we have extra torsion constraints on the vielbein. Unlike usual supergravity, the torsions and the world-volume metric 
are not in general invariant under background transformations, but in section \ref{section:5-3}, we show that both torsion and 
world-volume metric is invariant up to the ${\cal U}$-constraints under background field transformation. One important application for 
adding backgrounds in the non-degenerate Poincar\'e algebra is that not only the field content is in the background, but also 
the field strengths are part of the background. In section \ref{section:5-4}, we solve the torsion constraints and find the 
prepotential of D$=$3 ${\cal A}$-theory.

In section \ref{section:6}, we construct the Lagrangians of  ${\cal A}$-theory. There are currently no formulations for general dimensions.
For the D$=$3 case, the SL(5) $G$-symmetry covariant Lagrangian is given in section \ref{section:6-1} and the SL(6) $A$-symmetry covariant Lagrangian 
 is given in section \ref{section:6-2}.

We also discuss some approaches in constructing on-shell amplitude for ${\cal A}$-theory in section \ref{section:7}. In section 
\ref{section:7-1}, we discuss some generality of the 4-point tree amplitude and how one can construct a manifest gauge invariant
amplitude by using on-shell field strengths. In section \ref{section:7-2}, we introduce the twistor formalism for ${\cal A}$-theory which is a useful tool to construct the amplitude as on-shell objects in a manifestly symmetric way.
In section \ref{section:7-3} we explicitly construct the D$=$3 SL(4;$\mathbb{R}$) $H$-symmetry covariant amplitude based on previous sections. 

Here the current progress of generalization of ${\cal A}$-theory depending on D as well as ${\cal T}$-theory are listed below.
\vskip 5mm

\begin{minipage}{\textwidth}
	\begin{tabular}{ |c||p{2cm}|p{2cm}|p{2cm}|p{2cm}|p{2cm}|}
	\multicolumn{6}{c}{Generalizations of theories} \\
	 \hline
	 D (Rank)
	 &Lagrangian & $A/L$ & SUSY & Curved background & Critical dimensions 
	 \\
	 \hline
	${\cal A}$-theory&&&&&\\
	 General    & \cite{Linch:2016ipx}($\leq 7 ?$)    & \cite{Siegel:2018puf}($\leq 6$)& \cite{Linch:2016ipx}($\leq 7 ?$)  & & \cite{Linch:2015fca}($\leq 7?$)\\
	 $0\sim 3$&     &    & & &\\
	 4 & & \cite{Linch:2015fya}&  \cite{Siegel:2019wrr}& \cite{Siegel:2019wrr, Hatsuda:2012vm}&\\
	 5  &\cite{Linch:2015qva, Hatsuda:2021wpb} & \cite{Hatsuda:2021wpb} &  & \cite{Hatsuda:2021wpb, Hatsuda:2013dya}&\\
	 6&    & \cite{Siegel:2018puf}& &  &\\
	 \hline
	 ${\cal T}$-theory& \cite{Hatsuda:2018tcx} & & \cite{Hatsuda:2014qqa, Hatsuda:2014aza, Polacek:2014cva,  Hatsuda:2019xiz} & \cite{Hatsuda:2014qqa, Polacek:2014cva}&\\
	 \hline
	\end{tabular}
\end{minipage}

\vskip 6mm
\noindent
Many properties of ${\cal A}$-theory rely on its specific ranks and thus many of the directions listed above can be done only in specific dimensions. Some of them have already been worked out in lower dimensions. The Lagrangian column lists the references where the Lagrangian formalism is constructed. The ``$A/L$'' column lists the references where the explicit extension from $G/H$ to $A/L$ is formulated. The ``SUSY'' column lists the references where complete non-degenerate and supersymmetric Poincar\'e algebra is given, and the ``Curved backgrounds'' column list the references that applied curved background to current algebra, and in some of them, the torsion constraints are solved.  In the ``Critical dimensions'' column, we added the flat internal space generators to the current algebras, and the algebra needed to be extended to add generators with mixed indices. 

\subsection{Conventions}
We follow the indices conventions unless stated. 
\begin{enumerate}
	\item $M, N, P, \cdots$ are curved target space indices;  $A, B, C, \cdots$ are flat target space indices.
	\item $m, n, p, \cdots$ are curved world-volume indices;  $a, b, c, \cdots$ are flat world-volume indices.
	\item $\mu, \nu, \cdots$ are curved spinor indices (for supersymmetry); $\alpha, \beta, \cdots$ are 
	flat spinor indices.
	\item ${\cal M}, {\cal N}, {\cal P}, \cdots$ are curved indices for the full non-degenerate Poincar\'e algebra;
	${\cal A}, {\cal B}, {\cal C}, \cdots$ are flat indices for the full non-degenerate Poincar\'e algebra.
    \item $I, J, K \cdots$ correspond to the adjoint indices of $H$.
    \item We use index $\varpi$ to indicate the representation of ${\cal U}$ constraints. 
	\item D for target space dimensions and d for world-volume dimensions.
 
\end{enumerate}



\section{U-duality symmetry} 
\label{section:2}

In this section, we are going to discuss the construction of theories with manifest U-duality. 
The way these theories are constructed is in line with the first-order approach to regular Einstein Hilbert gravity. In first-order formalism, we can identify the vielbein as fields parametrizing a coset $G/H$
where $G$ is the general linear group ${\rm GL}$(D) and $H$ is the Lorenz group ${\rm SO}$(D$-$1,1). 
This idea extends naturally into string theory where the metric and $B$ field are naturally recast into a generalized metric which is again parametrizing a coset $G/H$ this time for $G$ being ${\rm SO}$(D,D) group and 
$H$ being double Lorentz group ${\rm SO(D}-1,1)^2$.
The group {SO}(D,D) contains ${\rm GL}$(D) as its subgroup, and hence there is well defined gravitational subsector. Advantage of this manifest ${\rm SO}$(D,D) formalism is that fields acquire a natural action of the former group in which some generators are identified with the action of T-duality.
One can add the R-R sector fields to the coset which leads to further extension of {SO}(D,D) to the exceptional group ${\rm E}_{{\rm D}+1}$.

 We shall call the manifest duality symmetries of ${\cal A}$, $\mathcal{T}$, $\mathcal{M}$ and S-theory 
``$G$-symmetries''. In our language, these ``$G$-symmetries" can be understood as the manifest symmetry we found when we try to 
formulate the current algebras (i.e. the Hamiltonian formalism) of the theories above.
\begin{itemize}
\item{${\cal A}$-theory has the exceptional group ${\rm E}_{{\rm D}+1}$ U-duality symmetry}
\item{${\cal T}$-theory has the O(D,D) T-duality  symmetry }
\item{${\cal M}$-theory has the GL(D+1) group including the SL(2) S-duality group to of S-theory which has the D-dimensional diffeomorphism symmetry generated by GL(D). }
\end{itemize}
In the next section, we show how Dynkin diagrams of the  exceptional algebra ${\rm E}_{\rm D+1}$ include both the {O}(D,D) T-duality symmetry algebra
and the {GL}(D+1) S-duality algebras as its subalgebras.

There is a distinguishing feature when one passes to a model with ${\rm E}_{\rm{D}}$ symmetry. It turns out that contrary to the string models even the world-volume coordinates are transforming in nontrivial representations of the exceptional group. This has a consequence of having the unnatural dimension of the world-sheet which is however dealt with via imposing a section constraint. 
same symmetry. The extra coordinates serve as the Lagrange multipliers for the  Gau\ss{}  law constraint $\mathcal{U}=0$.

$G$-symmetries of ${\cal A,T,M}$, S-theories do not mix the spacetime and the world-volume currents.
But $A$-symmetry of ${\cal A}$-theory mixes the spacetime and the world-volume in a way that
the extended vielbein fields are elements of the coset $A/L$
including both the spacetime and the world-volume vielbeins.\\

\subsection{Dynkin diagrams } \label{section:2-1}
\subsubsection{Dynkin diagrams of $G$-symmetries of theories}

$G$-symmetries of D-dimensional S-theory and its corresponding theories ${\cal M}$, ${\cal T}$, ${\cal A}$-theories have
following Dynkin diagrams of $A_{\textrm D-1}$,  $A_{\rm D}$, $D_{\rm D}$, $E_{\rm D+1}$ in Table \ref{table:data_type}.
Although we are working with the real forms of Lie algebra in this paper, for the remainder of this section, we consider the  ``complexified" algebra,
and the GL(1) factor of GL(D$+$1) is decoupled.

\begin{table}[hbtp]
	\caption{Dynkin diagram of $G$-symmetries of theories}
	\label{table:data_type}
	\begin{center}
		${\cal A}$-theory: ~${\rm E}_{\rm D+1}$~~~\\
		\setlength{\unitlength}{.1in}
		\thicklines
		\begin{picture}(25,9)
			\put(1,4){\circle{2}}	\put(0.65,3.7){\scriptsize 1}
			\put(2,4){\line(1,0){1}}
			\put(4,4){\circle{2}}  \put(3.65,3.7){\scriptsize 2}
			\put(5,4){\line(1,0){1}}
			\put(6.2,3.95){....}
			\put(8,4){\line(1,0){1}}
			\put(10,4){\circle{2}}	\put(9.1,3.7){\scriptsize D-3}
			\put(11,4){\line(1,0){1}}
			\put(13,4){\circle{2}} 	\put(12.1,3.7){\scriptsize D-2}
			\put(13,5){\line(0,1){1}}
			\put(13,7){\circle{2}}	\put(14.1,6.6){\scriptsize D+1}
			\put(14,4){\line(1,0){1}}
			\put(16,4){\circle{2}}	\put(15.1,3.7){\scriptsize D-1}
			\put(17,4){\line(1,0){1}}
			\put(19,4){\circle{2}}	\put(18.65,3.7){\scriptsize D}
		\end{picture}\\
		$\swarrow$~~~~~~~~~~~~~~~~~~~~~~~~~~~~~~~~~~~~~~~~~~~~~~~~~~~~~~$\searrow$\\
		~~~~~~~~~~~~\\
		${\cal M}$-theory:~GL(D+1)~~~~~~~~~~~~~~~~~~~~~~~~~~~~~~~~~~~~~~~~~~~${\cal T}$-theory:~O(D,D)\\~~~~~
		\setlength{\unitlength}{.1in}
		\thicklines
		\begin{picture}(20,9)
			\put(1,4){\circle{2}}	\put(0.65,3.7){\scriptsize 1}
			\put(2,4){\line(1,0){1}}
			\put(4,4){\circle{2}}  \put(3.65,3.7){\scriptsize 2}
			\put(5,4){\line(1,0){1}}
			\put(6.2,3.95){....}
			\put(8,4){\line(1,0){1}}
			\put(10,4){\circle{2}}	\put(9.1,3.7){\scriptsize D-3}
			\put(11,4){\line(1,0){1}}
			\put(13,4){\circle{2}} 	\put(12.1,3.7){\scriptsize D-2}
			\put(14,4){\line(1,0){1}}
			\put(16,4){\circle{2}}	\put(15.1,3.7){\scriptsize D-1}
			\put(17,4){\line(1,0){1}}
			\put(19,4){\circle{2}}	\put(18.65,3.7){\scriptsize D}
		\end{picture}
		~~~~~~~~~~~~~~~~~~~~~~~~~~~
		\setlength{\unitlength}{.1in}
		\thicklines
		\begin{picture}(20,9)
			\put(1,4){\circle{2}}	\put(0.65,3.7){\scriptsize 1}
			\put(2,4){\line(1,0){1}}
			\put(4,4){\circle{2}}  \put(3.65,3.7){\scriptsize 2}
			\put(5,4){\line(1,0){1}}
			\put(6.2,3.95){....}
			\put(8,4){\line(1,0){1}}
			\put(10,4){\circle{2}}	\put(9.1,3.7){\scriptsize D-3}
			\put(11,4){\line(1,0){1}}
			\put(13,4){\circle{2}} 	\put(12.1,3.7){\scriptsize D-2}
			\put(13,5){\line(0,1){1}}
			\put(13,7){\circle{2}}	\put(12.5,6.6){\scriptsize D}
			\put(14,4){\line(1,0){1}}
			\put(16,4){\circle{2}}	\put(15.1,3.7){\scriptsize D-1}
		\end{picture}\\
		$\searrow$~~~~~~~~~~~~~~~~~~~~~~~~~~~~~~~~~~~~~~~~~~~~~~~~~~~~~~$\swarrow$\\
		S-theory: ~GL(D)~~~\\
		\setlength{\unitlength}{.1in}
		\thicklines
		\begin{picture}(18,6)
			\put(1,4){\circle{2}}	\put(0.65,3.7){\scriptsize 1}
			\put(2,4){\line(1,0){1}}
			\put(4,4){\circle{2}}  \put(3.65,3.7){\scriptsize 2}
			\put(5,4){\line(1,0){1}}
			\put(6.2,3.95){....}
			\put(8,4){\line(1,0){1}}
			\put(10,4){\circle{2}}	\put(9.1,3.7){\scriptsize D-3}
			\put(11,4){\line(1,0){1}}
			\put(13,4){\circle{2}} 	\put(12.1,3.7){\scriptsize D-2}
			\put(14,4){\line(1,0){1}}
			\put(16,4){\circle{2}}	\put(15.2,3.7){\scriptsize D-1}
		\end{picture}\\
	\end{center}
\end{table}

\noindent
The $A_{\rm D}$ diagram for GL(D+1) of ${\cal M}$-theory is obtained from
the $E_{\rm D+1}$ diagram for ${\cal A}$-theory by removing the (D$+$1)-th node
in the $E_{\rm D+1}$ diagram attached to the long chain.
The $D_{\rm D}$ diagram for O(D,D) of ${\cal T}$-theory is obtained from
the $E_{\rm D+1}$ diagram for ${\cal A}$-theory by removing the D-th node
in the $E_{\rm D+1}$ diagram which is the right end of the long chain.
The $A_{\rm D-1}$ diagram for GL(D) of S-theory is obtained from
${\cal M}$-theory by removing the D-node
in the $A_{\rm D}$ diagram which is the right end of the long chain.
It is also obtained from
${\cal T}$-theory by removing the D-th node
in the $D_{\rm D}$ diagram attached to the long chain.

Dynkin diagrams, examples of simple roots, roots, and the number of roots of $A_n$, $D_n$, and $E_6$
are listed in the following Table \ref{table:data_type2}.
\begin{table}[h!]
	\caption{Roots of $G$-symmetries of theories}
	\label{table:data_type2}
	\centering
	~~~~~  \\
	\begin{tabular}{|c|c|l|l|}
		\hline
		Algebra                                                            & Dynkin                                                                                                              & Simple roots & Roots~\lbrack number of roots\rbrack \\\hline
		$A_{\rm n}$                                                        &
		\setlength{\unitlength}{.1in}
		\thicklines
		\begin{picture}(15,6)
			\put(1,2){\circle{2}}	\put(0.65,1.7){\scriptsize 1}
			\put(2,2){\line(1,0){1}}
			\put(4,2){\circle{2}}  \put(3.65,1.7){\scriptsize 2}
			\put(5,2){\line(1,0){1}}
			\put(6.2,1.95){....}
			\put(8,2){\line(1,0){1}}
			\put(10,2){\circle{2}}	\put(9.2,1.7){\scriptsize n-1}
			\put(11,2){\line(1,0){1}}
			\put(13,2){\circle{2}} 	\put(12.7,1.7){\scriptsize n}
		\end{picture}
		                                                                   &
		\begin{tabular}{l}
			$\ctext{1}~e_1-e_2$                 \\$\ctext{2}~e_2-e_3$\\
			~$\vdots$                           \\
			$\ctext{n-1}~e_{\rm n-1}-e_{\rm n}$ \\
			$\ctext{n}~e_{\rm n}-e_{\rm n+1}$
		\end{tabular}
		                                                                   & \begin{tabular}{l}
			                                                                     $e_{i}-e_{j}$,~	$_{1\le i\neq j\le n+1}$ \\
			                                                                     \lbrack n(n+1)\rbrack
		                                                                     \end{tabular}
		\\\hline
		$D_{\rm n}$                                                        &
		\setlength{\unitlength}{.1in}
		\thicklines
		\begin{picture}(18,7)
			\put(1,2){\circle{2}}	\put(0.65,1.7){\scriptsize 1}
			\put(2,2){\line(1,0){1}}
			\put(4,2){\circle{2}}  \put(3.65,1.7){\scriptsize 2}
			\put(5,2){\line(1,0){1}}
			\put(6.2,1.95){....}
			\put(8,2){\line(1,0){1}}
			\put(10,2){\circle{2}}	\put(9.2,1.7){\scriptsize n-3}
			\put(11,2){\line(1,0){1}}
			\put(13,2){\circle{2}} 	\put(12.2,1.7){\scriptsize n-2}
			\put(13,3){\line(0,1){1}}
			\put(13,5){\circle{2}}	\put(12.6,4.7){\scriptsize n}
			\put(14,2){\line(1,0){1}}
			\put(16,2){\circle{2}}	\put(15.2,1.7){\scriptsize n-1}
		\end{picture}
		                                                                   &
		\begin{tabular}{l}
			$\ctext{1}~	e_1-e_2$                  \\$\ctext{2}~e_2-e_3$\\
			~		$\vdots$                           \\
			$\ctext{n-2}~e_{\rm n-2}-e_{\rm n-1}$ \\
			$\ctext{n-1}~e_{\rm n-1}-e_{\rm n}$   \\
			$\ctext{n}~~e_{\rm n-1}+e_{\rm n}$
		\end{tabular}
		                                                                   & \begin{tabular}{l}
			                                                                     $\pm e_{i}\pm e_{j}$,~	$_{1\le i\neq j\le n}$
			                                                                     \\\lbrack	2n(n$-$1)\rbrack
		                                                                     \end{tabular}                                                           \\\hline
		$E_{\rm 6}$
		                                                                   & \setlength{\unitlength}{.1in}
		\thicklines
		\begin{picture}(15,7)
			\put(1,2){\circle{2}}	\put(0.65,1.7){\scriptsize 1}
			\put(2,2){\line(1,0){1}}
			\put(4,2){\circle{2}}  \put(3.65,1.7){\scriptsize 2}
			\put(5,2){\line(1,0){1}}
			\put(7,2){\circle{2}} 	\put(6.65,1.7){\scriptsize 3}
			\put(7,3){\line(0,1){1}}
			\put(7,5){\circle{2}}	\put(6.65,4.6){\scriptsize 6}
			\put(8,2){\line(1,0){1}}
			\put(10,2){\circle{2}}	\put(9.65,1.7){\scriptsize 4}
			\put(11,2){\line(1,0){1}}
			\put(13,2){\circle{2}}	\put(12.65,1.7){\scriptsize 5}
		\end{picture} &
		\begin{tabular}{l}
			$\ctext{1}~e_1-e_2$                   \\
			$\ctext{2}~e_2-e_3$                   \\$\ctext{3}~e_3-e_4$\\
			$\ctext{4}~e_4-e_5$                   \\
			$\ctext{5}~~\frac{1}{2}(-e_1-e_2-e_3$ \\
			~~~~~~$-e_4+e_5-\sqrt{3}e_6)$         \\
			$\ctext{6}~~\frac{1}{2}(-e_1-e_2-e_3$ \\
			~~~~~~$+e_4+e_5+\sqrt{3}e_6)$
		\end{tabular}
		                                                                   & \begin{tabular}{l}
			                                                                     $\pm e_{i}\pm e_{j}$,~	$_{1\le i\neq j\le 5}$ \\
			                                                                     \lbrack 5$\times$4$\times 2^2$/2=40\rbrack    \\\\
			                                                                     $\frac{1}{2}(\pm e_1\pm e_2\cdots
			                                                                     \pm e_5\pm\sqrt{3}e_6)$                       \\
			                                                                     odd number of $+$'s                           \\
			                                                                     \lbrack$2^6/2$=32\rbrack
		                                                                     \end{tabular}                                                          \\   \hline
	\end{tabular}
\end{table}
The dimension of an algebra is the sum of the rank and the number of roots.
Simple roots of the 5-th and 6-th nodes of the $E_6 $ Dynkin diagrams are spinor representation,
so removing these nodes reduce many dimensions of algebras.
Removing 5-th node brings ${\cal A}$-theory to ${\cal T}$-theory,
 reducing the R-R gauge fields.
Removing 6-th node brings ${\cal A}$-theory to ${\cal M}$-theory,
reducing the 3-form gauge field.
\vskip 6mm

\subsubsection{Dynkin diagram of $A$-symmetry of ${\cal A}$-theory}\label{section:2-2}
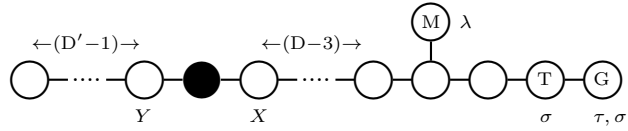
\begin{figure}[h!]
	\caption{Dynkin diagram of ${A}$-symmetry of ${\cal A}$-theory}
	\label{table:data_type3}
	\centering
	\begin{center}
		\setlength{\unitlength}{.1in}
		\thicklines
		\begin{picture}(35,10)
			\put(1,4){\circle{2}}		
			\put(2,4){\line(1,0){1}}
			\put(3.2,3.95){....}
			\put(1.1,5.6){\scriptsize $\leftarrow$(D$'-$1)$\rightarrow$}
			\put(5,4){\line(1,0){1}}
			\put(7,4){\circle{2}}		\put(6.5,1.8){\scriptsize $Y$}
			\put(8,4){\line(1,0){1}}
			\put(10,4){\circle*{2}}
			\put(11,4){\line(1,0){1}}
			\put(13,4){\circle{2}}		\put(12.5,1.8){\scriptsize $X$}
			\put(14,4){\line(1,0){1}}
			\put(15.2,3.95){....}
			\put(13.1,5.6){\scriptsize $\leftarrow$(D$-$3)$\rightarrow$}
			\put(17,4){\line(1,0){1}}
			\put(19,4){\circle{2}}		
			\put(20,4){\line(1,0){1}}
			\put(22,4){\circle{2}}		
			\put(22,5){\line(0,1){1}}
			\put(22,7){\circle{2}}
			\put(21.5,6.7){\scriptsize M}
			\put(23.5,6.7){\scriptsize $\lambda$}
			\put(23,4){\line(1,0){1}}
			\put(25,4){\circle{2}}	
			\put(26,4){\line(1,0){1}}
			\put(28,4){\circle{2}}
			\put(27.55,3.7){\scriptsize T}
			\put(27.7,1.8){\scriptsize  $\sigma$}
			\put(29,4){\line(1,0){1}}
			\put(31,4){\circle{2}}
			\put(30.5,3.7){\scriptsize G}
			\put(30.5,1.8){\scriptsize $\tau,\sigma$}
		\end{picture}
	\end{center}
\end{figure}

The Dynkin diagram of $A$-symmetry of ${\cal A}$-theory is given in Figure \ref{table:data_type3}. In general, the symmetry of Hamiltonian, $G$-symmetry, is enlarged to the symmetry of Lagrangian, $A$-symmetry,  where the timelike world-volume direction $\tau$ and the spacial world-volume directions $\sigma$'s are treated equally.
This extension increases one more node.
Manifestation of dualities, T and U-duality symmetries or S and U-dualities together in Lagrangian formulation, increases the number of nodes as shown in Figure \ref
{table:data_type}.
The 10-dimensional S-theory has GL(10) symmetry whose Dynkin diagram has 9 nodes,
then the manifestation of dualities increases them to $9+3=12$ nodes.

Fig. \ref{table:data_type3} is  the Dynkin diagram of 
A-symmetry of ${\cal A}$-theory with the D$'$-dimensional internal space
 \cite{Siegel:2018puf}.
Remove the node denoted as ``G", then
it becomes  $G$-symmetry of  ${\cal A}$-theory as shown in Table \ref{table:data_type}.
D+D$'$ = 10 is the critical dimension of S-theory which corresponds to 10 nodes. 
Together with nodes denoted by ``M" and ``T", the total number of nodes is 12.
The black node is dropped unless D$'$ = 0.
Removing node denoted by ``M" gives $G$-symmetry of ${\cal M}$-theory. 
Removing node denoted by ``T" gives $G$-symmetry of ${\cal T}$-theory.
Removing both nodes gives $G$-symmetry of S-theory.
By labeling nodes as $Y,X,\lambda$, $\sigma$, $(\tau,\sigma)$, we mean that these coordinates transform under the fundamental representations that correspond to the nodes. \footnote{ Each node corresponds to a simple root, and that simple root is the highest weight of its corresponding fundamental representation. }. $Y,X,\lambda$, $\sigma$ and  $(\tau,\sigma)$ are the spacetime coordinate, the internal coordinate, the parameter for $X$'s gauge invariance, the spacial world-volume coordinate and the $\tau\sigma$-covariant world-volume coordinate.

\subsection{Tables and diagrams}\label{section:2-3}

\subsubsection{Symmetry groups}
The $A, G, L, H$-symmetries of ${\cal A}$-theories in various dimensions D are listed in the following Table \ref{Table:2-1} and
Table \ref{Table:2-2}. Remember that $H$  ($G$) is the symmetries of current algebra in flat (curved) background, 
while  $L$ ($A$) is the symmetries of Lagrangian in flat (curved) background. $A$-symmetry reduces to $G$-symmetry in a fixed time direction $\tau$, while it reduces to $L$-symmetry in a flat background $|0\rangle$. Here D stands for the number of spacetime dimensions of S-theories, and d stands for the number of the world-volume dimensions.
\begin{table}[h!]
	\caption{Symmetry groups of ${\cal A}$-theories}\label{Table:2-1}
\begin{equation} 
\vcenter{\halign{\hfil#&\hfil#\hfil&#\hfil\cr
			& \textit{A} &\cr
			\raise.5em\hbox{$\tau$} $\swarrow$ & & $\searrow$ \raise.5em\hbox{$|0\rangle$}\cr
			\textit{G}\quad & & \quad  \textit{L} \cr
			$\searrow$\ & & $\swarrow$ \cr
			& \textit{H} &\cr}}
	\quad\quad\quad
	\vcenter{
		\halign{ # && \quad#\hfil \cr
			D & d & \textit{A} & \textit{G} & \textit{L} & \textit{H} \cr
			0 & 2 & GL(2) & GL(1) & GL(1, C) & I \cr
			1 & 3 & GL(3) & GL(2) & GL(2) & SO(1,1) \cr
			2 & 4 & SL(4)SL(2) & SL(3)SL(2) & SL(2)$^{2}$ & SL(2) \cr
			3 & 6 & SL(6) & SL(5) & SL(4) & Sp(4) \cr
			4 & 12 & SO(6,6) & SO(5,5) & SL(4,{\bf C}) & Sp(4,{\bf C}) \cr
			5 & 56 & E$_{7(7)}$ & E$_{6(6)}$ & U*(8) & USp(4,4) \cr
			6 & $\textrm{infinite}$ & $\textrm{infinite}$ & E$_{7(7)}$ & U*(8)$^2$ & SU*(8) \cr
			7 & $\textrm{infinite}$ & $\textrm{infinite}$ & E$_{8(8)}$ & U*(16) & SO*(16) \cr
		}
	}\nn
\end{equation}
\end{table}

\noindent As for $\mathcal{T}, \mathcal{M}$ and {$\mathcal{S}$ theories, their respective $A, G, L, H$-symmetries follow regular patterns which does not depend on dimensions. 
\begin{table}[h!]
	\caption{Symmetry groups of $\mathcal{T}, \mathcal{M}$ and $\mathcal{S}$-theories}\label{Table:2-2}
~~~~~~~~~~~~~~~~~~~~~~~~~~~~~~~~~	\begin{tabular}{c c c c c}
		& ~~~~~$A$ ~~~~~         & $G$   & $L$                 & $H$        \\ [0.5ex]
		$\mathcal{T}$ & O(D,D)SL(2)  & O(D,D) & O(D)$^{2}$SL(2)  & O(D)$^{2}$ \\
		$\mathcal{M}$  & GL(D+2)    & GL(D+1)& SO(D+2)          & SO(D+1)   \\
		$\mathcal{S}$  & GL(D)SL(2)   & GL(D)    & SO(D)SL(2)      & SO(D)
	\end{tabular} 
\end{table}

\subsubsection{Diamond diagram}
The closure of the brane extension of the Virasoro algebra 
gives rise to
the Gau\ss{} law constraint ${\cal U}=0$ and the  Laplace constraint ${\cal V}=0$.
Spacetime coordinate is reduced by solving 
Virasoro constraints ${\cal S}^m=0$ while 
world-volume coordinate is reduced by solving 
the Gau\ss{} law constraint ${\cal U}=0$,  
since the Virasoro operators are bilinears in the spacetime covariant derivatives
while the Gau\ss{} law operator is linear in both the spacetime and the world-volume derivatives.
Brane currents include the 0-modes and the non-zero modes,
then constraints of bilinears of 0-modes are section conditions
while constraints of linear of 0-modes are dimensional reductions. 
The section conditions  are the Virasoro constraints 
with replacing the spacetime covariant derivatives with their 0-modes as 
$\eta^{MNm}\partial_M\partial_N  \Phi(X)=0=
\eta^{MNm}(\partial_M \Phi)(\partial_N \Psi)$
on spacetime functions $\Phi(X)$ and $\Psi(X)$.
Solving the Virasoro constraints including the non-zero modes
reduces some dimensions.
Theories are related by sectionings as the following Diamond diagram.
\bea
&{\rm Diamond~diagram~of~}{\rm D}\mathchar`-{\rm dimensional~theories}&\nn\\\nn\\
&{{\cal A}\mathchar`-{\rm theory}}
&\nn\\
&\displaystyle\frac{G}{ H}=\frac{{\rm E}_{\rm D+1}}{ H}~&\nn\\
&{\cal S}~\swarrow  ~~~~~~~~~~~~~~~~~~~~\searrow ~{\cal U}&\nn\\
&{{\cal M}\mathchar`-{\rm theory}}
~~~~~~~~~~~~~~~~~~~~~~~~~~~~~~
{{\cal T}\mathchar`-{\rm theory}}
&\nn\\
&\displaystyle\frac{{\rm GL}({\rm D}+1)}{{\rm SO}({\rm D}+1)}~~~~~~~~~~~~~~~~~~~~~~~~~~
\displaystyle\frac{{\rm O}({\rm D},{\rm D})}{{\rm SO}({\rm D})^2}~&\nn\\
&{\cal U}~\searrow  ~~~~~~~~~~~~~~~~~~~~\swarrow~{\cal S}&\nn\\
&{{\rm S}\mathchar`-{\rm theory}}
&\nn\\
&\displaystyle\frac{{\rm GL}({\rm D})}{{\rm SO}({\rm D})}~&
\eea
\vskip 6mm

\subsubsection{Slug diagram}

The following slug-shaped diagram Fig. \ref{Slug} was introduced in \cite{Linch:2016ipx} indicating their index structure and the
corresponding $H$ groups for the physical spacetime dimension D mod 8. The bosonic spacetime coordinates $X$ and the world-volume coordinate $\sigma$ can be written as bispinors of $H$ group since fermions are spinor representations of $H$ group and bosonic coordinates appear on the right-hand side of the superalgebra.
The ovals indicate
the range of spinor indices, enclosing the dimensions (D) with equal ranges, before considering
reality properties; outside it increases by a factor of 2 when increasing D by 2. The range is
double that of the corresponding Lorentz group so starts with 2 for D = 1, 2.
The (spacetime) spinor size increases by a factor of 2 from one oval to the next. 
In the oval for 3 and 4, $X'$s are real and complex representations for D=3 and 4 theories respectively. 
$H'$ is the one for the D$'$-dimensional internal space with D+D$'$=10.  
 $Y$ is a bispinor of $H'$.
The primed indices are for extended supersymmetry for that dimension.

\begin{figure}[h]\caption{Slug diagram}\label{Slug}
	\includegraphics[width=8cm]{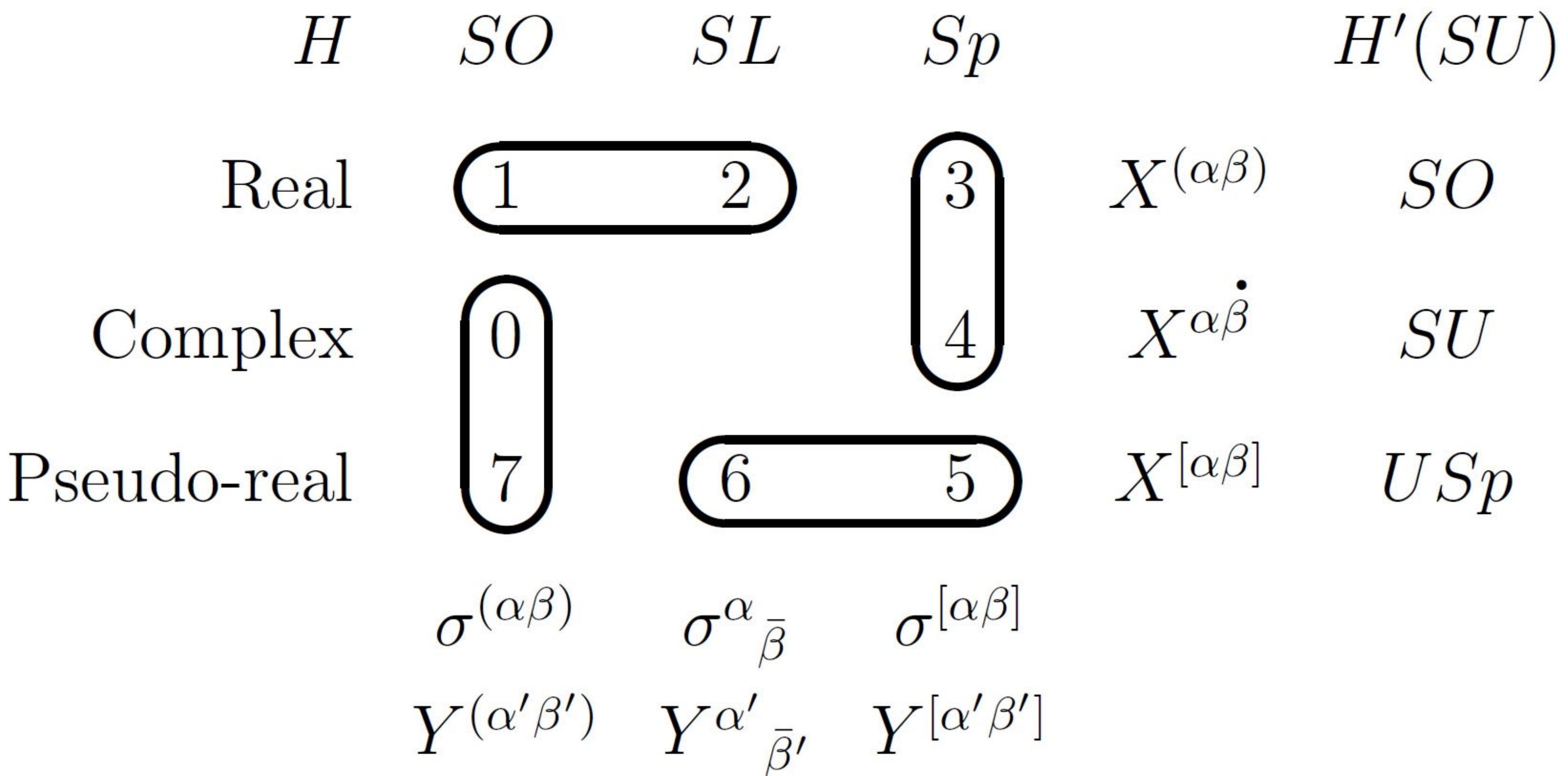}
	\centering
\end{figure}

As the exceptional group $G$-symmetry in the Hamiltonian formalism is enlarged to $A$-symmetry in the Lagrangian formalism, 
its tangent space symmetry $H$ is also enlarged to $L$-symmetry.
Going from the Lagrangian to the Hamiltonian formalism breaks $L$-symmetry (always a GL group) to $H$ 
by picking the matrix coefficient of $\tau$ in the matrix $\sigma$ to be a group metric for SO, SL, or Sp.

\subsection{Representations}\label{section:2-4}

The detailed formulations are highly dependent on their dimensions D. In this subsection, we list the representations of  
$A, L, G$-symmetry groups of ${\cal A}$-theory for world-volumes $\partial$, gauge parameters $\lambda$, spacetime coordinates $X$, and field strengths $F$ in various dimensions D.

The spacetime coordinate is the gauge field where the field strength $F=\partial X$ is invariant under the gauge transformation $\delta X=\partial \lambda$. 
Representations of $A$-symmetry of ${\cal A}$-theory world-volume fields in Lagrangian formulations are listed as follows. 
$${{\rm Representations~of}~{A}\mathchar`-{\rm symmetry ~of}~ {\cal A}\mathchar`-{\rm theory}}$$
$$
\vcenter{
\halign{ # & \quad# & \quad# & & \quad$# $\hfil \cr
D & $A$ & pattern & \partial& \lambda \mapright \partial& X \mapright \partial& F &  \cr
0 & GL(2) & forms & 2 & 0 & 1 & 2 & \cr
1 & GL(3) & forms & 3 & 0\oplus 1 & 1\oplus 3 & 3\oplus 3' & \cr
2 & SL(4)SL(2) & forms & (4,1) & (1,2) & (4,2) & (6,2) & \cr
3 & SL(6) & forms & 6 & 6 & 15 & 20 & \cr
4 & SO(6,6) & spinors & 12 & 32 & 32' & 32 & 1 \cr
5 & E$_{7(7)}$ & infinite & 56 & 912 & 133 & 56 & 133 \cr
}}
$$
\vskip 6mm

Choosing flat background breaks the $A$-symmetry to $L$-symmetry, analogously to that
GL(D) symmetry breaks to the Lorentz symmetry in a flat background.  
Representations of $L$-symmetry for ${\cal A}$-theory world-volume fields in Lagrangian formulations are listed as follows. 
$${{\rm Representations~of}~{L}\mathchar`-{\rm symmetry ~of}~ {\cal A}\mathchar`-{\rm theory}}$$
$$
\label{FtoL}
\vcenter{
\halign{ # & \quad# && \quad$#$\hfil \cr
D & $L$ & \partial& \lambda  & X & F \cr
0 & GL(1,{\bf C}) & 1\oplus 1^{-}&  0 & 1 & 1\oplus 1^{-} \cr
1 & GL(2) & 3 & 0\oplus 1 & 1\oplus 3 & 3\oplus 3 \cr
2 & GL(2)$^{2}$ & (2,2) & 2(1) & 2(2,2) & 2(1,3)\oplus 2(3,1) \cr
3 & GL(4) & 6 & 6 & 15 & 10\oplus 10' \cr
4 & GL(4,{\bf C}) & 6\oplus 6^{-} & 16\oplus 16' & 16_C \oplus  \overline{16}_C & 16\oplus 16' \cr
5 & U*(8) & 28\oplus 28' & 36\oplus 36'\oplus 420\oplus 420' & 63\oplus 70 & 28\oplus 28' \cr
}
} 
$$
 \vskip 6mm

In Hamiltonian formulation the world-volume covariance breaks to $\tau$ and $\sigma$ components, reducing $A$-symmetry to $G$-symmetry. The spacetime coordinate $X$ breaks to ($X_\tau$, $X_\sigma$)	as a representation of $G$-symmetry.
$X_\sigma$ is physical coordinate, while $X_\tau$ is auxiliary coordinate similar to $A_0$ of the usual Yang-Mills gauge field $A_m$. Representations of $G$-symmetry of ${\cal A}$-theory world-volume fields in Hamiltonian formulations are listed as follows.
$${{\rm Representations~of}~{G}\mathchar`-{\rm symmetry ~of}~ {\cal A}\mathchar`-{\rm theory}}$$
 $$ \vcenter{
	\halign{ # & \ #\hfil && \quad $#$\hfil \cr
		D & $G$ & \partial& \lambda  & X_{\tau} & X_{\sigma} & F \cr
		0 & GL(1) & 1\oplus 1 & 0 & 0 & 1 & 1\oplus 1 \cr
		1 & GL(2) & 1\oplus 2 & 1 & 1 & 1\oplus 2 & 2(1)\oplus 2(2) \cr
		2 & SL(3)SL(2) & 1\oplus (3,1) & (1,2) & (1,2) & (3,2) & (3,2)\oplus (3',2) \cr
		3 & SL(5) & 1\oplus 5 & 1\oplus 5 & 5 & 10 & 10\oplus 10' \cr
		4 & SO(5,5) & 2(1)\oplus 10 & 16\oplus 16' & 16 & 16' & 16\oplus 16' \cr
		5 & E$_{6(6)}$ & 2(1)\oplus 27\oplus 27' & 27\oplus 27'\oplus 2(78) & 27'\oplus 78 & 1\oplus 27 & 2(1)\oplus 27\oplus 27' \cr
		& & & 
  \oplus 351\oplus 351' & & \cr
	}
} $$
\vskip 6mm

For D$=$3 case, $G$-symmetry of ${\cal A}$-theory is SL(5), where representations of world-volume fields in Hamiltonian formulations are given as follows.
\bea 
&&
{\renewcommand{\arraystretch}{1.8}
\begin{array}{|c||c||c|c||c|}
	\hline
\partial&~~~~~\lambda~~~~~&~~~~X_\tau~~~~ &~~~~X_\sigma ~~~~&~~~~ F~~~~\\\hline
1\oplus 5&1\oplus 5&5&10&10\oplus 10'\\
\partial^{0}, \partial^{m},~_{m=1,\cdots,5}& \lambda^0, \lambda^m  &Y^m&X^{mn}&F_\tau{}^{mn}, F_\sigma{}_{mn}\\
\hline	\end{array}\nn}\\
&&{\renewcommand{\arraystretch}{1.8}
	\left\{\begin{array}{l}
F_\tau{}^{mn}=\dot{X}^{mn}-\partial^{[m}Y^{n]}\\
F_\sigma{}_{m_1m_2}=\frac{1}{2}\epsilon_{m_1\cdots m_5}\partial^{m_3}X^{m_4m_5}\end{array}\right. , ~
\left\{\begin{array}{l}
\delta_\lambda X^{mn}=\partial^{[m}\lambda^{n]}\\
\delta_\lambda Y^{m}=\dot{\lambda}^m-\partial^{m}\lambda^0 \end{array}\right. }
\label{SL5fields}
\eea
This is enlarged into the $A$-symmetry SL(6) covariant way. $A$-symmetry of D$=$3 ${\cal A}$-theory is SL(6), where representations of world-volume fields in Lagrangian formulations are given as follows.
\bea 
&&
{\renewcommand{\arraystretch}{1.8}
	\begin{array}{|c|c|c|c|c|}
		\hline
		\partial&~~~~~\lambda~~~~~&~~~~X ~~~~& ~~~~F~~~~\\\hline
	6&6&15&20\\
	\partial^{\hat{m}}, ~_{\hat{m}=0,\cdots,5}&\lambda^{\hat{m}}  &X^{\hat{m}\hat{n}}&F^{\hat{m}\hat{n}\hat{p}} \\		\hline	
 \end{array}}
\nn  \\
\nn\\
&&
F^{\hat{m}\hat{n}\hat{p}}=\frac{1}{2}\partial^{[\hat{m}}X^{\hat{n}\hat{p}]},~~
\delta_\lambda X^{\hat{m}\hat{n}}=\partial^{[\hat{m}}\lambda^{\hat{n}]}
\label{SL6fields}
\eea
\vskip 6mm

For D$=$4 case \cite{Hatsuda:2021wpb}, $G$-symmetry of ${\cal A}$-theory is SO(5,5), where representations of world-volume fields in Hamiltonian formulations are given as follows.
\bea 
&&
{\renewcommand{\arraystretch}{1.8}
	\begin{array}{|c||c||c|c||c|}
		\hline
		\partial&~~~~~\lambda~~~~~&~~~~X_\tau~~~~ &~~~~X_\sigma ~~~~&~~~~ F~~~~\\\hline
		2(1)\oplus 10&16\oplus 16&16&16'&16\oplus 16'\\
		\partial, \partial^{m},~_{m=1,\cdots,10}& \lambda_\mu, \bar{\lambda}^\mu  &Y_\mu&X^{\mu}, _{\mu=1,\cdots,16}.&	F_\tau{}^{\mu}, F_\sigma{}_{\mu}\\
		\hline	\end{array}\nn}\\
&&~~~~~~{\renewcommand{\arraystretch}{1.8}
		\left\{\begin{array}{l}
			F_\tau{}^{\mu}=\dot{X}^{\mu}-\gamma^{m\mu\nu}\partial_mY_\nu\\
			F_\sigma{}_{\mu}=\gamma_{m\mu\nu}\partial^m X^\nu\end{array}\right. , ~
		\left\{\begin{array}{l}
		\delta_\lambda X^{\mu}=\gamma^{m\mu\nu}\partial_m \lambda_\nu\\
		\delta_\lambda Y_{\mu}=\dot{\lambda}_\mu+
	\gamma_{m\mu\nu}\partial^m \bar{\lambda}^\nu \end{array}\right. }
\label{SO55fields}	
 \eea
$A$-symmetry of D$=$4 ${\cal A}$-theory is SO(6,6), where representations of world-volume fields in Lagrangian formulations are given as follows.
\bea 
&&
{\renewcommand{\arraystretch}{1.8}
	\begin{array}{|c|c|c|c|c|}

		\hline
		\partial&~~~~~\lambda~~~~~&~~~~X ~~~~& ~~~~F~~~~&~~~{\cal V}~~~\\\hline
		12&32&32'&32&1\\
		\partial^{\hat{m}}, ~_{\hat{m}=0,\cdots,11}&\lambda^{\underline{\mu}}, ~_{\underline{\mu}=1,\cdots,32} &X_{\underline{\mu}}&F^{\underline{\mu}}&{\cal V}\\
		\hline	\end{array}}\nn\\\nn\\
&&~~~~~
F^{\underline{\mu}}=\gamma^{\hat{m}\underline{\mu}\underline{\nu}}\partial_{\hat{m}}
X_{\underline{\nu}}
 , ~~\delta_\lambda X_{\underline{\mu}}=\gamma_{\hat{m}\underline{\mu}\underline{\nu}}\partial^{\hat{m}}
\lambda^{\underline{\mu}},~~
{\cal V}=\partial^{\hat{m}} \partial_{\hat{m}}
\eea

\section{Current algebras}\label{section:3}

In this section current algebras of gauge theory formulation of gravity theories with duality symmetries 
are described.	
Algebra of gauge covariant derivative presents a geometric description of a gauge theory
where the commutator of covariant derivatives gives the  field strength
and  the commutator of the covariant derivative and the gauge parameter gives  
the gauge transformation rule. 
After the brief review of ${\cal T}$-theory current algebras,
${\cal A}$-theory current algebras are explained.

\subsection{${\cal T}$-theory current algebras}\label{section:3-1}
	We begin by briefly reviewing the ${\cal T}$-theory current algebras. 
	The gauge theory of Einstein's gravity is based on covariance with respect to the general coordinate transformation.
	The local translation generator $p_m$ is the generator of the general coordinate transformation
	where  the Lie derivative of the vielbein field ${\mathfrak L}_{\xi} e_a{}^m$ 
	is given as 
	${\mathfrak L}_{\xi} e_a{}^m~p_m
	=i[\xi^n p_n, e_a{}^m p_m]$.
	This is generalized to the low energy effective gravity theory of string 
	by introducing the winding mode $\partial_\sigma x^m(\sigma)$ as the independent degrees of freedom to 
	the momentum $p_m(\sigma)$
	which make the covariant derivative $\dd_M(\sigma)$ for the doubled spacetime index $_M=(_m,{}^m)$. They satisfy the ${\cal T}$-theory current algebra which is O(D,D) covariant
	\bea
	\begin{array}{l}
		\\
		\left[\dd_{M}(1),\dd_{N}(2)\right]
		=2i\eta_{MN}\partial\delta(1-2)\\
	\end{array}
	\label{ODDCA}
	~~
	\begin{array}{c}
		\\,\end{array}
	~~
	\begin{array}{l}
		\quad\quad\quad~~~~~~~~~~~~~_n~~~~^n~~~~~~~~~~~~~~~~\\
		\eta_{MN}=
		\begin{array}{c}
			_m \\ ^m\end{array}
		\left(
		\begin{array}{cc}
			&\delta_m^n\\
			\delta_n^m& \end{array}
		\right)
	\end{array}.
	\eea
	where $\eta_{MN}$ is the O(D,D) invariant metric.
		Here $\sigma_1$, $\sigma_2$ are abbreviated as $1,2$ and 
	$\partial \delta(1-2)=\frac{\partial}{\partial \sigma_1}\delta(\sigma_1-\sigma_2)$.
	The generalized vielbein includes the NS-NS gauge field $B_{mn}$. 
	The new Lie derivative is obtained by  ${\mathfrak L}_{\Lambda} E_A{}^M~\dd_M(\sigma)
	=i[\int {\Lambda}^N \dd_N, E_A{}^M\dd_M(\sigma)]$ \cite{Siegel:1993bj}.

	The conformal symmetry of string theory  reflects the existence of Virasoro constraints where Virasoro operators generate world-volume 
	diffeomorphisms; generators of $\sigma$ and
	$\tau$ diffeomorphisms are    ${\cal S} = p_m\partial_\sigma x^m=0$ and
	${\cal H} = \frac{1}{2}(p_m){}^2+\frac{1}{2}(\partial_\sigma x^m)^2=0$ 
	respectively.
	The Virasoro constraints are generalized in ${\cal T}$-theory as
	\bea
	{\renewcommand{\arraystretch}{1.6}
		\left\{\begin{array}{ccl}
			{\cal S}&=&\frac{1}{4}\dd_M{\eta}^{MN}\dd_N=0\\
			{\cal H}&=&\frac{1}{4}\dd_M\hat{\eta}^{MN}\dd_N=0
		\end{array}\right.} ~~
	\eea
	with the doubled Minkowski metric $\hat{\eta}^{MN}$.
	The Virasoro algebra is given as follows:
	\bea
	{\renewcommand{\arraystretch}{1.6}
		\left\{\begin{array}{ccl}
			\left[{\cal S}(1),{\cal S}(2)\right]
			&=&{i}\left({\cal S}(1)+{\cal S}(2)\right)\partial\delta(1-2)\\
					\left[{\cal S}(1),{\cal H}(2)\right]
			&=&{i}\left({\cal H}(1)+{\cal H}(2)\right)\partial\delta(1-2)\\
					\left[{\cal H}(1),{\cal H}(2)\right]
			&=&{i}\left({\cal S}(1)+{\cal S}(2)\right)\partial\delta(1-2)
				\end{array}.\right.} 	
	\eea

	The spacetime coordinates and  momenta, $X^M=(x^m,~\tilde{x}_m)$ and $P_M=(p_m,~\tilde{p}^m)$  satisfy the following canonical commutator
	\bea
	[P_{M}(1),X^{N}(2)]&=&\frac{1}{i}\delta_{M}^{N}\delta(1-2)~~~.
	\eea
	The selfdual and anti-selfdual currents, $\dd_{M}$ and $\tilde{\dd}_{M}$, are written in terms of these coordinates as
	\bea
	{\renewcommand{\arraystretch}{1.8}
		\left\{\begin{array}{l}
			\dd_{M}=P_{M}+\eta_{MN}\partial_\sigma X^{N}\\
			\tilde{\dd}_{M}=P_{M}-\eta_{MN}\partial_\sigma X^{N}
		\end{array}.\right.}
	\eea
	The anti-selfdual current gives the selfduality condition
	\bea
	\tilde{\dd}_{M}=P_{M}-\eta_{MN}\partial_\sigma X^{N}=0
	\Leftrightarrow 
	\left\{
	\begin{array}{l}
		p_m=\partial_\sigma \tilde{x}^m\\
		\tilde{p}_m=\partial_\sigma {x}^m
	\end{array}
	\right.~~~.
	\eea
	The selfdual and anti-selfdual currents satisfy the following current algebras:
	\bea
	{\renewcommand{\arraystretch}{1.8}
		\left\{
		\begin{array}{lcl}
			\left[\dd_M(1),\dd_N(2)\right]
			&=&2i\eta_{MN}\partial\delta(1-2)\\
			\left[\dd_{M}(1),\tilde{\dd}_{N}(2)\right]
			&=&0
			\\
			\left[\tilde{\dd}_{M}(1),\tilde{\dd}_{N}(2)\right]
			&=&-2i\eta_{MN}\partial\delta(1-2)
		\end{array}\right.}.\label{CAODD}
	\eea

	The spacetime and the world-volume derivatives of a spacetime field $\Phi(X)$ are given by 
	\bea
	\frac{\partial}{\partial X^{M}}\Phi(X)=\partial_{M}\Phi(X)&=&
	i \displaystyle\int d\sigma'  [\dd_M(\sigma'),\Phi(X(\sigma))]\nn\\
	\frac{\partial}{\partial \sigma}\Phi(X(\sigma))=\partial_\sigma\Phi(X(\sigma))
	&=&i \displaystyle\int d\sigma'[({\cal S}-\tilde{\cal S})(\sigma'),\Phi(X(\sigma))]
	=\partial_\sigma X^M \left(\partial_M\Phi(X)\right)\label{sigmaderivative}
	\\
	&{\approx}&i \displaystyle\int d\sigma'[{\cal S}(\sigma'),\Phi(X(\sigma))]
	=\frac{1}{2}\dd_N\eta^{NM}\left(\partial_M\Phi(X)\right)~~~\nn
	\eea
	where the selfduality condition is used in the last line.

\subsection{D=3 ${\cal A}$-theory current algebras}\label{section:3-2}
In this section, we shall give a detailed description of D$=$3 current algebras of ${\cal A}$-theory.
Recall that the SL(5) $A$-symmetry covariant current in flat space is the 10-component  rank 2 anti-symmetric tensor $\dd_M(\sigma)=\dd_{m_1m_2}(\sigma)$, $M=1,\cdots, 10$ and $m_1, m_2=1,\cdots, 5$, which is a function on the 5-dimensional world-volume coordinate $\sigma_m$ with $m=1,\cdots, 5$.
The SL(5) current algebra is given by
\bea \left[\dd_{m_1m_2}(1),\dd_{m_3 m_4}(2)\right]
=2i\epsilon_{m_1\cdots m_5}\partial^{m_5}\delta(1-2)\label{SL5CA1}
\eea
with $\partial^m \delta(\sigma)=\frac{\partial}{\partial\sigma_m}\delta^{(5)} (\sigma)$.
$\epsilon_{m_1 \cdots m_5}$ is the SL(5) group invariant metric $\eta_{MN\rho}$.
The current algebra \bref{SL5CA1} is covariant under the SL(5) transformation
SL(5)$\ni{\cal A}_{m_1m_2}{}^{n_1n_2}$, $A_m{}^n$
\bea
\dd_{m_1m_2}\to\frac{1}{2}{\cal A}_{m_1m_2}{}^{n_1n_2}\dd_{n_1n_2}
~~,~~\partial^m\to \partial^n A_{n}{}^m~~,
\eea
with
\bea
\frac{1}{4}\epsilon^{m_1\cdots m_5}{\cal A}_{m_1m_2}{}^{n_1n_2}{\cal A}_{m_3m_4}{}^{n_3n_4}
=\epsilon^{n_1\cdots n_5}{A}_{n_5}{}^{m_5}~~~.
\eea
${\cal A}_{m_1m_2}{}^{n_1n_2}$ and $A_{m}{}^n$ are expressed by each other as
\bea
&{\cal A}_{m_1m_2}{}^{n_1n_2}=A_{[m_1}{}^{n_1} A_{m_2]}{}^{n_2}&\label{10to5}\\
&A_{m}{}^n=\frac{1}{32}\epsilon_{mm_1\cdots m_4}\epsilon^{nn_1\cdots n_4}{\cal A}_{n_1n_2}{}^{m_1m_2} {\cal A}_{n_3n_4}{}^{m_4m_4}~~~& \nn
\eea
with SL(5) condition on $A_m{}^n$
\bea
\epsilon^{m_1\cdots m_5}A_{m_1}{}^{n_1} A_{m_2}{}^{n_2}  A_{m_3}{}^{n_3} A_{m_4}{}^{n_4}A_{m_5}{}^{n_5}
=\det A ~\epsilon^{n_1\cdots n_5},~\det A=1,~A_{m}{}^{n}\in{\rm SL(5)}~~~.\label{SL5}
\eea

The currents $\dd$ and $\tilde{\dd}$ can be expressed in terms of spacetime coordinates $X^M=X^{m_1m_2}$ and $P_M=P_{m_1m_2}$. They satisfy the following canonical commutator
\bea
[P_{m_1m_2}(1),X^{n_1n_2}(2)]&=&\frac{1}{i}\delta_{[m_1}^{n_1}\delta_{m_2]}^{n_2}\delta(1-2)
\eea
The selfdual and anti-selfdual currents, $\dd_{mn}$ and $\tilde{\dd}_{mn}$, are written in terms of these coordinates as
\bea
{\renewcommand{\arraystretch}{1.8}
	\left\{\begin{array}{l}
		\dd_{m_1 m_2}=P_{m_1 m_2}+\frac{1}{2}\epsilon_{m_1\cdots m_5}\partial^{m_3} X^{m_4 m_5} \\
		\tilde{\dd}_{m_1 m_2}=P_{m_1 m_2}-\frac{1}{2}\epsilon_{m_1\cdots m_5}\partial^{m_3} X^{m_4 m_5}
	\end{array}\right.}
\eea
They satisfy the following current algebras
\bea
{\renewcommand{\arraystretch}{1.8}
	\left\{
	\begin{array}{lcl}
		\left[\dd_{m_1m_2}(1),\dd_{m_3 m_4}(2)\right]
		 & = & 2i\epsilon_{m_1\cdots m_5}\partial^{m_5}\delta(1-2)  \\
		\left[\dd_{m_1m_2}(1),\tilde{\dd}_{m_3 m_4}(2)\right]
		 & = & 0
		\\
		\left[\tilde{\dd}_{m_1m_2}(1),\tilde{\dd}_{m_3 m_4}(2)\right]
		 & = & -2i\epsilon_{m_1\cdots m_5}\partial^{m_5}\delta(1-2)
	\end{array}\right.}\label{CASL5}
\eea

Virasoro constraints ${\cal S}^m={\cal H}=0$ and the Gau\ss{} law constraint ${\cal U}_m=0$ which is required by the closure of the Virasoro algebra are given by 
\bea
{\renewcommand{\arraystretch}{1.8}
	\left\{\begin{array}{ccl}
		{\cal S}^{m} & = & \frac{1}{16}\dd_{m_1m_2}\epsilon^{mm_1\cdots m_4}\dd_{m_3m_4}=0        \\
		{\cal H}     & = & \frac{1}{16}\dd_{m_1m_2}\delta^{m_1[m_3}\delta^{m_4]m_2}\dd_{m_3m_4}=0 \\
		{\cal U}_{m} & = & \partial^{n}\dd_{mn}=0
	\end{array}\right.}\label{Virasoros}~~.
\eea
They are bilinears in the spacetime currents and the world-volume derivative
contracted with the ${\cal A}$-symmetry invariant tensor $\eta^{MNm}$, the $H$-symmetry invariant metric $\hat{\eta}^{MN}$ and their combination
$U^M{}_N{}^m{}_n=\eta^{MLm}\eta_{LNn}-\delta^{M}_N\delta{}^m_n$
 as
\bea
{\renewcommand{\arraystretch}{1.8}
	\left\{\begin{array}{cclcl}
		{\cal S}^{m} & = & \frac{1}{4}\dd_M\eta^{MNm}\dd_N=0                            & , & \eta^{MNm}=\epsilon^{m_1m_2n_1n_2m} \\
		{\cal H}     & = & \frac{1}{4}\dd_M\hat{\eta}^{MN}\dd_N=0                       & , & \hat{\eta}^{MN}
		=\delta^{m_1[n_1}\delta^{n_2]m_2}                                                                                         \\
		{\cal U}_{m} & = & \frac{1}{4}\partial^l U^{N}{}_M{}^n{}_l \dd_N \delta_n^{m_2}=0 & , &
		U^N{}_M{}^n{}_l=\delta^{[n_1}_l\delta^{n_2]}_{[m}\delta_{m_2]}^n
	\end{array}\right.}\label{Ddeta}~~.
\eea

The Virasoro algebra is given as follows:
\bea
\left[{\cal S}^m(1),{\cal S}^n(2)\right]
&=&\frac{i}{2}\left[\left({\cal S}(1)+{\cal S}(2)\right){}^{(m}\partial{}^{n)}\delta(1-2)\right.\nn\\
&&\left.~~~~+\delta(1-2)\left(\partial^{[m}{\cal S}^{n]}+\frac{1}{2}\epsilon^{mnl_1l_2l_3}\dd_{l_1l_2}{\cal U}_{l_3}\right)\right]\\
&=&i\left[{\cal S}^{(m}(1)\partial^{n)}\delta(1-2)\right.\nn\\
	&&\left.~~~~+\delta(1-2)\left(\partial^{n}{\cal S}^{m}+\frac{1}{4}\epsilon^{mnl_1l_2l_3}\dd_{l_1l_2}{\cal U}_{l_3}\right)\right]\nn\\
\left[{\cal S}^m(1),{\cal H}(2)\right]
&=&{i}\left[\left\{\left({\cal H}(1)+{\cal H}(2)\right)\partial^{m}
+\frac{1}{2}\left(\dd^{mn}\vec{\cal U}_n(1)-\dd^{mn}\vec{\cal U}_n(2)\right)\right\}\delta(1-2)\right.\nn\\
&&\left.~~~~-\frac{1}{2}\delta(1-2)\dd^{mn}{\cal U}_n\right]\\
&=&{i}\left[\left\{2{\cal H}(1)\partial^{m}
	+\dd^{mn}\vec{\cal U}_n(1)\right\}\delta(1-2)\right.\nn\\&&
	\left.~~~~+\delta(1-2)\left(\partial^m{\cal H}-\frac{1}{2}\vec{\cal U}_n\dd^{mn}\right)\right]\nn\\
\left[{\cal H}(1),{\cal H}(2)\right]
&=&{i}\left({\cal S}^m(1)+{\cal S}^m(2)\right)\partial_{m}\delta(1-2)\\
&=&{i}\left[2{\cal S}^m(1)\partial_{m}\delta(1-2)+\delta(1-2)\partial_m{\cal S}^m\right]\nn\\
\left[{\cal S}^m(1),{\cal U}_n(2)\right]
&=&0\nn\\\left[{\cal H}(1),{\cal U}_m(2)\right]
&=&0\nn\\
\left[{\cal U}_m(1),{\cal U}_n(2)\right]
&=&0\nn~~~.
\eea

In this computation, we have used the following derivative rules of spacetime and world-volume in order to obtain a set of closed algebras in the space
generated by only $\dd_M$ in \bref{SL5CA1}.
\bea
\frac{\partial}{\partial X^{mn}}\Phi(X)&=&\partial_{mn}\Phi(X)=i \displaystyle\int d^{5}\sigma'  [\dd_{mn}(\sigma'),\Phi(X(\sigma))  ]\nn\\
\frac{\partial}{\partial\sigma_m}\Phi(X(\sigma))&=&\partial^{m}\Phi(X(\sigma))\approx i \displaystyle\int d^5\sigma'[{\cal S}^m(\sigma'),\Phi(X(\sigma))]\nn\\
&=&\frac{1}{8}\epsilon^{mm_1\cdots m_4}\dd_{m_1m_2}\left(\partial_{m_3m_4}\Phi(X(\sigma))\right)~~~.\label{delsigma}
\eea
We now identify $\sigma$ derivative with the obtained constraint ${\cal U}_m=0$ as well as $\tilde{\cal S}^m=\frac{1}{4}\tilde{\dd}_M\eta^{MNm}\tilde{\dd}_N=0$  	as
\bea
\partial^{m}\Phi(X(\sigma))&=&i \displaystyle\int d^5\sigma'[
{\cal S}^m(\sigma')-
\tilde{\cal S}^m(\sigma')
,\Phi(X(\sigma))]-(\partial^{n}X^{ml})
\left(\partial_{nl}\Phi(X)\right)\nn\\
&=&i \displaystyle\int d^5\sigma'[
{\cal S}^m(\sigma')-
\tilde{\cal S}^m(\sigma')
,\Phi(X(\sigma))]-{\cal U}_l \left(X^{ml}(\sigma)\Phi(X)\right)
~~~.\nn\\
\eea

\vskip 6mm

\subsection{${\cal A}$-theory current algebra in general dimensions}\label{section:3-3}
In this section, we will start to formulate the current algebras and the constraints in general spacetime dimensions in (d+1)-dimensional world-volume. Our starting point will be the canonical target space coordinate and momentum:
\begin{equation}
	\left[ X^{M}(1), P_{N}(2) \right] = i\delta^{M}_{\enspace N}\delta (1-2) 
\end{equation}
with 
$\delta(1-2)=\delta^{\rm (d)}(\sigma_1-\sigma_2)$.
The current algebra generators are defined as 
\begin{equation}
	\dd_{N}(1) = P_{N} + \eta^{n}_{\enspace NM}\partial_{n}X^{M}(1), \quad \widetilde{\dd}_{N}(1) = P_{N} - \eta^{n}_{\enspace NM}\partial_{n}X^{M}(1).
\end{equation}
 $\eta^{n}_{\enspace NM}$ is an invariant tensor of $G$, which corresponds to the projection tensor $R_{1}\otimes R_{1} \rightarrow R_{2}$.
We shall impose the condition such that $\eta$ is symmetric in bosonic indices $NM$, which means $R_{2}$ needs to be in the symmetric product of $R_{1}\otimes R_{1} $ for $R_1$ index $M,N$.
Their commutators are
\begin{eqnarray}
	\left[ \dd_{N}(1), \dd_{M}(2)  \right] &=& 2i\eta^{n}_{\enspace NM}\partial_{n}\delta (1-2) \nonumber \\
	\left[\widetilde{\dd}_{N}(1), \widetilde{\dd}_{M}(2) \right] &=& -2i\eta^{n}_{\enspace NM}\partial_{n}\delta (1-2)\\
	\left[{\dd}_{N}(1), \widetilde{\dd}_{M}(2) \right] &=&0\nonumber \label{CAbrane}
\end{eqnarray}
with
$\partial_n\delta(1-2)=\frac{\partial}{\partial \sigma_1{}^n}\delta^{\rm (d)}(\sigma_1-\sigma_2)$.
\vskip 6mm

\subsection{Virasoro and Gau\ss{} law constraints for branes}\label{section:3-4}

The constraints of current algebra in principle could be derived from the ${\cal A}$-theory Lagrangian. But with certain constraints given, others can be derived by the consistency of the closure of 
the constraint algebra.
Here we begin by the U-duality covariant Virasoro constraints as a starting point.

In order for the ${\cal A}$-theory to be reduced to string theory, we need a generalization of the Virasoro constraints $\mathcal{S}_m$=0 which will generate the space-like diffeomorphisms of the world-volume, and reduce to ordinary Virasoro constraints on reduction to string theory. We also have 
${\cal H}=0$ constraint 
of ${\cal A}$-theory, which serves as a constraint that generates the time-like diffeomorphism of the world-volume.
Virasoro constraints are given as 
\begin{eqnarray}\label{vcons}
	\mathcal{S}_{m}  & = & \dd_{M}\eta_{m}^{ \; MN}\dd_{N}~=~0\\
	\mathcal{H}  & = & \dd_{M}\hat{\eta}^{{MN}}\dd_{N}~=~0
\end{eqnarray}
Note that here $\hat{\eta}^{{NM}}$ is the invariant metric for the isotropic group $H$. The Hamiltonian is sum of these Virasoro operators as the first class constraints with Lagrange multipliers. These multipliers become p+1 components of the world-volume metric. 

From the commutators of these constraints, one can find
Gau\ss{} law constraints:
\begin{eqnarray}\label{uvconstriants}
	\mathcal{U}^{n}_{\;P} & = & (U^{n}_{\;m})_{P}^{\;M}\partial^{m}\dd_{M}~=~0\\
	(\mathcal{V}_{n})_{\;MN} & = & (V_{n}^{\;\; pq})_{MN}\partial_{p}\partial_{q}~=~0~,
\end{eqnarray}
where the invariant tensors $U$ and $V$ are
\begin{eqnarray}
	(U^{n}_{\;m})_{P}^{\;M} & = & \delta^{n}_{\; m} \delta_{P}^{\;M} - \eta^{n}_{\enspace PQ}\eta_{m}^{\enspace QM}\\
(V_n{}^{pq})_{MN} & = & U^{p~~~P}_{\;nM}\eta^{q}_{\enspace NP}~.
\end{eqnarray}

There is also another set of constraints constructed written by anti-selfdual currents where we just replace every $\dd$ with $\tilde{\dd}$. But for now, we will focus on one sector of the constraints since they are independent of each other as shown in \bref{CAbrane}.

To verify all these constraints are closed, first calculate $[\dd_{N}, \mathcal{S}]$.
$$
	[\dd_{N}(1), \mathcal{S}_{m}(2)] = 2i\dd_{P}(2)\eta_{m}^{\enspace PQ}\eta^{n}_{\enspace QN}\partial_{n}\delta (1-2).
$$
Thus
\bea	\label{eq:10}
	[\mathcal{S}_{n}(1), \mathcal{S}_{m}(2)] &=& 4i \dd_{M}(1)(\eta_{m}\cdot \eta^{p} \cdot \eta_{n})^{MN}\dd_{N}(2)\partial_{p}\delta (1-2)\\
	\label{eq:11}
	[\mathcal{S}_{n}(1), \mathcal{H}(2)] &= &4i \dd_{M}(1)(\eta_{n}\cdot \eta^{p})^{M}_{\quad N} \dd^{N}(2)\partial_{p}\delta (1-2)\\
	\label{eq:12}
	[\mathcal{H}(1), \mathcal{H}(2)] &=& 4i \dd^{M}(1)\eta^{p}_{\enspace NM}  \dd^{N}(2)\partial_{p}\delta (1-2),
\eea
Since most invariant tensors $\eta^p{}_{NM}$ in the expression above have two $R_{1}$ indices, we view them as matrix $(\eta^p)_{NM}$ and avoid writing the explicit 
contractions. The contraction rule of $R_{1}$ indices should be obvious to the reader. The $R_{1}$ indices are raised the lowered by the 
$H$-invariant tensor $\hat{\eta}_{NM}$. So the $[\mathcal{H}, \mathcal{H}]$ and $[\mathcal{S}_{m}, \mathcal{H}]$ commutators are
are only $H$-invariant instead of $G$-invariant.

Note that when the derivative on the delta function is multiplied by a bi-local function 
(i.e. the function product of two functions on different points in space), the expression is ambiguous.  
 We have the following equation:
\begin{align}\label{eq:13}
	  & f(x)g(y)\delta^{'}(x - y) 	\nonumber                                                                      \\
	= & \left[ \frac{1}{2}(fg|_{x}+ fg|_{y}) + \frac{s}{2}(fg|_{x} -  fg|_{y}) \right]\delta^{'}(x - y) \nonumber \\
	+ & \left[\frac{1}{2}(g^{'}f - f^{'}g) + \frac{s}{2}\left(gf^{'} + fg^{'}\right) \right]\delta(x- y),
\end{align}
where $s$ is an arbitrary number and $\delta'(x-y)=\partial_x\delta(x-y)$. 
Fortunately in  our case at equation \bref{eq:10} and \bref{eq:12}, we know that the right-hand side is anti-symmetric with respect 
to interchange in $1 \leftrightarrow 2$, and the part that is dependent on $s$ are precisely symmetric in $1 \leftrightarrow 2$. 
Therefore, it is convenient to choose set $s = 0$.
Later we shall see that this ambiguity plays an important role 
for the calculation of the gauge transformation
which is the commutator of the gauge field and the integral of the current.
The left-hand side of the bracket is not necessarily anti-symmetric since one of the coordinates is integrated over.

A quick derivation of (\ref{eq:13}) might be helpful: The idea is that when a bi-local function is multiplied by $\delta{}'$, we could use the relation $\partial_{x}\delta (x- y) = -\partial_{y}\delta(x - y)$ to write the same expression in two different ways.
\begin{align*}
	  & f(x)g(y)\partial_{x}\delta (x - y)                                                           \\
	= & f(x)g(x)\partial_{x}\delta (x - y) + f(x)\partial_{x}g(x)\delta (x - y) \quad \textrm{(I)}   \\
	= & f(y)g(y)\partial_{x}\delta (x - y) - g(x)\partial_{x}f(x)\delta (x - y) \quad \textrm{(II)}.
\end{align*}
Therefore, the linear combination $\left(\frac{1}{2} + s\right) \cdot (\textrm{I}) + \left(\frac{1}{2} - s\right) \cdot (\textrm{II})$ gives 
equation  (\ref{eq:13}).

Going back to the commutators, using $\eta^{p}\eta_{n} = \delta^{p}_{n} - U^{p}_{n}$, we get
\bea
( \dd(1)\eta_{n}\cdot\eta^{p}\cdot\eta_{m}\dd(2)) \partial_{p} \delta 
= (\dd(1) \eta_{n} \dd(2)) \partial_{m} \delta - (\dd(1)\eta_{n} \cdot U^{p}_{m}\dd(2)) \partial_{p} \delta,
\eea
and
\bea
( \dd(1)\eta_{n}\cdot\eta^{p}\dd(2)) \partial_{p} \delta 
= (\dd(1) \dd(2)) \partial_{n} \delta - (\dd(1) U^{p}_{n} \dd(2)) \partial_{p} \delta .
\eea
Terms above that contain $U^{p}_{n}$ can be written in terms of the $\mathcal{U}$ and $\mathcal{V}$ constraints.
\bea
(U^{p}_{\enspace n})_{N}^{\enspace M}\dd_{M}(1)\partial_{p}\delta = 
(U^{p}_{\enspace n})_{N}^{\enspace M} \left[P_{N}(1)\partial_{p} + \eta^{q}_{\enspace MS}\partial_{q}X^{S}(1)\partial_{p} \right]\delta
= \mathcal{U}(P, \partial)\delta + \mathcal{V}(\partial, \partial)X\delta.
\eea
Here, the $\mathcal{U}$ and $\mathcal{V}$ constraints are weak constraints, thus the partial derivatives could act on anything.

Applying (\ref{eq:13}) to the terms that does not contain $U^{p}_{n}$, we get the commutators of $\mathcal{S}$ and $\mathcal{H}$ up to
$\mathcal{U}$, $\mathcal{V}$ constraints.
\bea
[\mathcal{S}_{n}(1), \mathcal{S}_{m}(2)] &\sim&
2i \left[ \mathcal{S}_{[m}(1) + \mathcal{S}_{[m}(2)\right]\partial_{n]}\delta (1-2) 
+ 4i\partial_{[m} \mathcal{S}_{n]}
\delta(1-2)
\\
\left[\mathcal{S}_{n}(1), \mathcal{H}(2)\right] &\sim& 2i\left[\mathcal{H}(1) + \mathcal{H}(2)  \right]\partial_{n}\delta(1-2)
\\
\left[\mathcal{H}(1), \mathcal{H}(2)\right] &=&
2i\left[\mathcal{S}^{m}(1) + \mathcal{S}^{m}(2)\right]\partial_{m}\delta(1-2).
\eea
 $\mathcal{H}$ and $\mathcal{S}_m$ constraints generate the world-volume diffeomorphisms. Although this is not the usual
diffeomorphisms that one might expect. The existence of $\mathcal{U}$ constraints in the commutators means that these diffeomorphisms
also transform the target space coordinate itself.

The infinitesimal transformation of world-volume diffeomorphisms is generated by 
\bea
\delta^{(\sigma)}_{\xi} = i\int d^{d}\sigma ~\xi^{m} \mathcal{S}_{m}, \quad \delta^{(\tau)}_{\zeta} = i\int d^{d}\sigma ~\zeta \mathcal{H}.
\eea
Their commutators are (up to $\mathcal{U}$, $\mathcal{V}$)
\bea
[\delta^{(\tau)}_{\zeta_{1}}, \delta^{(\tau)}_{\zeta_{2}}] \sim  \delta^{(\sigma)}_{\xi_{12}}, &\quad& 
\xi_{12}^{m}=\frac{1}{4}\left( \zeta_{1}\partial^{m}\zeta_{2}-\zeta_{2}\partial^{m}\zeta_{1}\right)\nn
\\
\left[\delta^{(\tau)}_{\zeta_{1}}, \delta^{(\sigma)}_{\xi_{2}}\right] \sim  \delta^{(\tau)}_{\zeta_{12}}, &\quad &
\zeta_{12} = \frac{1}{4}\left( \zeta_{1}\partial_{m}\xi_{2}^{m}-\xi_{2}^{m}\partial_{m}\zeta_{1}\right)
\\
\left[\delta^{(\sigma)}_{\xi_{1}}, \delta^{(\sigma)}_{\xi_{2}}\right] \sim  \delta^{(\sigma)}_{\eta_{12}}, &\quad &
\eta_{12}^{m} = \frac{1}{4}\left( \xi_{1}^{n}\partial_{n}\xi_{2}^{m} - \xi_{2}^{n}\partial_{n}\xi_{1}^{{m}} \right)~.\nn
\eea
\vskip 6mm

\subsection{Courant and Dorfman brackets }\label{section:3-5}
The Courant bracket, generalized derivatives, and generalized Lie derivatives can be found by adding background for the current algebra generators.
Let us define the generator
\begin{equation}
	Z(\Lambda) = i\int d^{\rm{d}}\sigma \Lambda^{N}(X)\dd_{N}
\end{equation}
Note that $Z(\Lambda)$ generates translation for the generalized coordinates.
\begin{equation}
	\left[ Z(\Lambda), \Phi(X) \right] = \Lambda^{N}\partial_{N}\Phi(X), \quad \partial_{N} = \frac{\partial}{\partial X^{N}}
\end{equation}
The commutators of the generator give Courant bracket. i.e.
\begin{equation}\label{ccommutator}
	[ Z(\Lambda_{1}), Z(\Lambda_{2})] \sim Z(\Lambda_{12}).
\end{equation}
Note that the $\sim$ implies that the equality is only up to the constraints. In the paragraphs below, if the context is clear, we will replace $\sim$ with the equal sign.
\begin{equation}
	\Lambda_{12}^{Q} \equiv [\Lambda_{1}, \Lambda_{2} ]_{C}^{Q} = \Lambda_{1}^{N}\partial_{N}\Lambda_{2}^{Q} + \frac{1}{2}Y^{PQ}_{\quad MN}\partial_{P}\Lambda_{1}^{M}\Lambda_{2}^{N} - (1 \leftrightarrow 2)
\end{equation}
is the Courant bracket in ${\cal A}$-theory. The invariant tensor is given as 
\begin{equation}
    Y^{\; PQ}_{\quad MN} = 2\eta_{m}^{PQ}\eta^{m}_{\enspace MN}.
\end{equation} 

The derivation is pretty straightforward, but one has to keep in mind that $\partial^m\delta(1-2)$ is defined as a distribution and needs special care.
\bea
	[\Lambda_{1}^{N}\dd_{N}, \Lambda_{2}^{M} \dd_{M}] =  -i\left( \Lambda_{1}^{N} \partial_{N} \Lambda_{2}^{M} - \Lambda_{2}^{N} \partial_{N} \Lambda_{1}^{M} \right)\dd_{M}\delta (1-2) + 2i\Lambda_{1}^{N}(1)\Lambda_{2}^{M}(2)(\eta^{m})_{NM} \partial_{m}\delta(1 - 2)\nn\\
\eea
The second term is multiplied by a derivative of the delta function. 
Apply equation \bref{eq:13} we get,
\bea\label{eq:1}
	\Lambda_{1}^{N}(1)\Lambda_{2}^{M}(2)\eta^{m}_{\enspace NM} \partial_{1m}\delta(1 - 2) &=& \left(\alpha \partial_{m}\Lambda_{1}^{N}\eta^{m}_{\enspace NM}\Lambda_{2}^{M}- \beta \partial_{m}\Lambda_{2}^{N}\eta^{m}_{\enspace NM}\Lambda_{1}^{M} \right)\delta( 1 - 2)   \\
 &&
 +\alpha \Lambda_{1}^{N}(1)\eta^{m}_{\enspace NM}\Lambda_{2}^{M}(1)\partial_{1m}\delta( 1 - 2)
 + \beta \Lambda_{2}^{N}(2)\eta^{m}_{\enspace NM}\Lambda_{1}^{M}(2) \partial_{1m}\delta( 1 - 2),
\nn
\eea
where $\alpha+\beta=1$. $\alpha = (1+ s)/2$ and $\beta = (1-s)/2$.

We shall rewrite $\Lambda \partial_{m} \Lambda$ terms
in such a way that
the derivative with respect to world-volume coordinates
is expressed by the target space derivative and currents. 
 Observe that
\bea
	\frac{1}{2}\left( \dd_{N} - \tilde{\dd}_{N}\right) & =  &\eta^{m}_{\enspace NM}\partial_{m}X^{M}                                                                                                                                                            \nn   \\
	&\Rightarrow                                        & \frac{1}{2}\eta_{m}^{\enspace MN}\left( \dd_{N} - \tilde{\dd}_{N}\right) = \eta_{m}^{\enspace MP}\eta^{n}_{\enspace PN}\partial_{n}X^{N} = \partial_{m}X^{M} +  (U^{n}_{\;m})_{P}^{\;M}\partial_{n}X^{P} \nn\\
	&\Rightarrow                                        & \partial_{m}X^{M} = \frac{1}{2}\eta_{m}^{\enspace MN}\left( \dd_{N} - \tilde{\dd}_{N}\right) -  (U^{n}_{\;m})_{P}^{\;M}\partial_{n}X^{P}.
\eea
And,
\bea
	\partial_{m}\Lambda_{1}^{N}\eta^{m}_{\enspace NM}\Lambda_{2}^{M} & =&  \partial_{m}X^{P}\partial_{P}\Lambda_{1}^{N}\eta^{m}_{\enspace NM}\Lambda_{2}^{M}               \nn \\
	&\Rightarrow                                                      & \frac{1}{2}\left( \dd_{N} -\tilde{\dd}_{N}\right)\partial_{M}\Lambda_{1}^{P}\eta_{m}^{\enspace MN}\eta^{m}_{\enspace PQ}\Lambda_{2}^{Q} - \partial_{n}X^{Q}(U^{n}_{\;m})_{Q}^{\;P}\partial_{P}\Lambda_{1}^{N}\eta^{m}_{\enspace NM}\Lambda_{2}^{M} \nn\\
	&\Rightarrow                                                      & \frac{1}{4}\left( \dd_{N} -\tilde{\dd}_{N}\right) \partial_{M}\Lambda_{1}^{P}Y^{MN}_{\quad PQ}\Lambda_{2}^{Q} -(\mathcal{U}_{\;m})_{Q}X^{Q} \Lambda_{1}^{N}\eta^{m}_{\enspace NM}\Lambda_{2}^{M},
\eea
where
\bea
	\partial_{m}\Lambda_{1}^{N}\eta^{m}_{\enspace NM}\Lambda_{2}^{M} \sim \frac{1}{4}\partial_{M}\Lambda_{1}^{P}Y^{MN}_{\quad PQ}\Lambda_{2}^{Q}\dd_{N}, \quad \textrm{mod } \mathcal{U}, \tilde{\dd}.
\eea
Expression of the commutator up to constraints are
\bea \label{commresults}
	[\Lambda_{1}^{N}\dd_{N}(1), \Lambda_{2}^{M} \dd_{M}(2)] & = &\frac{1}{i}\left( \Lambda_{1}^{N} \partial_{N} \Lambda_{2}^{M} - \Lambda_{2}^{N} \partial_{N} \Lambda_{1}^{M} \right)\dd_{M}\delta (1-2)                                                \nn\\
	&&-\frac{i}{2}Y^{MK}_{\quad NL} 
 \left( \alpha \Lambda_{1}^{N}\partial_{K}\Lambda_{2}^{L} - \beta \partial_{K}\Lambda_{1}^{N}\Lambda_{2}^{L}  \right)\dd_{M} \delta(1 - 2)\nn\\
 &&+2i\left( \alpha \Lambda_{1}^{N}\Lambda_{2}^{M}(1) + \beta \Lambda_{1}^{N}\Lambda_{2}^{M}(2)  \right)\eta^m_{~~NM}\partial_{1m}\delta(1 - 2)~.
 \eea

 To evaluate the right hand side of \bref{ccommutator}, we need to integrate the expression above with $\sigma_{1}$ and $\sigma_{2}$. Observe that the Schwinger term is always a total derivative and thus vanishes. Also the left hand side of \bref{ccommutator} is anti-symmetric with respect to $\Lambda_{1} \leftrightarrow \Lambda_{2}$ and $1\leftrightarrow 2$ That leads to $\alpha = \beta = \frac{1}{2}$. This is the Courant bracket analogously to \cite{Hatsuda:2012uk}.
 
If we add a curved background to the current algebra. i.e. $\dd_{A} \equiv E^{\; N}_{A}\dd_{N}$. The generator $Z(\Lambda)$ would generate the generalized Lie derivative
\begin{equation}
	[Z(\Lambda), \dd_{A}(2)] \equiv (\mathcal{L}_{\Lambda}E^{\; N}_{A})\dd_{N}(2), \quad
	\mathcal{L}_{\Lambda}E^{\; N}_{A} = \Lambda^{M} \partial_{M} E^{\; N}_{A} - E^{\; M}_{A} \partial_{M} \Lambda^{N} - \frac{1}{2}Y^{MN}_{\quad PQ} \partial_{M}\Lambda^{P}E^{\; Q}_{A}.\label{dEDorf}
\end{equation}
The above relation \bref{dEDorf} is nothing but the Dorfman bracket in ${\cal A}$-theory
\begin{equation}
	[\Lambda_{1}, \Lambda_{2} ]_{D}^{N} = \Lambda_{1}^{M}\partial_{M}\Lambda_{2}^{N} 
- \Lambda_{2}^{M}\partial_{M}\Lambda_{1}^{N} 
 - \frac{1}{2}Y^{MN}_{\quad PQ}\partial_{M}\Lambda_{1}^{P}\Lambda_{2}^{Q} .
\end{equation}
In this case, we need to take $\alpha = 0, \beta = 1$ in \bref{commresults}. Under this condition, the Schwinger term in \bref{commresults} is only a total derivative of $\sigma_{1}$, and $\sigma_{1}$ is integrated thus the Schwinger term dropped out again.

\vskip 6mm

\subsection{Non-degenerate Poincar\'e current algebra}\label{section:3-6}
${\cal A}$-theory is a gauge theory of the U-duality symmetry. 
The gauge field is the $G/H$ coset element, and the field strength is the commutator of the gauge covariant derivatives which include $H$-generator.
The Jacobi identity of string current algebras require that 
the group metric is non-degenerate.
The Jacobi identity of brane current algebras require non-flat world-volume vielbein.
In this section after explaining the non-degenerate algebras the consistency 
with the Virasoro operator is discussed.

 $G$-symemtry of Einstein's gravity is the D-dimensional general linear group while  $H$-symemtry is the Lorentz group. The field strength of this theory
is Riemannian curvature tensor.
The commutator of the covariant derivatives in curved backgrounds gives the Riemann curvature tensor as the
 coefficient of the Lorentz generator. Therefore it is natural to include the Lorentz generator $S$ to construct the gravity theory. (Not to be confused  with the Virasoro operator ${\cal S}$.)
The
Lorentz connection is included in the vielbein, so it is directly coupled to the string. 
The curvature tensor with manifest T-duality was given as the torsion in the current algebra in \cite{Siegel:1993th, Siegel:1993xq, Siegel:1993bj, Polacek:2013nla}, 
 and the construction of the ${A}$-symmetric current algebra including the Lorentz generators is first given in
\cite{Linch:2017eru}.

The Jacobi identity of the affine algebra requires that the group metric is non-degenerate. The non-degenerate partner of the 
Lonretz generator, denoted as $\Sigma$,  is introduced. Similarly, we will introduce supersymmetry, thus we introduce supersymmetry generators
 $D$, and by extension, we also have to introduce its non-degenerate partner $\Omega$. 
The concrete expression of non-degenerate pairs of the Lorentz generator ($S,\Sigma$) were introduced in the superstring theory 
 in the type II ${\cal T}$-theory \cite{Hatsuda:2015cia, Hatsuda:2014qqa}
and in the anti-de Sitter space \cite{Hatsuda:2001xf}.
We shall first recall the full-super Poincar\'e non-degenerate algebra in ${\cal T}$-theory, and then extended it to ${\cal A}$-theory
since both constructions are very similar. (Although there are crucial differences.)

\subsubsection{Non-degenerate  algebra in ${\cal T}$-theory} 

The non-degenerate super-Poincar\'e algebra in
 ${\cal T}$-theory is generated by $\dd_{\cal M}$  
 with the engineering dimensions  {$N_{\cal M}$ 
\begin{equation}\label{engineerd}
	\dd_{{\cal M}} = (S_{I}, D_{\mu}, P_{M}, \Omega^{\mu}, \Sigma^{I} )~~\textrm{and}~~N_{\cal M}= 0, 1/2, 1, 3/2, 2	
\end{equation}
Here $D, \Omega$ are fermionic
generators and $S, P, \Sigma$ are bosonic generators. Note that $S_{I}$ is in the adjoint representation of its corresponding $H$-symmetry
 indices $I, J, K \cdots$ to be the index for adjoint representations of $H$. Note that the adjoint representation can be always projected out from the anti-symmetric part vector indices $[NM]$, therefore sometimes we will use the anti-symmetric indices $[NM]$ and adjoint indices $I$ interchangeably. i.e.
\begin{equation}
S_{[MN]} \equiv \mathbb{P}^{I}_{[MN]}S_{I},~~
\Sigma^{[MN]} \equiv \mathbb{P}_{I}^{[MN]}\Sigma^{I},
\end{equation}
where $\mathbb{P}_{I}^{[MN]}$ is the projection tensor from the anti-symmetric pairs of vector indices to the adjoint indices. In fact, in ${\cal T}$-theories, the projection is just identity.

The algebra can then be collectively written as 
\begin{equation}
	[\dd_{\cal M}(1), \dd_{\cal N}(2)] = f_{{\cal MN}}^{\hspace{15pt} {\cal P}}\dd_{{\cal P}}(1)\delta(1-2) + i\eta_{\cal{MN}}\delta^{'}(1-2)
\end{equation}
with $\delta^{'}(1-2)=\frac{\partial}{\partial \sigma} \delta(\sigma_1-\sigma_2)$.
The Jacobi identity gives constraints to the structure constants.
\begin{equation}\label{TJacobi}
	f_{{\cal [ MN|}}^{\hspace{20pt} {\cal P}}\;f_{{\cal P |Q)}}^{\hspace{15pt} {\cal R }} = 0, \quad 
	f_{{\cal [ MN|}}^{\hspace{20pt} {\cal P}}\;\eta_{\cal{P|Q)}} = 0.
\end{equation}
The first equation above comes from the coefficient of the regular term of Jacobi identity, while the second equation comes from the coefficient of the Schwinger term. 

The non-degenerate group metric $\eta_{\cal{MN}}$  is given by
\bea
&\begin{array}{lcl}
\eta_{{\cal \underline{MN} }}&=&
\left(
\begin{array}{cc}
\eta_{{\cal MN} }&0\\
0&\eta_{{\cal M'N'}}
\end{array}\right)\\\\
&=&
\left(
\begin{array}{cc}
\eta_{\cal MN}&0\\
0&-\eta_{\cal MN}
\end{array}\right)\end{array}
,~
\eta_{{\cal MN}}
=
\begin{array}{c}_S\\_D\\_P\\_\Omega\\_\Sigma\end{array}
\left(\begin{array}{ccccc}
& & & &\delta_{[M}^{L}\delta_{N]}^{K}\\
 & & &\delta_\mu^{\; \nu}& \\
 & &\eta_{MN}& & \\
 &-\delta_\nu^{\; \mu}& & & \\
\delta^{[M}_L\delta^{N]}_{K}& & & & 
\end{array}\right)~~&\label{etaMN}
\eea
Note that in ${\cal T}$-theory, we can always decompose the full index ${\cal \underline{M}} = ({\cal M}, {\cal  M'}) $ to its left and right moving 
part in the SO(D) $\times$ SO(D) isotropic group.
By design, the group metric $\eta_{\cal{MN}}$ is invertible, thus the second condition of (\ref{TJacobi}) is just the statement that 
the structure constant $f_{\cal NMP}$ is graded anti-symmetric:   $f_{\cal [NMP)} = 0$, and this condition comes directly from the 
the first condition of the Jacobi identity (\ref{TJacobi}).
The $ff$ Jacobi identity can also be solved in ${\cal T}$-theory, the explicit expression of $f_{\cal NMP}$, are given in \cite{Hatsuda:2015cia, Hatsuda:2014qqa}. 

\subsubsection{Non-degenerate algebra in ${\cal A}$-theory}
To extend the current algebra to ${\cal A}$-theory, there are  several changes that need to be done. The most significant difference from ${\cal T}$-theory is that more world-volume coordinates need to be incorporated as we are working with a higher dimensional brane theory as opposed to two dimensional string model.

 This means that the dual generators $\Omega$ and $\Sigma$ need to have extra world-volume indices $n$ in order for the affine terms to be consistent
\begin{equation}
	\Omega^{\mu} \Rightarrow \Omega^{\mu}_{n}, \quad \Sigma^{I} \Rightarrow \Sigma^{I}_{n}.
\end{equation}\label{affinet}
The modified affine terms should be 
\begin{align}
	[P_{N}, P_{M}] &\sim \eta_{NM}^{\hspace{16pt}n}\partial_{n}\delta(1-2) \nn\\
	\{ D_{\mu},  \Omega^{\nu}_{n}\}&\sim  \delta_{\mu}^{\nu} \partial_{n}\delta(1-2)\\
	[S_{I},  \Sigma^{J}_{n} ] &\sim  \delta_{I}^{\;J} \partial_{n}\delta(1-2)\nn
\end{align}
with $\partial_n\delta(1-2)=\frac{\partial}{\partial \sigma^n} \delta^{({\rm d})}(\sigma_1-\sigma_2)$.
We only show the affine terms in the above commutator.

One might think that the obvious generalization of ${\cal A}$-theory current algebra is simply 
\begin{equation}
	[\dd_{\cal M}(1), \dd_{\cal N}(2)] = f_{{\cal MN}}^{\hspace{15pt} {\cal P}}\dd_{{\cal P}}(1)\delta(1-2) + i\eta_{\cal{MN}}^{\hspace{18pt}n}\partial_{n}\delta(1-2),
\end{equation}
where the affine terms $\eta_{\cal{MN}}^{\hspace{18pt}n}$ are given at (\ref{affinet}). The problem is that since the Lorentz generators should
generate the $H$-symmetry transformations, and $\partial_{n}\delta$ have a world-volume index $n$ that also transforms under $H$-rotations. On the other hand, since $\partial_{n}\delta$ doesn't depend on any dual coordinate of $S$, this means that
$[S, \partial_{n}\delta] = 0$. This is a contradiction because if we demand that $S$ and  $\partial_{n}\delta$ commutes, then the $f\eta$ Jacobi identity implies $\eta$ is not an invariant tensor of $H$ \cite{Ju:2016hla}.
To resolve this problem, even in flat backgrounds, we need to replace ordinary world-volume derivatives $\partial_{n}$ with covariant
derivatives ${\cal D}_{n}$. Where the derivatives obtain an non-flat but fixed background as
\bea
{\cal D}_{n} = \bar{e}_{n}^{\; m}\partial_{m} ~~~.
\eea
Therefore, the commutators of ${\cal D}_{n}$ and the Lorentz generators don't commute.

The commutators of $S_{I}$ and any other generators are fixed by demanding that $S$ generators the $H$-rotations. i.e.
\begin{equation}
	\delta_{\Lambda}\dd_{\cal O}  \equiv \int d^{d}\sigma \Lambda^{I}  [\; S_{I} , \dd_{\cal O}\;] = f(\Lambda)_{{\cal O}}^{\hspace{7pt}{\cal O}_{1}} \dd_{{\cal O}_{1}}
\end{equation}
where, 
\begin{equation}
	\dd_{\cal O}  =(S, D, P, \Omega, \Sigma ) ~\textrm{ and }  ~f(\Lambda)_{{\cal O}_{1}}^{\hspace{7pt}{\cal O}_{2}} \equiv \Lambda^{I}  f_{I{\cal O}_{1}}{}^{{\cal O}_{2}}. 
\end{equation}

Here $f(\Lambda)$ denotes the $H$-rotation matrix. Using the symbol $f$ to denote the rotation matrix is apt here since the structure constants between $S_I$ and any other generators $\dd_{\cal O}$ are just the rotation matrices.
The explicit form of $f_{I{\cal O}_{1}}{}^{{\cal O}_{2}}$ depends on the representations of ${\cal O}$, which could be constructed from invariant tensors of $H$.  

The commutator for $S_I$ and the covariant derivative should also generate a $H$-rotation. But there are also extra subtleties since the derivatives of delta functions appear.

\begin{equation}
    [S_{I}(1), {\cal D}_{n}(2)\delta(2-3)] = f_{In}{}^{m}\delta(1-2){\cal D}_{m}(2)\delta(2-3) .
\end{equation}

Using these structure constants, the $ff$ Jacobi identity of $SS\dd_{{\cal O}}$, and $S\dd_{{\cal O}_{1}}\dd_{{\cal O}_{2}}$ are satisfied automatically. The Jacobi identity of
$SS\dd_{{\cal O}}$  simplify means that
\bea
[\delta_{\Lambda_{1}}, \delta_{\Lambda_{2}}] = \delta_{[\Lambda_{1}, \Lambda_{2}]},
\eea
which is just the definition of the representation matrices.

Given arbitrary two generators $\dd_{{\cal O}_{1}}$ and $\dd_{{\cal O}_{2}}$, and there structure constants 
$[\dd_{{\cal O}_{1}}, \dd_{{\cal O}_{2}}] = f_{{\cal O}_{1}{\cal O}_{2}}^{\hspace{15pt}{\cal O}_{3}}\dd_{{\cal O}_{3}}$. 
The $S_I\dd_{{\cal O}_{1}}\dd_{{\cal O}_{2}}$ Jacobi identity implies that
\begin{equation}
	f_{I{\cal O}_{1}}^{\hspace{7pt}{\cal O}_{3}} f_{{\cal O}_{3}{\cal O}_{2}}^{\hspace{15pt}{\cal O}_{4}} 
	- (-1)^{|{\cal O}_{1}||{\cal O}_{2}|}
	f_{I{\cal O}_{2}}^{\hspace{7pt}{\cal O}_{3}} f_{{\cal O}_{3}{\cal O}_{1}}^{\hspace{15pt}{\cal O}_{4}} -
	f_{{\cal O}_{1}{\cal O}_{2}}^{\hspace{15pt}{\cal O}_{3}} f_{I{\cal O}_{3}}^{\hspace{7pt}{\cal O}_{4}} = 0,
\end{equation}
here $|{\cal O}|$ is the order of the corresponding generators (1 for fermionic generators, and 0 for bosonic generators). 
This condition just tell us that the structure constants $f_{{\cal O}_{1}{\cal O}_{2}}^{\hspace{15pt}{\cal O}_{3}}$ itself 
need to be an invariant tensor under $H$.
Since all $ff$ Jacobi identity involving the dimension 0 operators $S_I$ is satisfied, and the highest total dimension a 
Jacobi identity can have 2. The only non-trivial $ff$ Jacobi identity in (\ref{TJacobi}) is $DDD$ identity,
and $DDP$ identity.

By consulting \cite{Hatsuda:2015cia, Hatsuda:2014qqa}, we can see that the generalizations of non-degenerate Poincar\'e algebra
from ${\cal T}$-theory and ${\cal A}$-theory should be (Only the relevant ones for solving the Jacobi identities are listed here, and the affine terms are ignored.): non-degenerate Poincar\'e algebra
\begin{align}
	\{D_{\mu}(1), D_{\nu}(2) \} = \enspace & (\gamma^{N})_{\mu \nu} P_{N}(1)\delta(1-2)\\
	[D_{\mu}(1), P_{N}(2) ] = \enspace &(\gamma^{P})_{\mu \nu} \eta_{PN}^{\hspace{14pt}n}\Omega_{n}^{\;\nu}(1)\delta(1-2)\\
	[P_{M}(1), P_{N}(2)] = \enspace & \frac{1}{2}\eta_{[M|P}f_{I|N]}{}^{P}\Sigma_{n}^{\;I}(1)\delta(1-2) \\
	\{D_{\mu}(1), \Omega_{n}^{\;\nu}(2)\} = \enspace & \left[-\delta_{n}^{\; m}(f_{I})_{\mu}^{\;\nu} + p\,\delta_{\mu}^{\;\nu}(f_{I})_{n}^{\; m} \right]\Sigma_{m}^{I}\delta(1-2) 
\end{align}
Here $\gamma^{M}_{\mu\nu}$ is the ``gamma'' function for the isotropic group $H$. i.e. It is the intertwiner between 
the spinors and vectors and $p$ is a constant that needs to be fixed later.

To verify these commutators, the  $DDD$ Jacobi identities implies that,
\begin{equation}\label{jc1}
	\gamma^{M}_{\;(\mu\nu}\;\eta^{n}_{\;MN}\;\gamma_{\rho)}^{N\;\tau} = 0.
\end{equation}
The Fierz identities (\ref{jc1}) are verified for dimension $D=3, 4$. \cite{Linch:2015fca}

The $DDP$ Jacobi identity implies that
\begin{eqnarray}
    -\gamma_{v\rho}\eta^{n}{}_{PN}\left[\delta_{n}^{\; m}(f_{I})_{\mu}^{\;\rho} - p\,\delta_{\mu}^{\;\rho}(f_{I})_{n}^{\; m} \right] + (\mu \leftrightarrow \nu) = -\frac{1}{2}\gamma^{P}_{\mu\nu}\left(\eta_{[P|Q}f_{I|N]}{}^{Q}\right).
\end{eqnarray}
Using the fact that $\gamma_{\mu\nu}^{P}$ is a $H$-invariant tensor.
\begin{equation}
    (f_{I})^{Q}_{\;P}\gamma_{\mu\nu}^{P} = f_{I(\mu}{}^{\rho}\gamma^{Q}_{\rho)\nu}.
\end{equation}
We have the following identity.
\begin{equation}
    \frac{1}{2}\eta^{n}{}_{[N|Q}f_{I|P]}{}^{Q} = \eta^{n}{}_{NQ}(f_{I})^{\;Q}_{\;\;P} - \frac{1}{2}f_{Im}{}^{n}\eta^{m}{}_{NP}.
\end{equation}
Using the two equations above, the Jacobi identity is simplified to 
\begin{equation}
    \gamma^{P}_{\mu\nu}\left[\eta^{n}{}_{NQ}(f_{I})^{\;Q}_{\;\;P} - \frac{1}{2}f_{Im}{}^{n}\eta^{m}{}_{NP}\right] =  \gamma^{P}_{\mu\nu}\left[\eta^{n}{}_{NQ}(f_{I})^{\;Q}_{\;\;P} - 2pf_{Im}{}^{n}\eta^{m}{}_{NP}\right],
\end{equation}
which fix $p = \frac{1}{4}$.

Now, we also have to verify that the $f\eta$ Jacobi identities holds. Before doing that, there is a subtle issue in the current algebra we need to resolve. The affine term ${\cal D}\delta(1-2)$ is ambiguous in the sense that covariant derivatives could act on either the first or second coordinate. The most general expression should be linear combinations of both of them.

We find that the following linear combinations are consistent with the $f\eta$ Jacobi identities.
\begin{equation}
    [{\cal O}_{\cal N}(1), {\cal O}_{\cal M}(2)] \sim \frac{1}{2}\eta^{n}{}_{\cal NM}\left(N_{\cal  N}{\cal D}(1) - N_{\cal M}{\cal D}(2)\right)\delta(1-2),
\end{equation}
where $N_{\cal M}$ is the engineering dimensions defined in \bref{engineerd}.


Putting everything together, the final from of ${\cal A}$-theory non-degenerate current algebra is 
\begin{align}
    & [\,S_{I}(1), S_{J}(2)\,] = f_{IJ}{}^{K}S_{K}(1)\delta(1 -2) \\
    & [\,S_{I}(1), D_{\mu}(2)\,] = (f_{I})_{\mu}^{\;\nu}D_{\nu}(1)\delta(1 -2)\\
   &  [\,S_{I}(1), P_{N}(2)] = (f_{I})_{N}^{\; M}P_{M}(1)\delta(1 -2)\\
   &[\,S_{I}(1), \Omega^{\;\mu}_{n}(2)\,] = \left[(f_{I})^{m}_{n}\delta_{\mu}^{\nu} -(f_{I})_{\mu}^{\nu}\delta_{n}^{m}\right]\Omega^{\;\nu}_{m}(1)\delta(1 -2)\\
    &[\,S_{I}(1), \Sigma^{\;J}_{n}(2)\,] = \left[(f_{I})^{m}_{n}\delta_{J}^{K} -f_{IJ}{}^{K}\delta_{n}^{\;m}\right]\Sigma^{\;K}_{m}(1)\delta(1 -2) - i\delta_{I}^{\;J}{\cal D}_{n}(2)\delta(1-2)\\
    &[\,S_{I}(1), {\cal D}(2)_{n}\delta(2-3)\,] = \frac{1}{2}f_{In}{}^{m}{\cal D}(2)_{m}\delta(2-3)[\delta(1-2) + \delta(1-3)]\\
    &\{D_{\mu}(1), D_{\nu}(2) \} = \enspace  \gamma^{N}_{\mu \nu} P_{N}(1)\delta(1-2)\\
	&[D_{\mu}(1), P_{N}(2) ] = \enspace \gamma^{P}_{\mu \nu} \eta_{PN}^{\hspace{14pt}n}\Omega_{n}^{\;\nu}(1)\delta(1-2)\\
	&[P_{M}(1), P_{N}(2)] =  \frac{1}{2}\eta_{[M|P}f_{I|N]}{}^{P}\Sigma_{n}^{\;I}(1)\delta(1-2) + \frac{i}{2}\eta_{NM}^{\hspace{15pt}n}[{\cal D}(1)_{n} - {\cal D}(2)_{n} ]\delta(1-2)\\
	&\{D_{\mu}(1), \Omega_{n}^{\;\nu}(2)\} = \enspace  \left[-\delta_{n}^{\; m}(f_{I})_{\mu}^{\;\nu} + \frac{1}{4}\,\delta_{\mu}^{\;\nu}(f_{I})_{n}^{\; m}  \right]\Sigma_{m}^{I}(1)\delta(1-2) \nonumber\\
    & \hskip 25mm + i\delta_{\mu}^{\;\nu} \left[ \frac{1}{4}{\cal D}_{n}(1) -  \frac{3}{4}{\cal D}_{n}(2)\right]\delta(1-2).
\end{align}

We have found the algebra in terms of five $H$-tensor invariants. $\eta_{MN}^{n}, \gamma^{P}_{\mu\nu}, (f_{I})_{n}^{\;m}, (f_{I})_{\mu}^{\;\nu}, (f_{I})_{N}^{\;M}$. The exact forms of these tensors depend on the dimension.

\vskip 6mm
\subsubsection{Virasoro algebra}
The Virasoro operators in the full non-degenerate Poincar\'e  algebra should also be extended to include 
$D, \Omega, S$ and $\Sigma$:
\begin{equation}
	\mathcal{S}_{n}\equiv \frac{1}{2}\dd_{\cal N} \eta_{n}^{\cal NM} \dd_{\cal M}  =  \frac{1}{2}P_{N}P_{M}\eta^{NM}_{n}+S_{I}\Sigma^{I}_{n}+\Omega_{n}^{\mu}D_{\mu}.
\end{equation}
We wish to verify that the actions of the Virasoro operator $\mathcal{S}$ on some other operators behave like to world-volume diffeomorphism, i.e. up to some other constraints, $[\mathcal{S}_n,{\cal O}] \sim i{\cal O}\partial_{n}\delta$.   

First we shall calculate  $[{\cal S}_{n}, P_Q]$, we have
\begin{equation}
    \begin{split}
     [\frac{1}{2}P_N\eta^{NM}_{n}P_{M}(1),P_{Q}(2)]=&\eta^{NM}_{n}P_M(1)i\eta_{NQ}{}^{m}{\cal D}_{m}\delta(1-2) \\
    +&\eta^{NM}{}_{n}P_{M}\frac{1}{2}f_{I[N}{}^R\eta_{Q]R}{}^{m}\Sigma_{m}^{I}\delta(1-2) \\
    =&iP_Q(1){\cal D}_n\delta(1-2) -i(U^{m}_{\;n})_{Q}^{\;M}P_M{\cal D}_{m}\delta(1-2) \\
    -&f_{IQ}{}^{R}(\delta_{R}^{\;M}\delta_{n}^{\;m}-(U_{n}^{\;m})_{R}^{\;M})P_{M}\Sigma_{n}^{I}\delta(1-2) \\
    +&\frac{1}{2}f_{Ip}{}^{m}(\delta^{\;M}_{N}\delta^{\; p}_{m} - (U_{m}^{\;p})_{N}^{\; M})P_{M}\Sigma_m^{I}\delta(1-2).
    \end{split}
\end{equation}
To get the last line of the above, we have used the following relations.
\begin{equation}
    \begin{split}
    &f_{PP}{}^{\Sigma}\equiv f_{NM}{}^{n}_{\;\;I} =  \frac{1}{2}f_{I[N}{}^{P}\eta^{n}{}_{P\;|M]}\\
    & \frac{1}{2}f_{I(N}{}^{P}\eta^{n}{}_{P\;|M)} - f_{I\; m}{}^{n}\eta^{m}{}_{NM} =0
    \end{split}
\end{equation}
The first line can be directly seen from the algebra's structure constant itself, and the second line is just the statement that $\eta^{n}{}_{MN}$ is a tensor invariant. 
The other pieces of the commutators are also computed.
\begin{equation}
    \begin{split}
    [S_{I}\Sigma^{\; I}_{n}(1),P_{Q}(2)]=&f_{IQ}{}^{R}P_{R}\Sigma^{\; I}_{n}{}\delta(1-2)\\
    [\Omega_{n}^{\mu} D_{\mu}(1),P_{Q}(2)]=& \Omega_{n}^{\;\mu}\gamma_{\mu\nu}^{N}\eta_{NQ}{}^{m}\Omega_{m}^{\; \nu}\delta(1-2)\\
    \end{split}
\end{equation}
Observe that due to the symmetry of the gamma matrices,  we have $\Omega_{n}^{\; \mu}\Omega_{m}^{\; \nu}\gamma_{\mu\nu}^{N}=\frac{1}{2}\Omega_{[n}^{\; \mu}\Omega_{m]}^{\; \nu}\gamma_{\mu\nu}^{N}$. Replacing $\Omega_{n}^{\; \mu}$ with ${\cal D}_{n}\theta^{\mu}$ and applying additional anti-symmetric world-volume ${\cal V}$ constraint ${\cal D}_{[n}\otimes{\cal D}_{m]}\to 0$ can help us drop it. Then apart from the usual ${\cal U}$ constraint term, we can also drop $(U^{\; n}_{m})_{N}^{\; M}P_{M}\Sigma_{n}^{\; I}$, as we have $\Sigma^{I}_{n}=i{\cal D}_{n}\alpha^{I}$, where $\alpha$ labels the $S$ coordinates. Putting everything together, we get up to ${\cal U, V}$ constraints:
\begin{equation}
    [\mathcal{S}_{n}(1),  P_{Q}(2)]=iP_{Q}{\cal D}_{n}\delta(1-2) + \frac{1}{2}f_{In}{}^{m} P_{Q}\Sigma_{m}^{I}\delta(1-2)
\end{equation}
There is an extra term proportional to $f_{In}{}^{m} P_{Q}\Sigma_{m}^{I}$, we will denote this as $\Sigma_{n}$ from now on. 

We also compute the commutator of $[{\cal S}_{n}, S_{I}]$. Since the generator $S_{I}$ generate $H$-rotation, and ${\cal S}_{m}$ is explicitly $H$-covariant, the commutator will contain the term $-f_{In}{}^{m}{\cal S}_{m}\delta$, that is just the $H$-rotation on ${\cal S}_{n}$. The only other term in the commutator comes from the affine term between $S_{I}$ and $\Sigma_{n}^{\;I}$.

\begin{equation}
    \begin{split}
        [S_{I}\Sigma^{I}_n(1),S_{J}(2)]=&-f_{Jn}{}^{m}S_{I}\Sigma^{I}_{m}\delta(1-2)\\
        &+iS_{J}(1){\cal D}_{n}\delta(1-2),
    \end{split}
\end{equation}
and
\begin{equation}
    [\mathcal{S}_{n}(1),S_{I}(2)]=iS_{I}{\cal D}_{n}\delta(1-2)-f_{In}{}^{m}\mathcal{S}_{m} \delta(1-2) 
\end{equation}
There is an extra term $f_{In}{}^{m}\mathcal{S}_{m} \delta$. This term is expected, as this commutator can also be interpreted as $S_{I}$ acting on $\mathcal{S}_N$, performing a $H$-rotation on the world-volume index.

We compute the commutator of $[{\cal S}_{n}, \Sigma^{\; I}_{m}]$.
\begin{equation}
    \begin{split}
         [\mathcal{S}_n(1),\Sigma^{\; I}_{m}(2)]&=\Sigma^{\; I}_{n}(1)i{\cal D}_{m}\delta(1-2)+\Sigma_{n}^{\; J}(f_{Jm}{}^{p}\Sigma^{\; I}_{p}-f_{JK}{}^{I}\Sigma^{K}_{m})\delta(1-2)\\
         &=i\Sigma^{\; I}_{n}(1){\cal D}_{m}\delta(1-2) - f_{JK}{}^{I}\Sigma_n^{\; K}\Sigma_{m}^{\;J}\delta(1-2)\\
         &+f_{Jm}{}^{p}\Sigma_n^{\; J}\Sigma_{p}^{I}\delta(1-2)\\
    \end{split}
\end{equation}
The $f_{JK}{}^{I}\Sigma_n^{\; K}\Sigma_{m}^{\;J}$ can be dropped by the ${\cal V}$ constraint, similar to the trick applied to $\Omega$ in the previous computation.  But the last term remains, and the final result can be rearranged into 
\begin{equation}
 [{\cal S}_{n}, \Sigma^{\; I}_{m}] =i\Sigma^{\; I}_{n}(1){\cal D}_{m}\delta(1-2)+ \Sigma^{\; I}_{n}\Sigma_{m}\delta(1-2).
\end{equation}

For $[{\cal S}_{n}, \Omega^{\; \mu}_{m}]$ there is:
\begin{equation}\label{comsomega}
    \begin{split}
        [\mathcal{S}_{n}(1),\Omega_{m}^{\; \mu}(2)]&=\Sigma_{n}^I(f_{Im}{}^{p}\delta_{\nu}^{\;\mu}-f_{I\nu}{}^{\mu}\delta_m^{\; p})\Omega_{p}^{\; \nu}\delta(1-2)\\
        &+\Omega_{n}^{\rho}(f_{I\rho}{}^{\mu}\delta_{m}^{p}-\frac{1}{4}f_{Im}{}^{p}\delta_{\rho}^{\; \mu})\Sigma_p^{\; I}\delta(1-2)\\
        &+\Omega_{m}^{\; \mu}(1) i{\cal D}_{n}\delta(1-2)\\
        &=i\Omega_{n}^{\; \mu} i{\cal D}_{m}\delta(1-2) +\frac{3}{4}\Omega_m^{\mu} f_{In}{}^{p}\Sigma_p^{\; I}\delta(1-2).\\
    \end{split}
\end{equation}
\\

And finally for $[{\cal S}_{n}, D_{\mu}]$, we have:
\begin{equation}\label{comsd}
    \begin{split}
        [\mathcal{S}_n(1),D_{\mu}(2)]&=f_{I\mu}{}^{\nu} D_{\nu} \Sigma_n^{\; I}\delta(1-2)-(f_{I\mu}{}^{\nu}\delta_{n}^{m}-\frac{1}{4}f_{In}{}^{m}\delta_{\mu}^{\nu})\Sigma_{m}^{\; I}D_{\nu} \delta(1-2)\\
        &(- \Omega_{n}^{\; \nu}\gamma_{\nu\mu}^{N}P_{N} +\eta^{NR}{}_{n}P_{N}\gamma_{\mu\nu}^{M}\eta_{RM}{}^{m}\Omega_{m}^{\; \nu})\delta(1-2)\\
        &+iD_{\mu}(1){\cal D}_{n}\delta(1-2) \\
        &=iD_{\mu}(1){\cal D}_{n}\delta(1-2) + pD_{\mu}(f_{In}{}^{m})\Sigma_{m}^{\; I} \delta(1-2). \\
    \end{split}
\end{equation}
We used again the ${\cal U}$ constraints $(U_{n}^{\;m})_{N}^{\; M}P_{M}\Omega^{\;\mu}_{m} \sim 0 $ to eliminate the second line in the above equations.

Recalling that  $f_{In}{}^{m}\Sigma_{m}^{I}\equiv\Sigma_{n}$, we now conclude the behavior of Virasoro operator acting on other operators ${\cal O}=D,P,\Omega,\Sigma$:
\begin{equation}
    [\mathcal{S}_n,{\cal O}] = i{\cal O}{\cal D}_{n}\delta +\frac{N_{\cal O}}{2}\Sigma_{n}{\cal O}\delta \label{SDSigma}
\end{equation}
Since the engineering dimensions of ${\cal D}_n$ and $\Sigma_n$ are equal, the combination of the right hand side of \bref{SDSigma} suggests further covariantization. The new contribution of $\Sigma$ plays a role of the connection of $G/H$.
Among them, ${\cal O}=S$ is a little bit special, 
$[\mathcal{S}_n,S] = iS{\cal D}_{n}\delta -f_{na}{}^{b}{\cal S}_b\delta $ since the Lorentz acts on the Virasoro operator.
 In deriving the commutators above, we used the following ${\cal U}$ and ${\cal V}$ constraints:
\begin{equation}
    \begin{split}
        U^{Nn}_{\varpi}P_N\partial_{n}&=0 \\
      {\cal U}_{\varpi}~=~U^{Nn}_{\varpi}P_N\Sigma_{n}&=0 \\
     {\cal V}_{nm}~=~ {\cal D}_{[n}\otimes{\cal D}_{m]}&=0\\
        \Sigma_{[n}{\cal D}_{m]}&=0\\
    \end{split}
\end{equation}
Here we can also simplify the set of constraints by defining a so-called covariant world-volume differential $\hat{\cal D}_n$:
\begin{equation}
\begin{split}
    \hat{\mathcal{D}}_n&\equiv\bar{e}_n{}^m\partial_m + \Sigma_n\\
         {\cal U}_{\varpi}&=U_{\varpi}^{Nn}\dd_N\hat{\mathcal{D}}_{n}~=~0\\
      {\cal V}_{nm} &=\hat{\mathcal{D}}_{[n}\otimes \hat{\mathcal{D}}_{m]}~=~0\\
\end{split}
\end{equation}

The commutator $[\mathcal{S}_a(n),\mathcal{S}_{m}(2)]$ could be derived using the result above:
\begin{equation}
    \begin{split}
        [\mathcal{S}_{n}(1),\mathcal{S}_{m}(2)]=&\mathcal{S}_{(n}(\frac{1}{2}(1+2))i{\cal D}_{m)}\delta(1-2)-\frac{i}{2}\partial_{[n}\mathcal{S}_{m]}\delta(1-2) \\
        +& \frac{1}{2}\eta_{NQ}{}^{p}\eta^{NM}{}_{n}\eta^{QR}{}_{m}f_{Ip}{}^qP_{M}P_{R}\Sigma^{\; I}_q\delta(1-2)\\
        -& f_{In}{}^p\mathcal{S}_p\Sigma_m^{\; I}\delta(1-2)\\
        +&S_{I}f_{Jm}{}^{p}\Sigma_n^{\; J}\Sigma_p^{\; I}\delta(1-2)\\
    \end{split}
\end{equation}
Terms in the second line need some extra treatment. With the help of $\eta$ being invariant tensor, we have:
\begin{equation}
    \begin{split}
        \eta_{NQ}{}^{p}\eta^{NM}{}_{n}\eta^{QR}{}_{m}f_{Ip}{}^{q}&=f_{I(m}{}^r\eta_{NQ}{}^{q}\eta^{NM}{}_{n)}\eta^{QR}{}_{r}-f_{IS}{}^{(M|}\eta_{NQ}{}^q\eta^{NS}{}_n\eta^{Q|R)}{}_m\\
        &=f_{I(m}{}^{r}\eta^{QR}{}_{r}(\delta^M_Q\delta^q_{n)}-(U^{\;q}_{n})_{Q}^{\;M})\\
        &-f_{IS}{}^{(M|}\eta^{IN}{}_{n}(\delta^{|R)}_{N}\delta^{q}_{m}-(U_{m}^{\; q})^{|R)}_{N})\\
    \end{split}
\end{equation}
Then when coupled with $PP\Sigma$, we can drop the ${\cal U}$ term, and the result is:
\begin{equation}
    \frac{1}{2}P_M\eta^{MR}{}_{r}P_{R}f_{I(m}{}^r\Sigma^{\; I}_{n)}-\frac{1}{2}f_{In}{}^p\eta^{MR}{}_qP_{M}P_{R}\Sigma_{b}^M
\end{equation}
Then we can simplify the extra term in the commutator:
\begin{equation}
    \begin{split}
        [\mathcal{S}_n(1),\mathcal{S}_m(2)]=&\frac{1}{2}\mathcal{S}_{(n}({\cal D}(1)_{m)} - {\cal D}(2)_{m)})\delta(1-2)-\frac{i}{2}{\cal D}_{[n}\mathcal{S}_{m]}\delta(1-2) \\
        +& f_{I[n}{}^{p}\Sigma_{m]}^{\; I}\mathcal{S}_{q}\delta(1-2)\\
    \end{split}
\end{equation}
\vskip 6mm

\section{Sectioning}\label{section:4}

In order to recover the usual string theory with suitable dimensions of the source and target of the defining sigma model, the sectioning condition has to be solved.
While writing the general section condition for arbitrary dimension is possible, its solution is heavily dimension dependent as it depends on the structure constants of the underlying exceptional group which varies through dimensions.

The target space coordinates $X^{M}(\sigma)$ include zero-modes $x^{M}$ and non-zero modes. If we replace $P_{M}(\sigma)$ by its zero modes $p_{M} = \frac{\partial}{\partial x^{M}}$ in Virasoro and Gau\ss{} law constraints, we get the section condition. 
Eliminating both zero and non-zero modes of some target space coordinates by Virasoro and  Gau\ss{} law constraints is called dimensional reduction.

In this section, we discuss in detail examples of  $D=3$ and $D=4$ string theory. Once the sectioning is performed all the way down to the string model one is left with Virasoro algebra as a remaining algebra of constraints.

\vskip 6mm
\subsection{D$=$3 theories}\label{section:4-1}

${\cal M, T}$ and S-theories are obtained by different sections of ${\cal A}$-theory \cite{Linch:2015fya,Siegel:2020qef}.
Section conditions are the Virasoro constraints ${\cal S}^m = 0$ and the Gau\ss{} law constraints ${\cal U}_m = 0$  written by 0-modes  
in \bref{Virasoros}
which act on spacetime functions as follows:
\bea
{\rm Spacetime}&:&{\cal S}^m|_P=P_M\eta^{MNm}P_N=\epsilon^{mn_1\cdots n_4}P_{n_1n_2}P_{n_3n_4}=0\nn\\
&\Rightarrow& \epsilon^{mn_1\cdots n_4}\partial_{n_1n_2}\partial_{n_3n_4}\Psi(X)= \epsilon^{mn_1\cdots n_4}\partial_{n_1n_2}\Phi(X)\partial_{n_3n_4}\Psi(X)=0\nn\\\label{sction}\\
{\textrm {World-volume}}&:&{\cal U}_m|_{\partial,P}=\partial^l U^{N}{}_M{}^n{}_l P_N\delta_n^{m_2}=\partial^nP_{mn}=0\nn\\
&\Rightarrow& \partial^n\partial_{mn}\Psi(X(\sigma))
= \partial^n\Phi(X(\sigma)) \partial_{mn}\Psi(X(\sigma))=0\nn
\eea

D=3 theories with $G/H$ coset and coordinates are related by section conditions as;
\bea
&{\rm Diamond~diagram~of}~{\rm D=3~theories}&\nn\\&
{{\cal A}\mathchar`-{\rm theory}}
&\nn\\
&~~~~~~~~~~~~~~~~~~~~~~~~\displaystyle\frac{G}{ H}=\frac{{\rm SL}(5)}{{\rm SO}(5)}~
\left\{
\begin{array}{l}10:P_{mn}\cdots{\rm spacetime} \\~5:\partial^m~~\cdots{\textrm{world-volume}}\\
	m=1,\cdots,5
\end{array}
\right.
&\nn\\
&{\cal S}~\swarrow  ~~~~~~~~~~~~~~~~~~~~\searrow ~{\cal U}&\nn\\
&{{\cal M}\mathchar`-{\rm theory}}
~~~~~~~~~~~~~~~~~~~~~~~~~~~~~~~~~~~~~~~~~~~{{\cal T}\mathchar`-{\rm theory}}
&\nn\\
&\displaystyle\frac{{\rm GL}(4)}{{\rm SO}(4)}~\left\{
\begin{array}{l}4:P_{\underline{m}} \\4:\partial^{\underline{m}}
	\\
	\underline{m}=1,\cdots,4
\end{array}
\right.~~~~~~~~~~~~~~~~~~~~~~~~~\displaystyle\frac{{\rm O}(3,3)}{{\rm SO}(3)^2}~\left\{
\begin{array}{l}6:P_{\underline{mn}} \\1:\partial
\end{array}
\right.&\nn\\
&{\cal U}~\searrow  ~~~~~~~~~~~~~~~~~~~~\swarrow~{\cal S}&\nn\\
&{{\rm S}\mathchar`-{\rm theory}}
&\nn\\
&\displaystyle\frac{{\rm GL}(3)}{{\rm SO}(3)}~
\left\{
\begin{array}{l}3:P_{\bar{m}} \\1:\partial\\
	\bar{m}=1,\cdots,3
\end{array}
\right.
&
\eea
We write SO($n$) as the $H$-group for simplicity, but
the Wick rotation is necessary to identify $H$ group as the Lorentz group.

\begin{enumerate}
	\item {Section of ${\cal A}$-theory $\to$ ${\cal M}$-theory

	D=3 ${\cal M}$-theory has GL(4) $G$-symmetry and the background gauge field is GL(4)/SO(4) parameter which is
	the 4-dimensional metric.
	5-representation of SL(5) is decomposed by GL(4) as $5\to 4+1$, $m\to (\underline{m},~5)$.
	10 spacetime coordinate $P_M=P_{mn}$ is decomposed as $6+4$ as $(P_{\underline{m}\underline{n}},~P_{\underline{m}})$.
	The spacetime section conditions, $\epsilon^{\underline{m}_1\cdots \underline{m}_4}P_{\underline{m}_1\underline{m}_2}
		P_{\underline{m}_3\underline{m}_4}=0$ and
	$\epsilon^{\underline{m}_1\cdots \underline{m}_4}P_{\underline{m}_1\underline{m}_2}
		P_{\underline{m}_3}=0$, are solved in  $P_{\underline{m}}\neq 0$ as
	\bea
	P_{\underline{m}_1\underline{m}_2}= 0~~~.
	\eea
	Elimination of all oscillator modes of $P_{\underline{m}_1\underline{m}_2}$ leads to that
	$X^{\underline{m}_1\underline{m}_2}$ has only constant 0-mode.
	Then currents of D=3 ${\cal M}$-theory are momentum and the winding mode
	\bea
	\dd_{\underline{m}}=P_{\underline{m}}~~,~~
	\dd_{\underline{m}_1\underline{m}_2}=\epsilon_{\underline{m}_1\cdots \underline{m}_4}
	\partial^{\underline{m}_3}X^{\underline{m}_4}~~~.
	\eea
	Consistency of the current algebra $[\dd_{\underline{m}_1\underline{m}_2}, \dd_{\underline{m}_3\underline{m}_4}]\Rightarrow 0$ gives $\partial^5=0$.
	There are five Virasoro constraints and one Gau\ss{} law constraint as
	\bea
	&{\cal S}=\epsilon_{\underline{m}_1\cdots \underline{m}_4}\partial^{\underline{m}_1}X^{\underline{m}_2}
	\partial^{\underline{m}_3}X^{\underline{m}_4}~~,~~
	{\cal S}^{\underline{m}}=P_{\underline{n}}\partial^{[\underline{m}}X^{\underline{n}]}&\nn\\
	&{\cal U}=\partial^{\underline{m}}P_{\underline{m}}~~,~~
	\label{SUM}&
	\eea
where the commutator of ${\cal S}$ and any operator is 0.
	Among 5 directions of the 5-brane the 4-dimensional volume is spread 
		in the 4-dimensional spacetime while 1 direction may be spread in the internal space.
	${\cal U}_{\underline{m}}=0$ is trivial.}

	\item{Section of ${\cal M}$-theory $\to$ S-theory

	D=3 S-theory has GL(3) $G$-symmetry and the background gauge field is GL(3)/SO(3) parameter
	which is the 3-dimensional metric. 4-representation of GL(4) is decomposed by GL(3) as $4\to 3+1$, $\underline{m}\to (\bar{m},~4)$.
	World-volume section conditions, $\partial^{\bar{m}}P_{\bar{m}}+\partial^4P_4=0$
	and $\partial^5 P_{\underline{m}}=0$  in \bref{SUM}, are solved in $\partial^4\neq 0$
	as
	\bea
	\partial^{\bar{m}}=0~~,~~P_4=0~~~.
	\eea
	Then currents of D=3 S-theory are momentum and the winding mode
	\bea
	\dd_{\bar{m}}=P_{\bar{m}}~~,~~
	\dd_{\bar{m}_1\bar{m}_2}=\epsilon_{\bar{m}_1\bar{m}_2 \bar{m}_3}
	\partial^4X^{\underline{m}_3}~~~.
	\eea
	There is a Virasoro constraint as
	\bea
	{\cal S}^4=P_{\underline{n}}\partial^4X^{\underline{n}}
	~~,\label{SUS}
	\eea
	while others are trivial. This is the usual Virasoro operator for a string.
	}

	\item{Section of ${\cal A}$-theory $\to$ ${\cal T}$-theory

	D=3 ${\cal T}$-theory has O(3,3) $G$-symmetry and the background gauge field is
	O(3,3)/SO(3)$^2$ parameter which is the 3-dimensional metric and the $B$ field.
	The world-volume section conditions, $\partial^{\underline{m}}P_{\underline{m}}=0$
	and $\partial^{\underline{n}}P_{\underline{m}\underline{n}}+\partial P_{\underline{m}}=0$,
	are solved in  $P_{\underline{m}\underline{n}}\neq 0$ as
	\bea
	\partial^{\underline{m}}=0~~,~~ P_{\underline{m}}= 0~~,~~{\rm and}~~\partial=\partial_\sigma\neq 0~~.
	\eea
	D=3 ${\cal T}$-theory has the selfdual  current
	\bea
	\dd_{\underline{m}_1\underline{m}_2}&=&P_{\underline{m}_1\underline{m}_2}+\frac{1}{2}
	\epsilon_{\underline{m}_1\cdots \underline{m}_4}\partial X^{\underline{m}_3\underline{m}_4}
	~~~,~~\dd_{\underline{m}}=0~~~.
	\eea
	There is only one Virasoro constraint
	\bea
	{\cal S}&=&\frac{1}{8}\epsilon^{\underline{m}_1\cdots \underline{m}_4}\dd_{\underline{m}_1\underline{m}_2}\dd_{\underline{m}_3\underline{m}_4} \label{SVirasoroT}
	\eea
	while other Virasoro and Gau\ss{} law constraints are trivially 0.}

	\item{Section of ${\cal T}$-theory $\to$ S-theory

	D=3 S-theory is obtained by the spacetime  section condition,  $\epsilon^{4\bar{m}_1\bar{m}_2\bar{m}_3}P_{4\bar{m}_1}
		P_{\bar{m}_2\bar{m}_3}=0$ from \bref{SVirasoroT}.
	This is solved in  two ways:
	\begin{itemize}
		\item{
		$P_{\bar{m}}\neq 0$ case
		\bea
		&&P_{\bar{m}}\neq 0\Rightarrow P_{\bar{m}_1\bar{m}_2}= 0~~~\nn\\
		&&
		\dd_{\bar{m}}=P_{\bar{m}}~~,~~
		\dd_{\bar{m}_1\bar{m}_2}=\frac{1}{2}\epsilon_{\bar{m}_1\bar{m}_2 \bar{m}_3}
		\partial X^{\underline{m}_3}~~~\nn\\
		&&
		{\cal S}=P_{\bar{n}}\partial X^{\bar{n}}~~~.
		\eea}
		\item{$P_{\bar{m}_1\bar{m}_2}\neq 0$ case
		\bea
		&&P_{\bar{m}_1\bar{m}_2} \neq 0\Rightarrow P_{\bar{m}}= 0~~~\nn\\
		&&
		\dd_{\bar{m}}=\frac{1}{2}\epsilon_{\bar{m}\bar{m}_1 \bar{m}_2} \partial X^{\bar{m}_1\bar{m}_2}~~,~~
		\dd_{\bar{m}_1\bar{m}_2}=P_{\bar{m}_1\bar{m}_2}\nn\\
		&&
		{\cal S}=\frac{1}{2}P_{\bar{m}_1\bar{m}_2}\partial X^{\bar{m}_1\bar{m}_2}~~~.
		\eea}
	\end{itemize}
	The second case is T-dual to the first case with $\tilde{X}_{\bar{m_1}}=
		\frac{1}{2}\epsilon_{\bar{m}_1\bar{m}_2\bar{m}_3 }X^{\bar{m}_2\bar{m}_3}$,
	showing that the ${\cal T}$-theory is manifestly T-dual theory.}

\item{Section of ${\cal A}$-theory $\to$ the membrane theory in the supergravity

A non-perturbative membrane action in the 11-dimensional supergravity theory is given by \cite{Bergshoeff:1987cm}
\bea
I&=&\displaystyle\int d^3\sigma (L_0+L_{WZ})~,~
{\renewcommand{\arraystretch}{1.4}
\left\{\begin{array}{ccl}
	L_0        & = & -T\sqrt{-\det \partial_\mu x^m \partial_\nu x^n g_{mn}}                                                        \\
	L_{\rm WZ} & = & \frac{T}{3!}\epsilon^{\mu\nu\rho} \partial_\mu x^{m_1}\partial_\nu x^{m_2} \partial_\rho x^{m_3} C_{m_1m_2m_3}
\end{array}\right.}
\eea
with the spacetime index $m=0,1,\cdots,10$ and the world-volume index $\mu=0,1,2$.
In terms of the canonical coordinates $x^m$ and
$p_m$, and the spacial world-volume coordinate derivative
$\partial_i$ with $i=1,2$ the Hamiltonian is given by \cite{Hatsuda:2012vm}
\bea
H&=& \lambda_0{\cal T}+\lambda^i {\cal H}_i~,~
{\renewcommand{\arraystretch}{1.4}
\left\{\begin{array}{ccl}
	{\cal T}   & = & \frac{1}{2}\dd_{a}{\eta}^{ab}\dd_b+
	\frac{1}{8}\dd^{a_1a_2}\eta_{a_1[b_1}{\eta}_{b_2]a_2}\dd^{b_1b_2} \\
	{\cal H}_i & = & \partial_i x^mp_m
\end{array}\right.}
\eea
where $\dd_A=(\dd_a,~\dd^{ab})$ is related to
$\dd_M=(\dd_m=p_m,~\dd^{mn}=\epsilon^{ij}\partial_i x^{m_1}\partial_j x^{m_2})$
as
$\dd_A=E_A{}^M\dd_M$  for the background gauge field $E_A{}^M$.
$E_A{}^M$ includes  $g_{mn}$ and $C_{mnl}$.

The 4-dimensional part of the spacetime introduces the SL(4) index $\underline{m}=1,\cdots,4$.
The currents are written as
\bea
{\renewcommand{\arraystretch}{1.4}
	\begin{array}{ccl}
		\dd_{A}   & = & E_A{}^M\don\circ{\dd}_M                                         \\
		{\dd}_{M} & = & ({\dd}_{\underline{m}},~{\dd}_{\underline{m}_1\underline{m}_2})
	\end{array}}
~,~
{\renewcommand{\arraystretch}{1.4}
\left\{\begin{array}{ccl}
	{\dd}_{\underline{m}}                  & = & p_{\underline{m}}                                                                                                                     \\
	{\dd}_{\underline{m}_1\underline{m}_2} & = & \frac{1}{2}\epsilon_{\underline{m}_1\cdots \underline{m}_4}\epsilon^{ij} \partial_i x^{\underline{m}_3}\partial_j x^{\underline{m}_4}
\end{array}\right.} \label{M2currents}
\eea
The SL(4) index is enlarged to SL(5) $4+1 \to 5$ by adding the fifth direction as $m=(\underline{m},5)$.
The 10 currents $\dd_M=(\dd_{\underline{m}},~\dd_{\underline{m}_1\underline{m}_2})=\dd_{mn}$.
The background gauge field is given by
\bea
&E_{a}{}^{m}=
	{\renewcommand{\arraystretch}{1.8}
		\left(
		\begin{array}{cc}
				{\bf e}^{-3/5} & {\bf e}^{-3/5} \tilde{C}^{\underline{m}}         \\
				0              & {\bf e}^{2/5}e_{\underline{a}}{}^{\underline{m}}
			\end{array}
		\right)}~~~~~~~~~~~~~~~~~~~~~~~~~~~~~~~~~~~~~~~~~~~~~&\\
&G^{MN}=G^{m_1m_2; n_1n_2}=E_{a_1}{}^{m_1}E_{a_2}{}^{m_2}\eta^{a_1[b_1}\eta^{b_2]a_2}E_{b_1}{}^{n_1}E_{b_2}{}^{n_2}
~~~~~~~~~~~~~&\nn\\&~~~~~~~~~~
={\bf e}^{-2/5}
{\renewcommand{\arraystretch}{1.8}\left(
\begin{array}{cc}
		g^{\underline{m}_2\underline{n}_2}                                & -\tilde{C}^{[\underline{n}_1}g^{\underline{n}_2] \underline{m}_2} \\
		-\tilde{C}^{[\underline{m}_1}g^{\underline{m}_2] \underline{n}_2} &
		{\bf e}^{2}g^{\underline{n}_1[ \underline{m}_1}g^{\underline{m}_2] \underline{n}_2}
		+\tilde{C}^{[\underline{m}_1}
		g^{\underline{m}_2] [\underline{n}_1}
		\tilde{C}^{\underline{n}_2]}
	\end{array}
\right)}&\\
&{\bf e}=\det E_{\underline{m}}{}^{\underline{a}}~,~
\tilde{C}^{\underline{m}} =\frac{1}{3!}\epsilon^{\underline{m}\underline{m}_1\cdots \underline{m}_3}C_{\underline{m}_1\underline{m}_2 \underline{m}_3}&
\nn
\eea
Commutators of \bref{M2currents} are given as

\bea
\left[\dd_{\underline{m}}(1),\dd_{\underline{n}}(2)\right]
&=&0\nn\\
\left[\dd_{\underline{m}}(1),\dd_{\underline{m}_1 \underline{m}_2}(2)\right]
&=&2i\epsilon_{\underline{m}_1\cdots \underline{m}_4}\epsilon^{ij}\partial_ix^{\underline{m}_3}\delta_{\underline{m}}^{\underline{m}_4}
\partial_j\delta(1-2)\label{CA5check}\\
\left[\dd_{\underline{m}_1\underline{m}_2}(1),\dd_{\underline{m}_3 \underline{m}_4}(2)\right]
&=&0\nn~~~.
\eea

On the other hand the current algebra of ${\cal A}$-theory \bref{CASL5} in the SL(4) index is given as
\bea
\left[\dd_{\underline{m}}(1),\dd_{\underline{n}}(2)\right]
&=&0\nn\\
\left[\dd_{\underline{m}_1}(1),\dd_{\underline{m}_2 \underline{m}_3}(2)\right]
&=&2i\epsilon_{\underline{m}_1\cdots \underline{m}_4}
\partial^{\underline{m_4}}\delta(1-2)\label{SL5SL4}\\
\left[\dd_{\underline{m}_1\underline{m}_2}(1),\dd_{\underline{m}_3 \underline{m}_4}(2)\right]
&=&2i\epsilon_{\underline{m}_1\cdots \underline{m}_4}
\partial^{5}\delta(1-2)
\nn~~~.
\eea

The non-perturbative M2-brane algebra in  \bref{CA5check} coincides with 
the perturbative ${\cal A}$5-brane algebra in	\bref{SL5SL4} by the following conditions.

\bea
\begin{array}{cccl}
&{\rm Spacetime}&{\textrm{World-volume}}&{\rm Conditions}\\
{\rm M2}& x^{\underline{m}},~ _{\underline{m}=1,\cdots,4}&
\sigma^i,~_{i=1,2}&{\rm static~gauge}~ \partial_i x^{\underline{m}}=\delta_i^{\underline{m}}\\
{\cal A}{\rm 5}& X^{mn},~_{{m}=1,\cdots,5}& \sigma^{m},~_{m=(\underline{m},5)}&
	{\textrm{world-volume~sections}} \\
&&&\partial_5=0,~
\partial_{\underline{m}}=
\left\{\begin{array}{l}
\partial_i\neq 0\\
0~
{\rm for}~ _{\underline{m}\neq i},
\end{array}\right.
\end{array}
\eea
where 2 directions $i=1,2$ among 4 directions $\underline{m}=1,\cdots,4$ of the ${\cal A}$-theory 5-brane are selected.

 The static gauge  $\partial_i  x^{\underline{m}}=\delta_i^{\underline{m}}$ and
the world-volume derivative with the raised index $\partial^{i}=\epsilon^{ij}\partial_j$  in the second algebra in \bref{CA5check} are imposed,
then the right hand side becomes $2i\epsilon_{\underline{m}\underline{m}_1\underline{m}_2 i}\partial^i \delta(1-2)$.
The world-volume index is further decomposed as $\underline{m}=(i,~i')$ with $i=1,2$ and $i'=3,4$.
The following condition is imposed in the second line of \bref{SL5SL4}
\bea
\partial^{i'}=0 ~~,~~\partial^{i}\neq 0 ~~~,
\eea

The  non-perturbative M2-brane is related to
the perturbative 5-brane, analogously to the D=4 case where the non-perturbative 5-brane in the supergravity theory \cite{Hatsuda:2013dya}
is related to the perturbative 10-brane
\cite{Hatsuda:2021wpb}.}
\end{enumerate}
\vskip 6mm

\subsection{D$=$4 theories}\label{section:4-2}
For the D=4 case the SO(5,5) $G$-symmetry covariant current algebra
is realized on the 10 dimensional world-volume brane \cite{Hatsuda:2021wpb}.
The spacetime and world-volume coordinates  are $X^\mu$ with $\mu=1,\cdots,16$ and $\sigma_m$ with $m=1,\cdots,10$ respectively.  
Coordinates in the Lagrangian formulation are given in section \ref{section:2-4}.
Virasoro constraints ${\cal S}^m$, the Gau\ss{} law constraints ${\cal U}^\mu=0$ and ${\cal V}=0$-constraint  are given by 
\bea
{\cal S}^m&=&\frac{1}{4}\dd_\mu\gamma^{m\mu\nu}\dd_{\nu}=0\nn\\
{\cal U}^\mu&=&\dd_\nu \gamma_m{}^{\nu\mu}\partial^m 
=0\label{SUVD=4}\\
{\cal V}&=&\eta_{mn}\partial^m\partial^n=0~~~.\nn
\eea

D=4 theories with $G/H$ coset and coordinates are related by section conditions as;

\bea
&{\rm Diamond~diagram~of}~{\rm D=4~theories}&\nn\\&
{{\cal A}\mathchar`-{\rm theory}}
&\nn\\
&~~~~~~~~~~~~~~~~~~~~~~~~\displaystyle\frac{G}{ H}
=\frac{{\rm SO}(5,5)}{{\rm SO}(5)^2}~
\left\{
\begin{array}{l}16:P_{\mu}\cdots{\rm spacetime}\\10:\partial^m~~\cdots{\textrm{world-volume}}\\
	\mu=1,\cdots,16,~		m=1,\cdots,10
\end{array}
\right.
&\nn\\
&{\cal S}~\swarrow  ~~~~~~~~~~~~~~~~~~~~\searrow ~{\cal U}&\nn\\
&
{\cal M}\mathchar`-{\rm theory}
~~~~~~~~~~~~~~~~~~~~~~~~~~~~~~~~~~~~~~~~~~~
{\cal T}\mathchar`-{\rm theory}
&\nn\\
&\displaystyle\frac{{\rm GL}(5)}{{\rm SO}(5)}~\left\{
\begin{array}{l}5:P_{\underline{m}}\\5:\partial^{\underline{m}}
	\\
	\underline{m}=1,\cdots,5
\end{array}
\right.~~~~~~~~~~~~~~~~~~~~~~~~~\displaystyle\frac{{\rm O}(4,4)}{{\rm SO}(4)^2}~\left\{
\begin{array}{l}8:P_{\underline{\mu}}\\1:\partial\\
	{\underline{\mu}}=1,\cdots,8
\end{array}
\right.&\nn\\
&{\cal U}~\searrow  ~~~~~~~~~~~~~~~~~~~~\swarrow~{\cal S}&\nn\\
&{{\rm S}\mathchar`-{\rm theory}}
&\nn\\
&\displaystyle\frac{{\rm GL}(4)}{{\rm SO}(4)}~
\left\{
\begin{array}{l}4:P_{\bar{m}}\\1:\partial\\
	\bar{m}=1,\cdots,4
\end{array}
\right.
&
\eea
The spacetime currents $\dd_\mu$ is given by
\bea
\dd_\mu=P_\mu+\gamma_{m\mu\nu}\partial^mX^{\nu}~~~.
\eea
It is also written in terms of GL(5) tensors as
\bea
\left\{
\begin{array}{ccl}
\dd_{\underline{m}}&=&P_{\underline{m}}+\partial^{\underline{n}}
X_{\underline{mn}}
+\bar{\partial}_{\underline{m}}\bar{X}\\
\dd^{\underline{m}\underline{n}}&=&
P^{\underline{mn}}+\partial^{[\underline{n}}X^{\underline{m}]}
+\frac{1}{2}\epsilon^{\underline{mnl_1l_2l_3}}\bar{\partial}_{\underline{l}_1}
X_{\underline{l}_2\underline{l}_3}\\
\bar{\dd}&=&\bar{P}+\bar{\partial}_{\underline{m}} X^{\underline{m}}
X_{\underline{mn}}
\end{array}
\right.
\eea

\begin{enumerate}
	\item{Section of ${\cal A}$-theory $\to$ ${\cal M}$-theory 
		
		D=4 ${\cal M}$-theory has GL(5) $G$-symmetry and the background gauge field is GL(5)/SO(5) parameter which is
		the 5-dimensional metric.
		10-representation of SO(5,5) is decomposed by GL(5) as 
		$10\to 5+5$, $\partial^m\to(\partial^{\underline{m}},\bar{\partial}_{\underline{m}})$ with 
		$\underline{m}=1,\cdots,5$. 
		16 spacetime coordinate $P_M=P_{\mu}$ is decomposed as $5+10+1$ as $(P_{\underline{m}},~P^{\underline{mn}},~\bar{P})$. 
		The spacetime section conditions 
		\bea
		P_{\underline{n}}P^{\underline{mn}}=0=P_{\underline{m}}\bar{P}
		+\frac{1}{8}\epsilon_{\underline{m}\underline{m}_1\cdots \underline{m}_4}P^{\underline{m}_1\underline{m}_2}
		P^{\underline{m}_3\underline{m}_4}
		\eea
		are solved in 
		\bea
		P_{\underline{m}}\neq 0~~,~~	P^{\underline{m}_1\underline{m}_2}= \bar{P}=0~~~. 
		\eea
		The world-volume is chosen as
		\bea
		\partial^{\underline{m}}\neq 0~~,~~\bar{\partial}_{\underline{m}}=0~~~,
		\eea
		in such a way that  currents of D=4 ${\cal M}$-theory are the momentum and the winding mode
		\bea
		\dd_{\underline{m}}=P_{\underline{m}}~~,~~
		\dd^{\underline{m}_1\underline{m}_2}=\partial^{[\underline{m}_2}X^{\underline{m}_1]}~~
		\eea
		with the trivial covariant derivative $\bar{\dd}=0$.
		The nontrivial Virasoro operators and Gau\ss{} law operators are
		\bea
		&	{\cal S}^{\underline{m}}=P_{\underline{n}}\partial^{[\underline{m}}X^{\underline{n}]}~~,~~
		{\cal S}_{\underline{m}}=\epsilon_{\underline{m}\underline{m}_1\cdots \underline{m}_4}\partial^{\underline{m}_1}X^{\underline{m}_2}
		\partial^{\underline{m}_3}X^{\underline{m}_4}
		&\nn\\
		&{\cal U}_{\underline{m}_1\underline{m}_2}=\epsilon_{\underline{m}_1\cdots\underline{m}_5}
		\partial^{\underline{m}_3}X^{\underline{m}_4}\partial^{\underline{m}_5}
		~~,~~\bar{\cal U}=P_{\underline{m}}\partial^{\underline{m}}~~,
		\label{SUMD=4}&
		\eea
		while the remaining Gau\ss{} law operator and the world-volume section condition are trivial, ${\cal U}^{\underline{m}}=0$ and ${\cal V}=\partial^{\underline{m}}\bar{\partial}_{\underline{m}}=0$.}
	
	\item{Section of ${\cal M}$-theory $\to$ S-theory 
		
		D=4 S-theory has GL(4) $G$-symmetry and the background gauge field is GL(4)/SO(4) parameter
		which is the 4-dimensional metric field. 5-representation of GL(5) is decomposed by GL(4) as $5\to 4+1$, $\underline{m}\to (\bar{m},~5)$. 
		World-volume section conditions, $\partial^{\bar{m}}P_{\bar{m}}+\partial^5P_5=0$ 
		and 
		$\partial^{[\bar{m}_1}X^{\bar{m}_2]}\partial^5 +
		\partial^{5}X^{[\bar{m}_1}\partial^{\bar{m}_2]}
		-	 \partial^{[\bar{m}_1|}X^{5}\partial^{|\bar{m}_2]}
		=0$  in \bref{SUMD=4}, are solved 		as 
		\bea
		\partial^5\neq 0~~,~~	\partial^{\bar{m}}=0~~\Rightarrow~~P_5=0~~,~~P_{\bar{m}}\neq 0~~~.
		\eea
		Then the non-trivial Virasoro operator is written in terms of
		currents which are the momentum and the winding mode
		\bea
		{\cal S}^5=P_{\bar{m}}\partial^5 X^{\bar{m}}
		~~,\label{SUSD=4}	~~~	
		\dd_{\bar{m}}=P_{\bar{m}}~~,~~
		\dd^{\bar{m}5}=\partial^5 X^{\bar{m}}~~~.
		\eea
  ${\cal S}^5$ is the usual Virasoro operator for a string.
	}
	
	\item{Section of ${\cal A}$-theory $\to$ ${\cal T}$-theory
		
		D=4 ${\cal T}$-theory has O(4,4) $G$-symmetry and the background gauge field is
		O(4,4)/SO(4)$^2$ parameter which is the 4-dimensional metric and the B field.  
		The world-volume section condition  in \bref{SUVD=4}, 
		${\cal V}=\eta_{mn}\partial^{m}\partial^n=0$, 	is solved analogously in the lightcone gauge  
		\bea
		\partial^{+}\neq 0~~,~~\partial^m=0 ~~{\rm for}~  m\neq +~~~.
		\eea
		Then the Gau\ss{} law constraint in \bref{SUVD=4}, 
		$\partial^{m}\gamma_{{m}}{}^{\mu\nu}P_{{\mu}}=0$, 	is solved as
		\bea
		\gamma^- P=0~~\Rightarrow~~{\cal P}^+P=P_{\underline{\mu}}\neq 0 ~{\rm with}~\underline{\mu}=1,\cdots,8~~~
		\eea
		with the lightcone projection operator ${\cal P}^\pm$.	
		The  nontrivial Virasoro operator of the D=4 ${\cal T}$-theory is written in terms of the 8 component selfdual  current as
		\bea
		&{\cal S}^+=\frac{1}{4}\dd_{\underline{\mu}}\gamma^{+\underline{\mu}\underline{\nu}}\dd_{\underline{\nu}}\label{SVirasoroTD=4}~~,~~			\dd_{\underline{\mu}}=P_{\underline{\mu}}+\partial^+(\gamma^-  X)_{\underline{\mu}}~~~.&
		\eea}
	\item{Section of ${\cal T}$-theory $\to$ S-theory
		D=4 S-theory is obtained by the spacetime  section condition,  $\gamma^{+\underline{\mu}\underline{\nu}}
		P_{\underline{\mu}}P_{\underline{\nu}}=0$ from \bref{SVirasoroTD=4}.
		The 8-component lightcone spinor is decomposed by U(4) as $P_{\underline{\mu}}\to
		(P_{\bar{m}},\bar{P}^{\bar{m}})$ with $\bar{m}=1,\cdots,4$, resulting 
		the section condition $P_{\bar{m}}\bar{P}^{\bar{m}}=0$. 
		This is solved as
		\bea
		&P_{\bar{m}}\neq 0~~,~~\bar{P}^{\bar{m}}=0~~~&
		\eea
		The  nontrivial Virasoro operator of the D=4 S-theory is written in terms of the 4-dimensional momentum and the winding mode 
		\bea
		&					{\cal S}^+=P_{\bar{m}}\partial^+ X^{\bar{m}}~~,~~
		\dd_{\bar{m}}=P_{\bar{m}}~~,~~
		\dd^{\bar{m}}=\partial^+ X^{\bar{m}}~~~.&
		\eea
  ${\cal S}^+$ is the usual Virasoro operator for a string.}
\end{enumerate}
\vskip 6mm

\section{Curved background}\label{section:5}
In this section, we will describe a method of coupling flat space currents defining the brane theories to a curved background. This is done simply by making the flat currents couple to target space and worldline vielbeins $\dd_{\cal M} \rightarrow E_{\cal A}^{\hspace{5pt}{\cal M}}\dd_{\cal M} \equiv \dd_{\cal A}$, and 
{
${\partial}^{n} \rightarrow e_{n}^{\; a}{\partial}^{n} \equiv {\cal D}^{a}$}. 

By studying the current algebra in curved backgrounds, we can learn a lot about the dynamics of the theories.
With the curved background, the commutators are modified by torsion terms:
\begin{equation}
	[\dd_{\cal A}(1), \dd_{\cal B}(2)] = T_{\cal A B}^{\hspace{9pt}\cal C}\dd_{\cal C}(1)\delta(1-2) + G_{\;\;\cal A B}^{\; a}{\cal D}_{a}(1)\delta(1-2).
\end{equation}
The Bianchi identities could be solved by specifying the torsion constraints. Solving the torsion constraints will single out various components of the vielbein $ E_{\cal A}^{\hspace{5pt}{\cal M}}$, giving us the physical fields and their field strengths. 

Before going into the details of how the general formalism works, we would like to work out a simple case in $D = 3$ ${\cal A}$-theory, and only the backgrounds of the translations (The $P$ generators of the full non-degenerate Poincar\'e algebra.) are considered.

\vskip 6mm
\subsection{$G/H$=SL(5)/SO(5) background gauge fields}\label{section:5-1}

The SL(5) current algebra in \bref{SL5CA} gives the following commutator of two vectors $\Lambda_1{}^{m_1m_2}(X)$, $\Lambda_2{}^{m_1m_2}(X)$
\bea
&&[\frac{1}{2}\Lambda_1{}^{m_1m_2}\dd_{m_1m_2}(1),\frac{1}{2}\Lambda_2{}^{m_3m_4}\dd_{m_3m_4}(2)]\nn\\
&&~~~~~~~~~~=\frac{i}{4}\left[\epsilon_{m_1\cdots m_5}\left(\Lambda_1{}^{m_1m_2}\Lambda_2{}^{m_3m_4}(1)+\Lambda_1{}^{m_1m_2}\Lambda_2{}^{m_3m_4}(2)\right)
	\partial^{m_5}\delta(1-2)\right.\nn\\&&\left.
	~~~~~~~~~~~~-\delta(1-2)\left(\Lambda_{[1|}{}^{m_1m_2}\partial_{m_1m_2}\Lambda_{|2]}{}^{n_1n_2}
	-\frac{1}{8}\Lambda_{[1|}{}^{[m_1m_2}\partial_{m_1m_2}\Lambda_{|2]}{}^{n_1n_2]}
	\right)\dd_{n_1n_2}\right]\nn\\\label{SL5Counrant}
\\
&&~~~~~~~~~~=\frac{i}{2}\left[\epsilon_{m_1\cdots m_5}\Lambda_1{}^{m_1m_2}\Lambda_2{}^{m_3m_4}(1)\partial^{m_5}\delta(1-2)\right.\nn\\&&\left.
	~~~~~~~~~~~+\frac{1}{2}\delta(1-2)\left(\Lambda_{2}{}^{m_1m_2}\partial_{m_1m_2}\Lambda_{1}{}^{n_1n_2}
	+\Lambda_{1}{}^{n_1n_2}\partial_{m_1m_2}\Lambda_{2}{}^{m_1m_2}\right.\right.\nn\\&&
	~~~~~~~~~~~~~~~~~~~~~~~~~~~~~~~~~~~~~~~~~~~~~~~~~~~~~~~~
	\left.\left.
	+\Lambda_{1}{}^{[n_1|[m_1}\partial_{m_1m_2}\Lambda_{2}{}^{m_2]|n_2]}
	\right)\dd_{n_1n_2}\right]\nn\\~~~.\label{SL5Dorfman}
\eea
The regular terms in the right hand side of \bref{SL5Counrant} gives the SL(5) Courant bracket
\bea
[\Lambda_1,\Lambda_2{}]_{\rm C}&=&\Lambda_{12}\nn\\
\Lambda_{12}{}^{n_1n_2}&=&\Lambda_{[1|}{}^{m_1m_2}\partial_{m_1m_2}\Lambda_{|2]}{}^{n_1n_2}
-\frac{1}{8}\Lambda_{[1|}{}^{[m_1m_2}\partial_{m_1m_2}\Lambda_{|2]}{}^{n_1n_2]}~~~
\eea
which is anti-symmetric in $1$ and $2$ of $\Lambda_1$ and $\Lambda_2$ \cite{Hatsuda:2012uk}.
On the other hand, the ones of \bref{SL5Dorfman} gives
the SL(5) Dorfman bracket  as
\bea
[\Lambda_1,\Lambda_2{}]_{\rm D}&=&\delta_{\Lambda_2}\Lambda_{1}\nn\\
\delta_{\Lambda_2}\Lambda_{1}{}^{n_1n_2}&=&\Lambda_{2}{}^{m_1m_2}\partial_{m_1m_2}\Lambda_{1}{}^{n_1n_2}
+\Lambda_{1}{}^{n_1n_2}\partial_{m_1m_2}\Lambda_{2}{}^{m_1m_2}
+\Lambda_{1}{}^{[n_1|[m_1}\partial_{m_1m_2}\Lambda_{2}{}^{m_2]|n_2]}~~~\nn\\\label{SL5Dorfmantransf}.
\eea
which is useful to compute gauge transformation rules \cite{Hatsuda:2012uk}.
The SL(5) gauge transformation of
${\cal O}^{m_1m_2}$ is given by the Dorfman bracket with
$\Lambda_1{}^{m_1m_2}={\cal O}{}^{m_1m_2}$ and $\Lambda_2{}^{m_1m_2}=\lambda^{m_1m_2}$ in \bref{SL5Dorfmantransf}
\bea
&&\frac{1}{2}\delta_\lambda {\cal O}^{m_1m_2}\dd_{m_1m_2}=-i\displaystyle\int d^5\sigma'
[\frac{1}{2}{\cal O}^{m_1m_2}\dd_{m_1m_2}(\sigma),
\frac{1}{2}\lambda^{n_1n_2} \dd_{n_1n_2}(\sigma')]\\
&&\delta_\lambda {\cal O}^{m_1m_2}=\frac{1}{2}
\left[
\lambda^{n_1n_2}\partial_{n_1n_2}{\cal O}^{m_1m_2}
+{\cal O}^{m_1m_2}\partial_{n_1n_2}\lambda^{n_1n_2}
+{\cal O}^{[m_1|[n_2}\partial_{n_1n_2}\lambda^{n_2]|m_2]}
\right]\nn
\eea
There is the gauge symmetry of gauge symmetry
by using \bref{delsigma} as
\bea
&&\delta_\lambda \lambda^{m_1m_2}=\frac{1}{2}\epsilon^{m_1\cdots m_5}\partial_{m_3m_4}\lambda_{m_5}~~.\label{gaugegauge}\\
&&\delta_\lambda\displaystyle\int d^5\sigma  \frac{1}{2}\lambda^{m_1m_2}\dd_{m_1m_2}(\sigma)
=\displaystyle \int d^5\sigma \frac{1}{4}\epsilon^{m_1\cdots m_5}(\partial_{m_3m_4}\lambda_{m_5})\dd_{m_1m_2}(\sigma) \nn\\
&&~~~~~~~~~~~~~~~~~~~~~~~~~~~~~~~~~~~=\displaystyle 2\int d^5\sigma  \partial^{m}\lambda_m=0~~~.
\eea
\vskip 6mm

$G/H$=SL(5)/SO(5) background gauge fields are $E_{a_1a_2}{}^{m_1m_2}=E_{[a_1}{}^{m_1}E_{a_2]}{}^{m_2}$ and
$E_{a}{}^{m}(X)$ representing the spacetime and world-volume vielbein. $E_{a}{}^{m}$ satisfies the SL(5) condition \bref{SL5} as
\bea
\epsilon_{{m}_1\cdots {m}_5}
E_{{a}_1}{}^{{m}_1}E_{{a}_2}{}^{{m}_2}E_{{a}_3}{}^{{m}_3}E_{{a}_4}{}^{{m}_4}E_{{a}_5}{}^{{m}_5}
=\det E_a{}^m~\epsilon_{\hat{a}_1\cdots \hat{a}_5}~~,~~\det E_a{}^m=1~~~. \label{SL5bg}
\eea
It is transformed as $E_{a}{}^{m}\to h_a{}^b E_b{}^nA_n{}^m$ with $A_m{}^n\in$SL(5) and $h_a{}^b\in$SO(5).
Currents and derivatives in the background are given by
\bea
&\dd_{a_1a_2}={E}_{a_1a_2}{}^{m_1m_2}\dd_{m_1m_2}=
	{E}_{a_1}{}^{m_1}{E}_{a_2}{}^{m_2}\dd_{m_1m_2}~~,~~{\cal D}^a=E_m{}^a\partial^m&~~~.
\eea
with $E_a{}^mE_m{}^b=\delta_a^b$.

The SL(5) gauge transformation of
$E_{ab}{}^{m_1m_2}$ is given by the second line of \bref{gaugegauge} as
\bea
&&\frac{1}{2}\delta_\lambda E_{ab}{}^{m_1m_2}\dd_{m_1m_2}=-i\displaystyle\int d^5\sigma'
[\frac{1}{2}E_{ab}{}^{m_1m_2}\dd_{m_1m_2}(\sigma),
\frac{1}{2}\lambda^{n_1n_2} \dd_{n_1n_2}(\sigma')]\nn\\
&&\delta_\lambda E_{ab}{}^{m_1m_2}=\frac{1}{2}
\left[
\lambda^{n_1n_2}\partial_{n_1n_2}E_{ab}{}^{m_1m_2}
+E_{ab}{}^{m_1m_2}\partial_{n_1n_2}\lambda^{n_1n_2}
+E_{ab}{}^{[m_1|[n_2}\partial_{n_1n_2}\lambda^{n_2]|m_2]}
\right]~~.\nn\\
\eea
The one of $E_a{}^m$ is obtained by replacing as
$E_{ab}{}^{m_1m_2}=E_{[a}{}^{m_1}E_{b]}{}^{m_2}$
\bea
\delta_\lambda (E^{1/3}E_{a}{}^{m})=\frac{1}{2}
\lambda^{n_1n_2}\partial_{n_1n_2}(E^{1/3}E_{a}{}^{m})+
E^{1/3}E_{a}{}^{(m|}\partial_{n_1n_2}\lambda^{|n_1)n_2}~~
\eea
with $E=\det E_a{}^m$ and $E=1$ for SL(5).

The algebra of the curved background current is given by
\bea
[\dd_{a_1a_2}(1),\dd_{a_3a_4}(2)]&=&2i\epsilon_{a_1\cdots a_5}{\cal D}^{a_5}\delta(1-2)
-\frac{i}{2}T_{a_1a_2~a_3a_4}{}^{c_1c_2}\dd_{c_1c_2}\delta(1-2)).
\label{SL5CA}
\eea
with the torsion $T_{AB}{}^C=T_{a_1a_2~b_1b_2}{}^{c_1c_2}$.
It is given as
\bea
&&T_{a_1a_2~b_1b_2}{}^{c_1c_2}=
\frac{1}{4}\left\{E_{a_1a_2}{}^{m_1m_2}(\partial_{m_1m_2}E_{b_1b_2}{}^{n_1n_2})
-E_{b_1b_2}{}^{m_1m_2}(\partial_{m_1m_2}E_{a_1a_2}{}^{n_1n_2})\right\}E_{n_1n_2}{}^{c_1c_2}\nn\\
&&~~-\frac{1}{32}\left\{E_{a_1a_2}{}^{[m_1m_2}(\partial_{m_1m_2}E_{b_1b_2}{}^{n_1n_2]})
-E_{b_1b_2}{}^{[m_1m_2}(\partial_{m_1m_2}E_{a_1a_2}{}^{n_1n_2]})\right\}E_{n_1n_2}{}^{c_1c_2}~~~.\nn\\\label{TorsionEE}
\eea
It is convenient to introduce $c_{a_1a_2b}{}^c$ in terms of  $E_a{}^m$ as
\bea
c_{a_1a_2b}{}^c=(\partial_{a_1a_2}E_{b}{}^m)E_m{}^c~~,~~c_{a_1a_2b}{}^c=-c_{a_2a_1b}{}^c~~~\label{cterm}.
\eea
with $\partial_{a_1a_2}={E}_{a_1}{}^{m_1}{E}_{a_2}{}^{m_2}\partial_{m_1m_2}$.
The torsion in \bref{TorsionEE} is rewritten in terms of $E_a{}^m$ using with \bref{cterm} as
\bea
&T_{a_1a_2~b_1b_2}{}^{c_1c_2}=
\frac{1}{4}\left[3c_{a_1a_2[b_1}{}^{[c_1}\delta_{b_2]}^{c_2]}-3c_{b_1b_2[a_1}{}^{[c_1}\delta_{a_2]}^{c_2]}
-c_{[a_1|[b_1b_2]}{}^{[c_1}\delta_{|a_2]}^{c_2]}+c_{[b_1|[a_1a_2]}{}^{[c_1}\delta_{|b_2]}^{c_2]}\right.&\nn\\
&\left.
-c_{d[a_1a_2]}{}^d \delta_{b_1}^{[c_1}\delta_{b_2}^{c_2]}+c_{d[b_1b_2]}{}^d \delta_{a_1}^{[c_1}\delta_{a_2}^{c_2]}
+c_{d[a_1|[b_1}{}^d \delta_{b_2]}^{[c_1}\delta{}^{c_2]}_{|a_2]}
-c_{d[b_1|[a_1}{}^d \delta_{a_2]}^{[c_1}\delta{}^{c_2]}_{|b_2]}
		\right].
&
\eea

Torsion constraints are imposed for superfield in superspace formulation in order to suppress the degrees of freedom. 
The torsion in eq.(5.13) is just a result of the computation. $E$ must satisfy the SL(5) condition, $E^5\epsilon=\epsilon$ which is equal to $\det E=1$ as in \bref{SL5bg}.

The current algebra in the curved background in
\bref{SL5CA} is SL(5) covariant with use of  
the SL(5) invariant tensor as $ E^2\epsilon E^2 E=\epsilon$ brings back to $\epsilon$ by 
transformation of ${\cal D}\to E {\cal D}$.  

In a curved background the $\tau$-component Virasoro constraint ${\cal T}=0$ is modified as
\bea
&&{\cal T}=\frac{1}{16}\dd_{m_1m_2}G^{m_1m_2; m_3m_4}\dd_{m_3m_4}=\frac{1}{16}\dd_{a_1a_2}\delta^{a_1[a_3}\delta^{a_4]a_2}
\dd_{a_3a_4}\\
&&
G^{m_1m_2; m_3m_4}=E_{a_1}{}^{m_1}E_{a_2}{}^{m_2}\delta^{a_1[a_3}\delta^{a_4]a_2}  E_{a_3}{}^{m_3}E_{a_4}{}^{m_4}~~~,
\nn
\eea
while  ${\cal S}^m=0$ and ${\cal U}_m=0$ are preserved because of
${\cal S}^a={\cal S}^m{E}_m{}^a=0$ and ${\cal U}_a={\cal U}_mE_a{}^m=0$.
\vskip 6mm

\subsection{Vielbein, torsion and gauge transformations}\label{section:5-2}
Now returning to the general treatment of ${\cal A}$-theory in curved backgrounds.
The expression for the torsion follows from its definition in the current algebra with background:
\begin{equation}
        T_{\cal AB}{}^{\cal C} = E_{\cal A}{}^{\cal M} E_{\cal B}{}^{\cal N} f_{\cal MN}{}^{\cal P} E_{\cal P}{}^{\cal C} +(\delta_{[{\cal A}|}^{\cal E} \delta_{\cal D}^{\cal C}- \frac{1}{2}G_{[{\cal A}|{\cal D}}{}^a G^{\cal CE}{}_a)(\nabla_{\cal E} E_{|\cal B)}{}^{\cal M})E_{\cal M}{}^{\cal D}.
\end{equation}
We have used the following relations: for any superspace function $f$
\begin{equation}
    \begin{split}
         i[\dd_{\cal A}(1),f(2) \}& \equiv (\nabla_{\cal A}f)\delta(1-2)\\
         \mathcal{D}_a f &= \frac{1}{2}G^{\cal BC}{}_a (\nabla_{\cal B} f)\dd_{\cal C}\\
    \end{split}
\end{equation}
and $G^{\cal AB}{}_a$ is the curved analog of $\eta^{\cal NM}{}_{n}$.
\begin{equation}
	G^{\cal AB}{}_a = E_{\cal M}{}^{\cal A} E_{\cal N}{}^{\cal B} \eta^{\cal MN}{}_p {e}
 _a{}^p.
\end{equation}

 For the rest of the section, we shall only work in the $PD{\Omega}$ superspace, that is we restrict to functions that commute with $S$ and $\Sigma$. i.e. $\nabla_{\Omega} = 0$. Therefore, we exclude fields that have spins (transform non-trivially under $H$-rotations.). Although we could still identify the prepotential from the torsion constraints in $PD{\Omega}$ superspace, we could not get the expression of the generalized Riemann tensor in terms of the prepotential. The treatment of full superspace is left for future investigation.

The Bianchi identities are
\bea
	0&=&T_{[\cal AB}{}^{\cal E} T_{\cal C)E}{}^{\cal D} + (\delta_{[\cal A}^{\cal E} \delta_{\cal F}^{\cal D} - \frac{1}{2}G_{\cal F[A|}{}^a G^{\cal DE}{}_a) \nabla_{\cal E} T_{|\cal BC)}{}^{\cal F}\\
        T_{\cal C(A}{}^{\cal E} G_{\cal B]E}{}^d& =& F_{\cal ABC}{}^d - \frac{1}{2}F_{\cal C(AB]}{}^d\\
        -\frac{3}{4} F_{\cal C[AB)}{}^d&=&T_{\cal AB}{}^{\cal E} G_{\cal CE}{}^d - \frac{1}{2}T_{\cal C[A}{}^{\cal E} G_{\cal B)E}{}^d 
\eea
where $F_{\cal ABC}{}^d$ may be non-zero in a curved background in general .
\begin{equation}
	F_{\cal ABC}{}^d \equiv ( \nabla_{\cal C} G_{\cal AB}{}^m)\epsilon_m{}^d = \nabla_{\cal C} G_{\cal AB}{}^d + f_{{\cal C}e}{}^d G_{\cal AB}{}^e
\end{equation}
\begin{equation}
	 f_{{\cal A}b}{}^c \equiv (\nabla_{\cal A} \epsilon_b{}^m)\epsilon_m{}^c 
\end{equation}

Gauge transformations are generated by
\begin{equation}
 \delta = i[\int d^{\rm d}\sigma \Lambda^A \dd_{\cal A} ,\;\cdot\;].
\end{equation}
Acting on another $\dd$, the former gives
\begin{equation}\label{fullvariation}
	(\delta E_{\cal A}{}^{\cal M})E_{\cal M}{}^{\cal B} = \Lambda^{\cal C} T_{\cal CA}{}^{\cal B} - (\delta_{\cal A}^{\cal D} \delta_{\cal C}^{\cal B} - G_{\cal AC}{}^e G^{\cal BD}{}_e)\nabla_{\cal D} \Lambda^{\cal C} 
\end{equation} 
Due to the world-volume gauge symmetry, there also exists ${\cal U}$ and ${\cal V}$ constraints. 
\bea
		\mathcal{U}^{n}_{\;N} & =&  (U^{n}_{\;m})_{N}^{\;M}P_M \mathcal{D}_n \approx 0\\
        (U^{n}_{\;m})_{N}^{\;M}&=& \delta_N^M \delta_m^n - \eta_{PN}{}^n \eta^{PM}{}_m \equiv U_{Nm}^\varpi U^{Mn}_\varpi \\
		{\mathcal{V}}_{\varpi M}& =&U^{Nn}_\varpi \eta_{NM}{}^m \mathcal{D}_m\mathcal{D}_n\approx 0
\eea
These constraints are slightly modified compared with (\ref{uvconstriants}); the bare world-volume derivative $\partial_{n}$ is replaced with the covariant derivative ${\cal D}_{n}$. 
Here $U_{Nm}^\varpi$ stands for the decomposition of $(U^{n}_{\;m})_{M}^{\;N}$ with respect to the representation of world-volume gauge symmetry. For example, in this D=3 case, we have $\varpi$ living in a 5-representation of SL(5)
\begin{equation}
	(U^{n}_{\;m})_{M}^{\;N}=\delta_{[{n}_1}^{{n}}\delta_{{n}_2]}^{[{m}_1}\delta_{{m}}^{{m}_2]}= \delta_{[{n}_1}^{{n}}\delta_{{n}_2]}^{\varpi}\delta_\varpi^{[{m}_1}\delta_{{m}}^{{m}_2]}.
\end{equation}

For future convenience, we will denote
\begin{equation}
    \begin{split}
    \mathcal{U}_{Mn} f &\equiv (U^{n}_{\;m})_{M}^{\;N}\mathcal{D}_n \nabla_N f \\
	\mathcal{U}_{Mn}(f,g)&\equiv (U^{n}_{\;m})_{M}^{\;N}(\nabla_N f)(\mathcal{D}_n g). \\
    \end{split}
\end{equation}
for the two derivatives in $\mathcal{U}$ acting on the same function or different functions. We also define the action of $\mathcal{U}_\varpi f, \; \mathcal{U}_\varpi(f, g)$ and $\mathcal{V}_{\varpi M} f, \; \mathcal{V}_{\varpi M}(f, g) $ in a similar fashion.

Since we already have the gauge transformation of the vielbein (\ref{fullvariation}), one could check whether torsion $T_{\cal AB}{}^{\cal C}$ and the affine term $G_{\cal AB}{}^a$ are gauge covariant. Indeed, we find that they are covariant up to ${\cal U}$ constraints.

$G_{\cal AB}{}^a$ is covariant only up to $\mathcal{U}$ terms.
$$ (\delta G_{\cal AB}{}^{{a}})\mathcal{D}_a \approx \mathcal{U} $$
Writing every component explicitly, we have:
\begin{equation}
\begin{split}
\delta G_{DD}{}^{a}\mathcal{D}_a&=0 \\
\delta G_{DP}{}^{a}\mathcal{D}_a &=\delta  G_{\alpha A}{}^{a}\mathcal{D}_a=(U^{a}_{\;b})_{A}^{\;B}\nabla_{B}\Lambda_{\alpha}{}^{{b}}\mathcal{D}_a= \mathcal{U}_{A{{b}}}(\Lambda_{\alpha}{}^{{b}}, \;\cdot \;)\\
\delta G_{PP}{}^a\mathcal{D}_a&=0\\
\delta G_{D\Omega}{}^a\mathcal{D}_a&=\delta G_{\alpha}{}^\beta{}_{{b}}{}^{{a}} \mathcal{D}_a= \delta_\alpha^\beta (U^{a}_{\;b})_{A}^{\;B}\nabla_B\Lambda^A\mathcal{D}_a=\delta_\alpha ^\beta\mathcal{U}_{B{b}}(\Lambda^B,\;\cdot \;)\\
\delta G_{P\Omega}{}^a\mathcal{D}_a&=\delta G_{A{b}}{}^{\alpha}{}^a\mathcal{D}_a= (U^{a}_{\;b})_{A}^{\;B}\nabla_{B}\Lambda^\alpha \mathcal{D}_a=\mathcal{U}_{A{{b}}}(\Lambda^{\alpha},\;\cdot \;)\\
\delta G_{\Omega\Omega}{}^a\mathcal{D}_a&=0. \\    
\end{split}
\end{equation}
Torsions are also covariant only up to $\mathcal{U}$ terms. They can be discarded when the torsion is coupled with 
$\dd_\Omega( =\mathcal{D}\theta )$ or $\dd_P$ \cite{Linch:2015fca,Linch:2016ipx}.

Here we list the gauge transformations of torsions with extra pieces up to dimension 1.
\begin{equation}
	\begin{split}
        \delta T_{DD}{}^\Omega\dd_\Omega&=\delta T_{\alpha\beta\gamma}{}^{{a}}\mathcal{D}_a\theta^\gamma=-(U^{a}_{\;b})_{B}^{\;C}\nabla_C\Lambda_{(\alpha}{}^{b}\gamma_{\beta)\gamma}^B \mathcal{D}_a\theta^\gamma  \\
&=-\gamma_{\gamma(\beta}^B\mathcal{U}_{B{b}}(\Lambda_{\alpha)}{}^{{b}},\theta^\gamma)\\
\delta T_{DP}{}^\Omega\dd_\Omega&=\delta T_{\alpha A\beta}{}^{{a}}\mathcal{D}_a\theta^\beta=(U^{a}_{\;b})_{A}^{\;C}\mathcal{D}_a\theta^\beta\nabla_C(\Lambda^D\eta_{DE}{}^{{b}}\gamma_{\alpha\beta}^E+\nabla_{(\alpha}\Lambda_{\beta)}{}^{{b}}) \\
&=\mathcal{U}_{A{b}}(\gamma_{\alpha\beta}^E\eta_{DE}{}^{{b}}\Lambda^D+\nabla_{(\alpha}\Lambda_{\beta)}{}^{{b}},\theta^\beta) \\
\delta T_{DP}{}^P\dd_P&=\delta T_{\alpha A}{}^B\dd_B=\eta^{BC}{}_{{a}}(U^{a}_{\;b})_{A}^{\;D}\nabla_{C}\nabla_D \Lambda_\alpha{}^{{b}} \dd_b=(U^{b}_{\;a})_{A}^{\;D}\mathcal{D}_a\nabla_D\Lambda_\alpha{}^{b} \\
&=\mathcal{U}_{A{b}}\Lambda_\alpha{}^{{b}}\\
\delta T_{PP}{}^\Omega\dd_\Omega&=\delta T_{AB}{}^{{a}}{}_\alpha\mathcal{D}_a\theta^\alpha=-\nabla_\alpha \eta_{C[A|}{}^{{b}}
(U^{a}_{\;\;b})_{|B]}^{\;D}\mathcal{D}_a\theta^\alpha\nabla_{D}\Lambda^C \\
&= -\eta_{C[A|}{}^{{b}}\mathcal{U}_{|B]{b}}(\nabla_\alpha\Lambda^C,\theta^\alpha) \\
	\end{split}
\end{equation}

\begin{equation}
\begin{split}
\delta T_{D\Omega}{}^\Omega\dd_\Omega&=\delta T_{\alpha{a}}{}^{\beta{b}}{}_\gamma\mathcal{D}_b\theta^\gamma= -\gamma_{\alpha\gamma}^A(U^{a}_{\;b})_{A}^{\;B}\nabla_B\Lambda^\beta \mathcal{D}_b\theta^\gamma \\
&= -\gamma_{\alpha\gamma}^A\mathcal{U}_{A{a}}(\Lambda^\beta,\theta^\gamma)\\
\delta T_{PP}{}^P\dd_P&=\delta T_{AB}{}^C\dd_C=\eta_{D[A|}{}^{{b}}\eta^{FC}{}_{{a}}(U^{a}_{\;\;b})_{|B]}^{\;E}\nabla_{E}\nabla_{F}\Lambda^D\dd_C=\eta_{D[A|}{}^{{b}}(U^{a}_{\;\;b})_{|B]}^{\;E}\mathcal{D}_a\nabla_E\Lambda^D  \\
&=\eta_{D[A|}{}^{{b}}\mathcal{U}_{|B]{b}}\Lambda^D \\
\delta T_{P\Omega}{}^\Omega\dd_\Omega&=\delta T_{A{a}}{}^{\alpha{b}}{}_\beta\mathcal{D}_b\theta^\beta=\frac{1}{2}(U^{b}_{\;a})_{A}^{\;B}\nabla_\beta\nabla_B\Lambda^\alpha\mathcal{D}_b\theta^\beta  \\
&=\frac{1}{2}\mathcal{U}_{A{a}}(\nabla_\beta\Lambda^\alpha,\theta^\beta) 
 \end{split}
\end{equation}
When calculating $\delta T_{DP}{}^P$ or $\delta T_{PP}{}^P$ above, we have used the fact that we can convert a spacetime derivative, together with a current, into a world-volume derivative up to ${\cal S}$ section conditions.
\bea
\eta^{CD}{}_{{a}}\nabla_D \Lambda\dd_C=2[ \frac{1}{2}\eta^{CD}{}_{{a}}\dd_{{C}}\dd_D , \Lambda ]\approx \mathcal{D}_a\Lambda 
\eea

\vskip 6mm
\subsection{Gauge transformations of the gauge parameters}\label{section:5-3}

Gauge transformations of the gauge parameters are defined as variations on spacetime gauge transformation parameters $\Lambda$ that leave the generator $\delta =i[\int d\sigma  \Lambda^{\cal A} \dd_{\cal A},] $ invariant, up to $\mathcal{S}$ or $\mathcal{U}$ constraints.

	Identifying the gauge transformation of gauge transformation is important for several purposes. First, they usually correspond to symmetries of the theory with physical significance, as we shall see below. Also, in order to identify the physical components of the vielbein, one needs to see which pieces of the vielbein could be gauged away by the gauge parameters, therefore knowing the pieces of the gauge parameters that could be used to eliminate the various components of the vielbein are important. In terms of covariant quantization, the gauge corresponds to the additional ghosts that need to be considered.  

Currently, there are three kinds of the gauge of gauge transformation are identified. 

\begin{itemize}
    \item The first kind of gauge of gauge transformation is 
    \begin{equation}
		\delta \Lambda^A\dd_A =-\eta^{BA}{}_{{a}}\nabla_B \lambda^{{a}}\dd_A =-2\mathcal{D}_a\lambda^{{a}}.
	\end{equation} 
	Up to $\mathcal{S}$ constraint, this transformation corresponds to the diffeomorphism of the world-volume. The transformation of linearized vielbein $H_{\cal A}^{\; B}$ ($\equiv E_{\cal A}^{\; B} - \delta_{\cal A}^{\; B}$)) under this transformation is listed below. 

\begin{equation}
    \begin{split}
\delta H_D{}^\Omega \dd_\Omega&=\delta H_{\alpha\beta}{}^{{a}}\mathcal{D}_a\theta^\beta=\gamma_{\alpha\beta}^A U_{A{b}}^{B{a}}\nabla_B \Lambda^{{b}} \mathcal{D}_a\theta^\beta\\
&= \gamma_{\alpha\beta}^A\mathcal{U}_{A{b}}(\lambda^{{b}},\theta^\beta) \\
\delta H_P{}^\Omega\dd_\Omega&=\delta H_{A\alpha}{}^{{a}} \mathcal{D}_a\theta^\alpha=U_{A{b}}^{B{a}}\nabla_\alpha\nabla_B\Lambda^{{b}}\mathcal{D}_a\theta^\alpha \\
&=\mathcal{U}_{A{b}}(\nabla_\alpha\lambda^{{b}},\theta^\alpha) \\
\delta H_P{}^P\dd_P&=\delta H_a{}^B\dd_B= U^{E{a}}_{A{b}}\eta^{BD}{}_{{a}}\nabla_{D}\nabla_{E}\Lambda^{{b}}\dd_B\\
&=\mathcal{U}_{A{b}}\lambda^{{b}} \\
\delta (\rm{others})&=0 \\
    \end{split}
\end{equation}
One can see that all the transformations are proportional to the $\mathcal{U}$ constraints.

\item The second kind of the gauge of gauge transformation is a $\kappa$-symmetry-like transformation.
\begin{equation}
	\delta \Lambda^\Omega \dd_\Omega =\delta\Lambda^{{a}}{}_\alpha{\mathcal{D}_a}\theta^\alpha=-U^{A{a}}_\varpi\nabla_A \Lambda^\varpi_\alpha {\mathcal{D}_a}\theta^\alpha =-\mathcal{U}_{\varpi}(\Lambda_\alpha^\varpi,\theta^\alpha) 
\end{equation}

Up to $\mathcal{U}$ constraint on $\theta$:
\begin{equation}
    \begin{split}
\delta H_D{}^\Omega \dd_\Omega &=\delta H_{\alpha\beta}{}^{{a}}{\mathcal{D}_a}\theta^\beta=U^{A{a}}_\varpi\nabla_A\nabla_{(\beta} \Lambda^\varpi_{\alpha)}{\mathcal{D}_a}\theta^\beta  \\
&= \mathcal{U}_\varpi ( \nabla_{(\beta}\Lambda_{\alpha)}^\varpi,\theta^\beta) \\
\delta H_D{}^P\dd_P&= \delta H_{\alpha}{}^A\dd_A=-\eta^{AB}{}_{{a}}\nabla_BU^{C{a}}_\varpi\nabla_C \Lambda^\varpi_\alpha \dd_A\\
&=-\mathcal{U}_\varpi \Lambda_\alpha ^\varpi \\
\delta H_P{}^\Omega\dd_\Omega&=\delta H_{A\alpha}{}^{{a}}\mathcal{D}_a\theta^\alpha =\nabla_A U^{B{a}}_\varpi\nabla_B \Lambda^\varpi_\alpha \mathcal{D}_a\theta^\alpha \\
&=\mathcal{U}_\varpi (\nabla_A \Lambda^\varpi _\alpha, \theta^\alpha) \\
\delta ({\rm others})&=0 \\
    \end{split}
\end{equation}
Again, all the transformations are proportional to the $\mathcal{U}$ constraints.
\item  The third kind of the gauge of gauge transformation generates a world-volume gauge transformation on $X$.
\begin{equation}
	\delta \Lambda^P \dd_P =\delta \Lambda^A\dd_A=-U^{A{a}}_\varpi\mathcal{D}_a\Lambda^\varpi \dd_A = -\mathcal{U}_\varpi(\quad, \Lambda^\varpi)
\end{equation}

Up to $\mathcal{U}$ sectioning on $X$, which generates
\begin{equation}
	\begin{split}
\delta H_D{}^P\dd_P&= \delta H_{\alpha}{}^A\dd_A=\nabla_\alpha U^{A{a}}_\varpi\mathcal{D}_a\Lambda^\varpi \dd_A\\
&=\mathcal{U}_\varpi(\quad, \nabla_\alpha\Lambda^\varpi) \\
\delta H_P{}^\Omega\dd_\Omega&=\delta H_{B\alpha}{}^{{b}}\mathcal{D}_b\theta^\alpha =-\eta_{AB}{}^{{b}}\nabla_\alpha U^{A{a}}_\varpi\mathcal{D}_a\Lambda^\varpi \mathcal{D}_b\theta^\alpha  \\
&= -\mathcal{V}_{B\varpi}(\nabla_\alpha\Lambda^\varpi,\theta^\alpha)\\
\delta H_P{}^P \dd_P&=\delta H_A{}^B\dd_B=\nabla_A U^{B{a}}_\varpi\mathcal{D}_a\Lambda^\varpi\dd_B -\eta_{AC}{}^{{b}}\eta^{BD}{}_{{b}}\nabla_D U^{C{a}}_\varpi\mathcal{D}_a\Lambda^\varpi\dd_B \\
&=\mathcal{U}_\varpi(\quad, \nabla_A \Lambda^\varpi) -\mathcal{V}_{A\varpi}\Lambda^\varpi \\
\delta ({\rm others})&=0 
    \end{split}
\end{equation}
This time, all the variations are proportional to $\mathcal{U}$, $\mathcal{V}$.
\end{itemize}
\vskip 6mm

\subsection{Prepotentials}\label{section:5-4}
The general procedure to solve the torsion constraints in superspace is to classify torsion constraints with their dimensions, and solve them from the lowest dimension to the highest one.
We assume
certain pieces of torsion $T$ and the metric $G$ are their flat values, and also some other parts of $T$ and $G$ could be gauged away using the gauge parameter $\Lambda$. We should constrain them as much as possible so the actual degrees of freedom in the theories are as small as possible. 

The dimension of a tensor is determined by subtracting the dimensions associated with upper indices from that for lower indices. Recall that the engineering dimensions of the indices are $[(D,P,\Omega,a)]=(1/2,1,3/2,2)$ where $a$ is the world-volume derivative index of ${\cal D}_a$.  

The relevant gauge parameters $\Lambda^{\cal A}$, linearized potentials $H_{\cal A}{}^{\cal B}$ ($E_{\cal A}{}^{\cal M}=\delta_{\cal A}^{\cal M}+H_{\cal A}{}^{\cal M}$) \footnote{under this decomposition, there is no need to distinguish between the flat and curved indices, so we will use all flat indices for this section.}, metrics $G_{\cal AB}{}^a$, and torsions $T_{\cal AB}{}^{\cal C}$ are labeled by the indices as follows:
$${\rm Engineering~dimensions~of~fields}$$
\begin{center}
    \begin{tabular}{|c|c|c|c|c|}
    \hline
dim & $\Lambda^{\cal A}$ & $H_{\cal A}{}^{\cal B}$\hfil & $G_{\cal AB}{}^a$ & $T_{\cal AB}{}^{\cal C}$ \\\hline
$-\frac{3}{2}$ & $\Omega$ & & & \\
$-1$ & $P$ & $_D{}^\Omega$ & $DD$ & \\
$-\frac{1}{2}$ & $D$ & ${}_D{}^P$ , ${}_P{}^\Omega$ & $DP$ & $_{DD}{}^\Omega$ \\
0 & $S$ & $_D{}^D$ , ${}_P{}^P$ , ${}_\Omega{}^\Omega$ , $({}_a{}^b)$ & $D\Omega$ , $PP$ & $_{DD}{}^P$ , ${}_{DP}{}^\Omega$ \\
$\frac{1}{2}$ & & $_P{}^D$ , ${}_\Omega{}^P$ & $P\Omega$ & $_{DD}{}^D$ , ${}_{DP}{}^P$ , ${}_{D\Omega}{}^\Omega$ , ${}_{PP}{}^\Omega$ \\
1 & & $_\Omega{}^D$ & $\Omega\Omega$ & $_{DP}{}^D$ , ${}_{D\Omega}{}^P$ , ${}_{PP}{}^P$ , ${}_{P\Omega}{}^\Omega$ \\\hline
    \end{tabular}
\end{center}

\noindent 
Where $S$ is a non-derivative $H$-transformations, but we won't use much of it. 
$h_a{}^b$ is the linearized world-volume metric ${e}_a{}^b$.
Without the currents $S$ and $\Sigma$, only the torsions contribute, until the dimension 5/2 $T_{\Sigma\Sigma}{}^D$.
The metric ($G_{\cal AB}{}^a$) constraints ``orthogonality", apart from $G_{PP}{}^{a}$ and $G_{D\Omega}{}^{a}$, are included in torsion constraints now. They are easier to solve than the full torsion 
($T_{\cal AB}{}^{\cal C}$) constraints because they are purely algebraic, and should be solved first at any particular dimension.

\subsubsection{Dimension $-$1/2 torsion constraints}
We use $\Lambda^D$ and $T_{DD}{}^\Omega = 0 $ to algebraically fix $H_D{}^P$ and $\Lambda^P$ as $D$ derivatives of $H_D{}^\Omega$ and $\Lambda^\Omega$.
\begin{equation}
    \begin{split}
        T_{\alpha\beta\gamma}{}^{a}=&-H_{(\alpha}{}^A \gamma_{\beta)\gamma}^B\eta_{AB}{}^{a}-\gamma_{\alpha\beta}^A H_{A}{}^{{a}}{}_\gamma+\nabla_{(\alpha}H_{\beta)\gamma}{}^{{a}}+\nabla_\gamma H_{\alpha\beta}{}^{{a}} \\
=& -H_{(\alpha}{}^A \gamma_{\beta\gamma)}^B\eta_{AB}{}^{{a}}+\nabla_{(\alpha}H_{\beta\gamma)}{}^{{a}}\\
&+ H_{\gamma}{}^A \gamma_{\alpha\beta}^B\eta_{AB}{}^{{a}}-\gamma_{\alpha\beta}^A H_{A}{}^{{a}}{}_\gamma\\
&+\frac{1}{2}(\nabla_{\alpha}H_{[\beta\gamma]}{}^{{a}}+\nabla_\beta H_{[\alpha\gamma]}{}^{{a}}) = 0\\
    \end{split}
\end{equation}
From the second term and the third term, we read off the set of conditions for $G_{DP}{}^a$ and $G_{DD}{}^a$: 
\begin{equation}
    \begin{split}
         G_{DP}{}^a& = G_{\alpha A}{}^a = H_{\gamma}{}^B \eta_{AB}{}^{{a}}-H_{A}{}^{{a}}{}_\gamma = 0\\
         G_{DD}{}^a& = G_{\alpha\beta}{}^a = H_{[\alpha\beta]}{}^a = 0 .
    \end{split}
\end{equation}
from which we can completely find the solution for $H_P{}^\Omega$ in terms of $H_D{}^P$ and have $ H_D{}^\Omega =H_{(\alpha\beta)}^a$. These solutions means that the following components inverse metric are not flat: 
\begin{equation}
    \begin{split}
        G^{\Omega_1 \Omega_2}{}_{a} &=H_{D}{}^{[\Omega_1}\eta^{\Omega_2] D}{}_a  =H_{(\alpha\beta)}^{[b}\delta^{c]}_a\\
     G^{\Omega P}{}_a&= H_{P^\prime}{}^\Omega \eta^{P^\prime P}{}_a+\eta^{\Omega D}{}_a H_D{}^P=-\mathcal{U}^{Ab}_{B a}H_{\alpha}{}^{B}.\\
    \end{split}
\end{equation}
The non-flatness is acceptable since it doesn't effect the ${\cal S}$ constraints. Then when coupled with $\dd_{\Omega_1}\dd_{\Omega_2}=\Omega_b^\alpha \Omega_c^\beta$ or $\dd_\Omega \dd_P=\Omega_b^\alpha P_A$, with $\Omega_a^\alpha\to \mathcal{D}_a \theta ^\alpha$, the influence on Virasoro operator vanishes upon $\mathcal{U}$ constraint, since:
\begin{equation}
    \begin{split}
        H_{(\alpha\beta)}^{[b}\delta^{c]}_a\mathcal{D}_{[b} \theta ^\alpha \Omega_{c]}^\beta&=0\\
        \mathcal{U}^{Ab}_{B a}H_{\alpha}{}^{B}P_A\mathcal{D}_b \theta&=0.\\
    \end{split}
\end{equation}

The $\Lambda^P$ can algebraically gauge away part of $H_D{}^\Omega$:
\begin{equation}
    \gamma_{A}^{\alpha\beta}\eta^{AB}{}_{{a}} H_{\alpha\beta}{}^{{a}}=0 \to \Lambda^A=-\gamma_{B}^{\alpha\beta}\eta^{AB}{}_{{a}} \nabla_\alpha \Lambda_{\beta}{}^{{a}}.
\end{equation}
And $\Lambda ^D$ can correspondingly gauge away part of $H_D{}^P$:
\begin{equation}
    \gamma^{\alpha\beta}_A H_{\alpha}{}^A=0 \to \Lambda^\beta =\gamma^{\alpha \beta}{}_A(\nabla_\alpha \Lambda^A -\eta^{AB}{}_a\nabla_B\Lambda_\alpha^a ).
\end{equation}

\subsubsection{Dimension 0 torsion constraint }

The constraints of dimension $0,\frac{1}{2},1$ are then solved to find the off-shell field strength, 
whose vanishing on shell gives the field equations. The dimension 0 objects are $T_{DD}{}^P$, $T_{DP}{}^\Omega$, $G_{PP}{}^{{a}}$, $G_{D\Omega}{}^{{a}}$.

First, Assuming $G_{PP}{}^{{a}}$ and $G_{D\Omega}{}^{{a}}$ are flat at linearized level
\begin{equation}
    \begin{split}
        0&=H_{(A}{}^M\eta_{|M|B)}{}^{{a}}-\eta_{AB}{{{}^m h_m{}^a}} \\
0&=H_{\alpha}{}^{\beta}\delta{{}_b ^a}+H_{{b}}{}^\beta {}_\alpha {}^{{a}}-\delta_\alpha^\beta{h_b{}^a}. \\
    \end{split}
\end{equation}
This gives us:
\begin{equation}
    \begin{split}
        H{{}^{ab}{}_{mn}}&=-{h_{[m}{}^{[a}\delta_{n]}^{b]}} \\
H_{{a}}{}^\beta {}_\alpha {}^{{m}}&= \delta_\alpha^\beta{h_m{}^a} -H_{\alpha}{}^{\beta}\delta{{}_a ^m}. \\
    \end{split}
\end{equation}
Furthermore, The solution above guarantees the flatness of the torsion $T_{DP}{}^{\Omega}$, up to an extra $\mathcal{U}$ operator term.
\begin{equation}
    \begin{split}
        T_{DP}{}^{\Omega}-f_{DP}{}^{\Omega}& = H_{\alpha}{}^\gamma \gamma_{\gamma\beta}^B \eta_{BA}{}^{{a}}+H_{A}{}^B\gamma_{\alpha\beta}^C\eta_{CB}{}^{{a}}-H_{\beta}{}^{{a}\gamma}{}_{{b}}\gamma_{\alpha\gamma}^B\eta_{BA}{}^{{b}} \\
& + \nabla_{\alpha}H_{A\beta}{}^{{a}}-\nabla_{A}H_{\alpha\beta}{}^{{a}}+\eta_{AB}{}^{{a}}\nabla_{\beta}H_{\alpha}{}^B\\
& = \eta_{BA}{}^{{a}} (H_{(\alpha}{}^\gamma \gamma_{\beta)\gamma}^B-\gamma_{\alpha\beta}^C H_C{}^B +\nabla_{(\alpha}H_{\beta)}{}^B -\eta^{BC}{}_{{b}}\nabla_{C} H_{\alpha\beta}{}^{{b}}) \\
&-\mathcal{U}_{A{b}}^{C{a}}\nabla_{C} H_{\alpha\beta}{}^{{b}}\\
&= \eta_{PP}{}^{{a}}(T_{DD}{}^P-f_{DD}{}^P) +\mathcal{U}( H_{D}{}^\Omega ) \\
    \end{split}
\end{equation}
which can be dropped as the torsion is coupled to $\dd_\Omega$.
\par
Now, the solution of $H_{\Omega}{}^\Omega$ may have different behavior under gauge transformation. Due to the world-volume vielbein, now its gauge transformation has an extra piece:
\begin{equation}
    \delta H_{{a}}{}^\beta {}_\alpha {}^{{m}} =-\delta{{}_a^m}\nabla_\alpha\Lambda^\beta+\delta_\alpha^\beta U_{A}^{B}{{}_a^m}\nabla_BRed{\Lambda}^A
\end{equation}
Yet this extra term may not be too harmful, since this $H_\Omega{}^\Omega$ term would only show up in $\dd_\Omega$ or $G_{D\Omega}{{{}^a\mathcal{D}_a}}$. Noting that we can find a world-volume differential in both of these two terms (keeping in mind that $\Omega = \partial\theta$), the extra term can be dropped via $\mathcal{U}$ constraint.
As a check, the Bianchi identity of the torsions gives:
\begin{equation}
\begin{split}
        \nabla_{(D_1}T_{D_2D_3)}{}^\Omega &=0 =T_{(D_1D_2}{}^PT_{D_3)P}{}^\Omega \\
        &\to \gamma_{(\alpha\beta}^A\gamma_{\theta)\psi}^B\eta_{AB}{}^a=0\\
\end{split}
\end{equation}
which is the Fierz identity (\ref{jc1}).

\subsubsection{Dimension 1/2 torsion constraints}
Here we have 4 different pieces $T_{DD}{}^D$, $T_{DP}{}^P$, $T_{D\Omega}{}^\Omega$ and $T_{PP}{}^\Omega$. In detail, we have:
\begin{equation}
    \begin{split}
        T_{\alpha\beta}{}^\gamma & =-\gamma_{\alpha\beta}^A H_{A}{}^\gamma +\nabla_{(\alpha} H_{\beta)}{}^\gamma \\
        T_{\alpha A}{}^B &= -\gamma_{\alpha \beta}^C \eta_{CA}{}^a H_{a}^\beta{}^B-H_A{}^\beta \gamma_{\alpha\beta}^B +\nabla_\alpha H_A{}^B -\nabla _A H_\alpha{}^B -\eta^{BC}{}_a\eta_{AD}{}^a \nabla_C H_\alpha {}^D\\
        T_{AB}{}_\alpha^a &= H_{[A}{}^\theta \gamma_{\theta\alpha}^C \eta_{B]C}^a +\nabla_{[A} H_{B]}{}_\alpha^a -\frac{1}{2}\eta_{C[A}{}^a\nabla_\alpha H_{B]}{}^C\\
        T_{\beta}{}_b^\theta{}_\psi^a&=-H_b^\theta{}^A\gamma_{\beta\psi}^B\eta_{BA}{}^a+\nabla_{\beta}H_b^\theta{}_\psi^a-\frac{1}{2}\nabla_\psi H^\theta_b{}_\beta^a +\frac{1}{2}\nabla_\psi \delta_b^a H_\beta{}^\theta\\
    \end{split}
\end{equation}
As of right now, the Lorentz currents $S$ and $\Sigma$ are not included in the algebra, therefore all 4 of the torsions cannot be put into 0 at the same time. Yet we could take a shortcut from the self-consistency of Bianchi identity. We can impose some linear combination of them being 0, upon $\mathcal{S}$ constraint.
\begin{equation}
    \begin{split}
        &\nabla_{(D_1}T_{D_2D_3)}{}^P\dd_P=0\\
        \to & (T_{(D_1D_2}{}^{D_4}T_{D_3)D_4}{}^P+T_{(D_1D_2}{}^{P^\prime}T_{D_3)P^\prime}{}^P)\dd_P=0\\
    \end{split}
\end{equation}
Because of the Fierz identity, $H_\Omega{}^P$ does not show up, and we can then find a solution for $H_P{}^D$:
\begin{equation}
    \begin{split}
        &\gamma_{(\alpha\beta}^A H_A{}^\gamma \gamma_{\theta)\gamma}^B\\
        =&\frac{1}{2}(\nabla_{(\alpha}H_\beta{}^\gamma \gamma_{\theta)\gamma}^B +\nabla_{(\theta}\gamma_{\alpha\beta)}^A H_A{}^B -(\delta_C^E\delta^B_F-\eta^{BE}{}_a\eta_{CF}{}^a)\nabla_E \gamma^C_{(\alpha\beta}H_{\theta)}{}^F)
    \end{split}
\end{equation}
The other identity involving $\nabla_D T_{DP}{}^\Omega$ would fix $H_\Omega {}^P$:
\begin{equation}
    \begin{split}
        &\nabla_{[D_1}T_{D_2P)}{}^\Omega \dd_\Omega=0\\
        \to 0=&(-T_{D_1D_2}{}^{D_3}T_{PD_3}{}^\Omega-T_{D_1D_2}{}^{P^\prime}T_{PP^\prime}{}^\Omega \\
        &+T_{D_2P}{}^{P^\prime}T_{D_1P^\prime}{}^\Omega +T_{D_2 P}{}^{\Omega^\prime}T_{D_1\Omega^\prime}{}^\Omega\\
        &-T_{PD_1}{}^{P^\prime}T_{D_2P^\prime}{}^\Omega -T_{PD_1}{}^{\Omega^\prime}T_{D_2\Omega^\prime}{}^{\Omega})\dd_\Omega\\
    \end{split}
\end{equation}
or more specifically: 
\begin{equation}
    (\gamma_{\theta \psi}^A\eta_{AB}{}^aT_{\alpha\beta}{}^\theta+\gamma_{\alpha\beta}^AT_{AB}{}_\psi^a+\gamma_{\psi(\alpha}^A\eta_{AC}{}^aT_{\beta)B}{}^C+\gamma_{\theta(\alpha}^A\eta_{AB}{}^bT_{\beta)}{}_b^\theta{}_\psi^a)\mathcal{D}_a\theta^\psi=0
\end{equation}
For each term, we have:
\begin{equation}
    \begin{split}
        \gamma_{\theta \psi}^A\eta_{AB}{}^a T_{\alpha\beta}{}^\theta  =&\gamma_{\theta \psi}^A\eta_{AB}{}^a(\nabla_{(\alpha} H_{\beta)}{}^\theta-\gamma_{\alpha\beta}^C H_{C}{}^\theta ) \\
        \gamma_{\alpha\beta}^A T_{AB}{}_\psi^a =&\gamma_{\alpha\beta}^A( H_{[A}{}^\theta \gamma_{\theta\psi}^C \eta_{B]C}^a +\nabla_{[A} H_{B]}{}_\psi^a -\frac{1}{2}\eta_{C[A}{}^a\nabla_\psi H_{B]}{}^C)\\
        \gamma_{\psi(\alpha}^A\eta_{AC}{}^aT_{\beta)B}{}^C=& -\gamma_{\psi(\alpha}^A\eta_{AC}{}^a \gamma_{\beta)\theta}^D\eta_{DB}{}^b H_b^\theta{}^C+\frac{1}{2}\gamma_{\psi(\alpha}^A\nabla_{\beta)}\eta_{C[A}{}^aH_{B]}{}^C\\
        &-\frac{1}{2}\gamma_{\psi(\alpha}^A\nabla_{\beta)}\eta_{AB}{}^bH_{b}{}^a-\gamma_{\psi(\alpha}^B\eta_{C[A}{}^a\nabla_{B]} H_{\beta)}{}^C+\mathcal{U}\\
        \gamma_{\theta(\alpha}^A\eta_{AB}{}^b T_{\beta)}{}_b^\theta{}_\psi^a=&- \gamma_{\theta(\alpha}^A\eta_{AB}{}^b H_b^\theta{}^D\gamma_{\beta)\psi}^C\eta_{CD}{}^a\\
        &-\gamma_{\alpha\beta}^A\eta_{AB}{}^b\nabla_{\psi} H_b{}^a +\gamma_{\psi(\alpha}^A\eta_{AB}{}^b\nabla_{\beta)}H_b{}^a\\
        &+\gamma_{\theta(\alpha}^A\eta_{AB}{}^a\nabla_{|\psi|}H_{\beta)}{}^\theta-\gamma_{\theta(\alpha}^A\eta_{AB}{}^a\nabla_{\beta)}H_\psi{}^\theta\\
    \end{split}
\end{equation}
Then we find a solution upon $\mathcal{U}$ constraint:
\begin{equation}
    \begin{split}
        \gamma_{\theta(\alpha}^A\eta_{AB}{}^b H_b^\theta{}^D\gamma_{\beta)\psi}^C\eta_{CD}{}^a\mathcal{D}_a\theta^\psi&=\frac{1}{2}(
        \gamma_{\theta \psi}^A\eta_{AB}{}^a(\nabla_{(\alpha} H_{\beta)}{}^\theta-\gamma_{\alpha\beta}^C H_{C}{}^\theta )\\
        &+\gamma_{\alpha\beta}^A H_{[A}{}^\theta \gamma_{\theta\psi}^C \eta_{B]C}^a \\
        &+\frac{1}{2}\gamma^A_{\alpha\beta}\eta_{C[A}{}^a\nabla_\psi H_{B]}{}^C-\frac{1}{2}\gamma_{\psi(\alpha}^A\nabla_{\beta)}\eta_{C[A}{}^aH_{B]}{}^C\\
        &+\gamma_{\psi(\alpha}^B\eta_{C[A}{}^a\nabla_{B]} H_{\beta)}{}^C-\gamma_{\alpha\beta}^A \nabla_{[A} \eta_{B]C}{}^aH_\psi{}^C\\
        &+\frac{1}{2}\gamma_{\psi(\alpha}^A\eta_{AB}{}^b\nabla_{\beta)}H_b{}^a-\frac{1}{2}\gamma_{\alpha\beta}^A\eta_{AB}{}^b\nabla_{\psi} H_b{}^a \\
        &+\gamma_{\theta(\alpha}^A\eta_{AB}{}^a\nabla_{|\psi|}H_{\beta)}{}^\theta-\gamma_{\theta(\alpha}^A\eta_{AB}{}^a\nabla_{\beta)}H_\psi{}^\theta)\mathcal{D}_a\theta^\psi\\
    \end{split}
\end{equation}
Although it appears complicated, the redundancy can be simplified when putting the solution of vielbein in terms of prepotential into it.

\subsubsection{Dimension 1 torsion constraints}
Now to solve $H_{\Omega}{}^D$, we can consider a $\nabla_P T_{DD}{}^P$ type Jacobi identity. The differential term gives $\nabla_{(\alpha}T_{\beta)A}{}^B$ and the related $TT$ terms are $T_{DP}{}^P$, $T_{D\Omega}{}^P$ and $T_{PP}{}^P$, with 
\begin{equation}
    \begin{split}
        T_{AB}{}^C&=\nabla_{[A}H_{B]}{}^C-\frac{1}{2}\eta^{CD}{}_a\eta_{E[A|}\nabla_DH_{|B]}{}^D\\
        T_\alpha{}_a^\beta {}^A&=H_a^{\beta\theta} \gamma_{\alpha\theta}^A +\nabla_\alpha H_a^\beta{}^A+\eta^{AB}{}_{a}\nabla_B H_\alpha{}^\beta-\frac{1}{2}\delta_\alpha^\beta\eta^{AB}{}_b\nabla_B H_a{}^b\\
    \end{split}
\end{equation}
Then there is
\begin{equation}
    \begin{split}
        0=&-\nabla_{(\alpha}T_{\beta)A}{}^B-\gamma_{\alpha\beta}^CT_{CA}{}^B-T_{\alpha\beta}{}^\gamma T_{\gamma A}{}^B\\
        &+T_{(\alpha|A}{}^CT_{|\beta)C}{}^B+T_{A(\alpha}{}^\theta \gamma_{\alpha)\theta}^B +\gamma_{\theta(\beta}^C\eta_{CA}{}^aT_{\alpha)}{}_a^\theta {}^B
    \end{split}
\end{equation}
from where we can algebraically solve $H_{\Omega}{}^D=H^{\alpha\beta}_a$.

\section{Lagrangians}\label{section:6}
In this section, Lagrangians with duality symmetries are presented for 
the D=3 ${\cal A}$-theory as a simplest non-trivial example. 
As shown in the section 3, the D=3 current algebra with SL(5) U-duality symmetry
is realized on a 5-brane world-volume.
The closure of the brane Virasoro algebra requires the Gau\ss{} law constraint
which is the gauge symmetry generator of the spacetime coordinate $\delta_\kappa X=[\int {\cal U}{\kappa}, X]$$\approx \partial \kappa$. 
As a result, the spacetime current becomes the field strength 
of the spacetime coordinate.
For the D=3 case the field strengths are SL(5) pseudo rank two anti-symmetric tensors,
where the spacetime coordinate is a SL(5) two rank anti-symmetric tensor
and the gauge parameter is a SL(5) vector and a scalar as listed in \bref{SL5fields}.   
Analogously to the usual Abelian gauge theory, the gauge symmetry transformation rule in
The Lagrangian formulation has larger Lorentz symmetry than the one in Hamiltonian formulations. 
Furthermore, the D=3 ${\cal A}$-theory Lagrangian with SL(6) enlarged duality symmetry 
is derived.
Since the U-duality symmetries $E_{D+1}$ depend on dimensions,
representations of field strengths, spacetime coordinates, gauge parameters 
are listed in \bref{SL6fields} of section 2.4.

\subsection{D$=$3 $G=$SL(5) covariant Lagrangian}\label{section:6-1}
	
	The Hamiltonian is given by the sum of Virasoro operators \bref{Virasoros} with multipliers.
	To obtain the world-volume covariant action  Virasoro constraints in terms of the anti-selfdual currents 
	$\tilde{\cal S}^m=\frac{1}{16}\tilde{\dd}_{m_1m_2}\epsilon^{mm_1\cdots m_4}\tilde{\dd}_{m_3m_4}$ 
	and $\tilde{\cal H}=\frac{1}{8}\tilde{\dd}_{m_1m_2}\tilde{\dd}^{m_1m_2}$ are also included \cite{Hatsuda:2018tcx,Hatsuda:2019xiz,Hatsuda:2021wpb}.
	The Hamiltonian form action is given analogously to the D=4 case 
	\cite{Hatsuda:2021wpb} as
	\bea
	I&=&\displaystyle \int d\tau d^5\sigma~L~,~L=\dot{X}^{mn}P_{mn}-H\nn\\
	H&=&g{\cal H}+s_m{\cal S}^m+\tilde{g}\tilde{\cal H}+\tilde{s}_m\tilde{\cal S}^m
	+{\cal U}_mY^m\nn\\
	&=&\frac{1}{16}\dd_{m_1m_2}(g\delta^{m_1[m_3}\delta^{m_4]m_2}+s_m\epsilon^{mm_1\cdots m_4} )\dd_{m_3m_4}\nn\\
	&&+\frac{1}{16}\tilde{\dd}_{m_1m_2}(\tilde{g}\delta^{m_1[m_3}\delta^{m_4]m_2}+\tilde{s}_m\epsilon^{mm_1\cdots m_4} )\tilde{\dd}_{m_3m_4}+\frac{1}{2}\dd_{mn}\partial^{[m} Y^{n]}\label{Hamiltoniangs}\\
	&=&\frac{1}{4}P_M M^{MN}P_N+\frac{1}{2}P_MN^{MN}\eta_{NLl}\partial^l X^L+
	\frac{1}{4} \eta_{MLn}\partial^l X^L M^{MN}  \eta_{NKk}\partial^k X^K\nn\\&&+ P_{mn}\partial^m Y^n\nn
	\eea
	where matrices $M^{MN}$ and $N^{MN}$  are given by
	\bea
	&&{\renewcommand{\arraystretch}{1.8}\left\{\begin{array}{ccl}
			M^{m_1m_2;m_3m_4}&=&\frac{1}{2}(g+\tilde{g})\delta^{m_1[m_3}\delta^{m_4]m_2}+\frac{1}{2}(s_m+\tilde{s}_m)\epsilon^{mm_1\cdots m_4}\\
			N^{m_1m_2;m_3m_4}&=&\frac{1}{2}(g-\tilde{g})\delta^{m_1[m_3}\delta^{m_4]m_2}+\frac{1}{2}(s_m-\tilde{s}_m)\epsilon^{mm_1\cdots m_4}\\
		\end{array}\right.}~~~.
	\eea
	The Lagrangian is given as
	\bea
	&&L=\varphi  \don\circ{F}_{+}{}^{m_1m_2}M^{-1}{}_{m_1m_2;n_1n_2}  \don\circ{F}_{-}{}^{n_1n_2} \label{Lplusminus}
	\eea
	with
	\bea
	&&M^{-1}{}_{m_1m_2;m_3m_4}=2\varphi
	\left\{
	\delta_{m_1[m_3}\delta_{m_4]m_2}-\epsilon_{m_1\cdots m_4m}\frac{s^m+\tilde{s}^m}{g+\tilde{g}}\right.\nn\\&&~~~~~~~~~~~~~~~~~~~~~~~~~~~~\left.
	+\frac{1}{(g+\tilde{g})^2}(s+\tilde{s})_{[m_1}\delta_{m_2][m_3}(s+\tilde{s})_{m_4]}
	\right\}
	\\
	&&~~~~~~~~~~~~~~~~~~\varphi=\frac{g+\tilde{g}}{(g+\tilde{g})^2-(s^l-\tilde{s}^l)^2}~~~.\nn
	\eea
	The matrix $M^{-1}$ has three terms, $\hat{\eta}^{MN}$, $\eta^{MNm}$ and $U^M{}_N{}^m{}_n$ in \bref{Ddeta}.
	The $U$ matrix, $U^M{}_N{}^m{}_n$,
	is defined from square of the SL(5) invariant metric $\eta^{MNl}=\epsilon^{m_1m_2n_1n_2 l}$ 
	as
	\bea
	\frac{1}{2}\epsilon^{m_1m_2n_1n_2m}a_m\epsilon_{n_1n_2l_1l_2l}b^l=a_mb^m ~\delta^{m_1}{}_{[l_1}\delta^{m_2}{}_{l_2]}
	+b^{[m_1}\delta^{m_2]}{}_{[l_1}a{}_{l_2]}
	\eea
	or symbolically
	\bea
	&&(\eta^{MN}\cdot a)(\eta_{NL}\cdot b)=a\cdot b~{\delta^{M}{}_{L}}+b\cdot {U}^M{}_L\cdot a \label{etaetaP}
	\eea
	with
	$\eta^{MN}\cdot a=\eta^{MNm}a_m$.
	In the second term of the right hand side of \bref{etaetaP}  $b\cdot {U}^M{}_L\cdot a=b^m {U}^M{}_L{}_m{}^la_l$ is an 
	abbreviation for
	${U}^M{}_L{}_m{}^l=\delta^{[m_1}{}_m \delta^{m_2]}{}_{[l_1} \delta^{l}{}_{l_2]}$.
	There are some useful relations among $\eta^{MNm}$ and the $U$ matrix such as
	\bea
	&&(\eta^{MN}\cdot a)(\eta_{NL}\cdot b)(\eta^{LK}\cdot a)=a\cdot b (\eta^{MK}\cdot a)\label{etaP}\\
	&&(\eta^{MN}\cdot a)(b\cdot {U}_{NL}\cdot b)+
	(b\cdot {U}^{MN}\cdot b)(\eta_{NL}\cdot a)=a\cdot b (\delta^{M}{}_L\cdot b)-b^2 (\delta^M{}_L\cdot a)\nn
	\eea

	Field strengths $\don\circ{F}_{\pm}{}^{mn}$ are given by
	\bea
	&&{\renewcommand{\arraystretch}{1.8}\left\{\begin{array}{ccl}
			\don\circ{F}_{+}{}^{mn}&=&\don\circ{F}_\tau{}^{mn}+\frac{1}{2}(M-N){}^{mn;l_1l_2}\don\circ{F}_\sigma{}_{;l_1l_2}\\
			\don\circ{F}_{-}{}^{mn}&=&\don\circ{F}_\tau{}^{mn}-\frac{1}{2}(M+N){}^{mn;l_1l_2}\don\circ{F}_\sigma{}_{;l_1l_2}
		\end{array}\right.}~\\
	&&
	{\renewcommand{\arraystretch}{1.8}
		\left\{\begin{array}{ccl}
			\don\circ{F}_{\tau}{}^{mn}&=&\dot{X}^{mn}-\partial^{[m}Y^{n]}\\
			\don\circ{F}_{\sigma}{}_{;mn}&=&\frac{1}{2}\epsilon_{mnl_1l_2l_3}\partial^{l_1}X^{l_2l_3}
		\end{array}\right.}
	\label{FtFs}
	\eea
	Field strengths $\don\circ{F}_\tau$ and $\don\circ{F}_\sigma$ have the gauge symmetry generated by the Gau\ss{} law constraint ${\cal U}_m=0$
	\bea
	&\delta_\kappa X^{mn}(\sigma)=i\displaystyle \int d^5\sigma' [X^{mn}(\sigma) ,\kappa^l~{\cal U}_l(\sigma')]&~~~\\
	&\delta_\kappa X^{mn}=\partial^{[m}\kappa^{n]}~~,~~\delta_\kappa Y^m=\dot{\kappa}^m+\partial^m \kappa & ~~~\nn
	\eea
	with further gauge symmetry of the gauge symmetry $\delta \kappa^m=\partial^m\bar{\kappa}$ and  $\delta \kappa=-\dot{\bar{\kappa}}$.
	
	In the SL(5) covariant action the spacetime coordinate  $X^{mn}$ and $Y^m$ with $m=1\cdots,5$ 
	and the  world-volume coordinate $\sigma_{m}$ and $\tau$ are
	$10\oplus 5$ and $5\oplus 1$ representations of SL(5)  
	\bea
	5\oplus 10&:& Y^m,~X^{mn}\\
	1\oplus 5&:&\tau, ~\sigma_{m}\nn~~~.
	\eea
	The field strengths are two rank tensors and the gauge symmetry parameters are a vector and a scalar as 
	\bea
	10\oplus 10'&:&\don\circ{F}_\tau{}^{{m}_1{m}_2},~\don\circ{F}_\sigma{}_{{m}_1{m}_2}
	\\
	1\oplus 5&:& \kappa,~\kappa^{m}~~~.\nn
	\eea

	\vskip 6mm
	
	Now let us rewrite the Lagrangian in such a way that 
	the kinetic term is the product of the selfdual current and the anti-selfdual current
	for the world-volume covariance, and constraint terms are
	bilinears of anti-selfdual currents with Lagrange multipliers.
	Using with \bref{etaetaP} and \bref{etaP}
	we have obtained the Lagrangian in terms of the selfdual and anti-selfdual currents 
	with Lagrange multipliers $\lambda$'s as
	\bea
	L&=&\frac{\phi}{2}\don\circ{F}_{\rm SD}{}^{mn}\don\circ{F}_{\overline{\rm SD}}{}_{mn}
	+\frac{\bar{\phi}}{2}\don\circ{F}_{\overline{\rm SD}}{}^{mn}\don\circ{F}_{\overline{\rm SD}}{}_{mn}
	+\frac{1}{4}\lambda{}^{m_1}\epsilon_{m_1\cdots m_5}\don\circ{F}_{\overline{\rm SD}}{}^{m_2m_3}
	\don\circ{F}_{\overline{\rm SD}}{}^{m_4m_5}\nn\\
	&&+\lambda{}_{mn}
	\don\circ{F}_{\overline{\rm SD}}{}^m{}^{l}\don\circ{F}_{\overline{\rm SD}}{}_{l}{}^{n}~~~.\label{SL5Lagrangian}
	\eea
	The selfdual current $\don\circ{F}_{\rm SD}{}^{mn}$ is physical, while
	anti-selfdual current $\don\circ{F}_{\overline{\rm SD}}{}^{mn}$ is unphysical;
	\bea
	{\renewcommand{\arraystretch}{1.8}
		\left\{\begin{array}{ccl}
			\don\circ{F}_{\rm SD}{}^{mn}&=&\don\circ{F}_\tau{}^{mn}-\frac{1}{2}\epsilon^{mnll_1l_2}s_{l}\don\circ{F}_{\sigma;l_1l_2}
			+g\don\circ{F}_\sigma{}^{mn}
			\\
			\don\circ{F}_{\overline{\rm SD}}{}^{mn}&=&\don\circ{F}_\tau{}^{mn}
			-\frac{1}{2}\epsilon^{mnll_1l_2}s_{l}\don\circ{F}_{\sigma;l_1l_2}
			-g\don\circ{F}_\sigma{}^{mn}
		\end{array}\right.}\label{selfdualcurrent}
	\eea
	The Hamiltonian form Lagrangian \bref{Lplusminus}
	is related to this Lagrangian with multipliers as
	\bea
	{\renewcommand{\arraystretch}{1.8}
		\left\{	\begin{array}{ccl}
			\phi&=&\displaystyle\frac{1}{2g}\\
			\bar{\phi}&=&\displaystyle\frac{g^2-\tilde{g}^2+(s+\tilde{s})^2}{2g\{(g+\tilde{g})^2-(s+\tilde{s})^2\}}\\
			\lambda{}^m&=&-\displaystyle\frac{1}{\{(g+\tilde{g})^2-(s+\tilde{s})^2\}}(s+\tilde{s})^{m}\\
			\lambda{}^{mn}&=&\displaystyle\frac{1}{(g+\tilde{g})\{(g+\tilde{g})^2-(s+\tilde{s})^2\}}(s+\tilde{s})^{m} (s+\tilde{s})^{n}
		\end{array}\right.}\label{SL5philambda}
	\eea
	\vskip 6mm
	
	In curved backgrounds the SL(5) background $G_{mn}$ field strengths  are written as
	\bea
	G_{mn}=E_m{}^a\delta_{ab}E_n{}^b~~,~~
	{\renewcommand{\arraystretch}{1.8}
		\left\{\begin{array}{ccl}
			{F}_{\rm SD}{}^{ab}&=&	\don\circ{F}_{\rm SD}{}^{mn}E_m{}^aE_n{}^b\\
			{F}_{\overline{\rm SD}}{}^{ab}&=&	\don\circ{F}_{\overline{\rm SD}}{}^{mn}E_m{}^aE_n{}^b
		\end{array}\right.}~~~.
	\eea
	The SL(5) covariant Lagrangian in a curved background is given by
	\bea
	L&=&\frac{\phi}{4}\don\circ{F}_{\rm SD}{}^{m_1m_2}G_{m_1m_2;m_3m_4}\don\circ{F}_{\overline{\rm SD}}{}^{m_3m_4}
	+\frac{\bar{\phi}}{4}\don\circ{F}_{\overline{\rm SD}}{}^{m_1m_2}
	G_{m_1m_2;m_3m_4}
	\don\circ{F}_{\overline{\rm SD}}{}_{m_3m_4}\nn\\
	&&+\frac{1}{4}\don\circ{\lambda}{}^{m_1}\epsilon_{m_1\cdots m_5}\don\circ{F}_{\overline{\rm SD}}{}^{m_2m_3}
	\don\circ{F}_{\overline{\rm SD}}{}^{m_4m_5}+\don\circ{\lambda}{}_{mn}
	\don\circ{F}_{\overline{\rm SD}}{}^m{}^{l_1}G_{l_1l_2}\don\circ{F}_{\overline{\rm SD}}{}^{l_2n}~~~\\
	&=&\frac{\phi}{2}{F}_{\rm SD}{}^{ab}{F}_{\overline{\rm SD}}{}_{ab}
	+\frac{\bar{\phi}}{2}{F}_{\overline{\rm SD}}{}^{ab}{F}_{\overline{\rm SD}}{}_{ab}+\frac{1}{4}\lambda{}^{a_1}\epsilon_{a_1\cdots a_5}{F}_{\overline{\rm SD}}{}^{a_2a_3}
	{F}_{\overline{\rm SD}}{}^{a_4a_5}+\lambda{}_{ab}
	{F}_{\overline{\rm SD}}{}^a{}^{c}{F}_{\overline{\rm SD}}{}_{;c}{}^b~~~.\nn
	\eea
	Lagrange multipliers $\lambda$'s and $\don\circ{\lambda}$'s are arbitrary, 
	where the variation of the multipliers gives the selfduality condition ${F}_{\overline{\rm SD}}=0$.
	
	\vskip 6mm
	
	\subsection{D$=$3 $A=$SL(6) covariant Lagrangian}\label{section:6-2}
	
	The SL(5) $G$-symmetry is enlarged to the SL(6) $A$-symmetry by
	requiring the (1+5)-dimensional world-volume covariance. 
	The spacetime coordinate  $X^{\hat{m}\hat{n}}=-X^{\hat{n}\hat{m}}$,  $\hat{m}=0,\cdots,5$ is 15 representation of SL(6), and the  world-volume coordinate $\sigma_{\hat{m}}$ is 6 representation of SL(6)  
	\bea
	15&:&X^{\hat{m}\hat{n}}=(X^{0m}=Y^m,~X^{mn})\\
	6&:&\sigma_{\hat{m}}=(\tau,~\sigma_m)\nn~~~.
	\eea
	The field strength becomes a rank three tensor and the gauge symmetry parameter becomes a vector as 
	\bea
	20&:&{F}{}^{\hat{m}_1\hat{m}_2 \hat{m}_3}=(
	{F}{}^{0m_2m_3}={F}_{\tau}{}^{m_2m_3},~{F}{}^{m_1m_2m_3}= \frac{1}{2}\epsilon^{m_1\cdots m_5}{F}_{\sigma;}{}_{m_4m_5})\\
	6&:& \kappa^{\hat{m}}=(\kappa,~\kappa^m).\nn~~
	\eea
	The field strength in the flat background is given by
	\bea
	\don\circ{F}{}^{\hat{m}_1\hat{m}_2 \hat{m}_3}=\frac{1}{2}\partial^{[\hat{m}_1}X^{\hat{m}_2\hat{m}_3]}~~~.
	\eea
	There is a gauge invariance of the field strength as
	$\delta_\kappa \don\circ{F}{}^{\hat{m}_1\hat{m}_2 \hat{m}_3}=0$ with 
	$\delta_\kappa X^{\hat{m}\hat{n}}=\partial^{[\hat{m}}\kappa^{\hat{n}]}$ and $\delta_\kappa \kappa^{\hat{m}}=\partial^{\hat{m}}\bar{\kappa}$.
	The selfdual and anti-selfdual field strengths in \bref{selfdualcurrent} with $g=1, s^m=0$ become as
	\bea
	{\renewcommand{\arraystretch}{1.8}
		\left\{\begin{array}{ccl}
			\don\circ{F}_{\rm SD;}{}_{\hat{m}_1\hat{m}_2\hat{m}_3}&=&\eta_{\hat{m}_1\hat{n}_1}\eta_{\hat{m}_2\hat{n}_2}\eta_{\hat{m}_3\hat{n}_3}\don\circ{F}{}^{\hat{n}_1\hat{n}_2\hat{n}_3}+ \frac{1}{3!}
			\epsilon{}_{\hat{m}_1\cdots \hat{m}_6}\don\circ{F}{}^{\hat{m}_4\hat{m}_5\hat{m}_6}\\
			\don\circ{F}_{\overline{\rm SD};}{}_{\hat{m}_1\hat{m}_2\hat{m}_3}&=&\eta_{\hat{m}_1\hat{n}_1}\eta_{\hat{m}_2\hat{n}_2}\eta_{\hat{m}_3\hat{n}_3}\don\circ{F}{}^{\hat{n}_1\hat{n}_2\hat{n}_3}- \frac{1}{3!}
			\epsilon{}_{\hat{m}_1\cdots \hat{m}_6}\don\circ{F}{}^{\hat{m}_4\hat{m}_5\hat{m}_6}
		\end{array}\right.}
	\eea
	
	The ${\cal A}$/L=SL(6)/SO(6)  background vielbein field $E_{\hat{m}}{}^{\hat{a}}$ includes both the spacetime and the world-volume vielbeins.   
	It is transformed as $E_{\hat{m}}{}^{\hat{a}}\to A_{\hat{m}}{}^{\hat{n}}E_{\hat{n}}{}^{\hat{b}}h_{\hat{b}}{}^{\hat{a}}$ with $A_{\hat{m}}{}^{\hat{n}}\in$SL(6) and $h_{\hat{b}}{}^{\hat{a}}\in$SO(6).
	The SL(6) gauge field, $E_{\hat{m}}{}^{\hat{a}}$, satisfies the following condition
	\bea
	\epsilon_{\hat{a}_1\cdots \hat{a}_6}
	E_{\hat{m}_1}{}^{\hat{a}_1}E_{\hat{m}_2}{}^{\hat{a}_2}E_{\hat{m}_3}{}^{\hat{a}_3}E_{\hat{m}_4}{}^{\hat{a}_4}E_{\hat{m}_5}{}^{\hat{a}_5}E_{\hat{m}_6}{}^{\hat{a}_6}&=&\det E_{\hat{m}}{}^{\hat{a}}~ \epsilon_{\hat{m}_1\cdots \hat{m}_6}~,~\det E_{\hat{m}}{}^{\hat{a}}=1~~.\nn\\\label{SL6E}
	\eea
	The SL(5)/SO(5) background in \bref{SL5bg} is embedded in the SL(6)/SO(6) gauge fields.
	The dimension of SL(6)/SO(6) is $20$ which is sum of 
	the number of the spacetime vielbein ${E}_{{m}}{}^{{a}}\in$ SL(5)/SO(5),
	14, and the number of the world-volume vielbein ${\cal E}_{\hat{m}}{}^{\hat{a}}$, $6$. 
	
	Field strengths in curved spacetime backgrounds are given by
	\bea
	&F{}^{\hat{a}_1\hat{a}_2\hat{a}_3}~=~ 
	\don\circ{F}{}^{\hat{m}_1\hat{m}_2\hat{m}_3}{E}_{\hat{m}_1}{}^{\hat{a}_1}{E}_{\hat{m}_2}{}^{\hat{a}_2}{E}_{\hat{m}_3}{}^{\hat{a}_3}.&
	\eea
	SL(6) covariant selfdual and anti-selfdual field strengths in \bref{selfdualcurrent} are givens as
	\bea
	{\renewcommand{\arraystretch}{1.8}
		\left\{\begin{array}{ccl}
			{F}_{\rm SD;}{}_{\hat{a}_1\hat{a}_2\hat{a}_3}&=&\eta_{\hat{a}_1\hat{b}_1}\eta_{\hat{a}_2\hat{b}_2}\eta_{\hat{a}_3\hat{b}_3}{F}{}^{\hat{b}_1\hat{b}_2\hat{b}_3}+ \frac{1}{3!}
			\epsilon{}_{\hat{a}_1\cdots \hat{a}_6}{F}{}^{\hat{a}_4\hat{a}_5\hat{a}_6}\\
			{F}_{\overline{\rm SD};}{}_{\hat{a}_1\hat{a}_2\hat{a}_3}&=&\eta_{\hat{a}_1\hat{b}_1}\eta_{\hat{a}_2\hat{b}_2}\eta_{\hat{a}_3\hat{b}_3}{F}{}^{\hat{b}_1\hat{b}_2\hat{b}_3}- \frac{1}{3!}
			\epsilon{}_{\hat{a}_1\cdots \hat{a}_6}{F}{}^{\hat{a}_4\hat{a}_5\hat{a}_6}
		\end{array}\right.}.
	\eea
	The selfduality condition is vanishing the anti-selfdual current.
	The SO(6) covariant selfduality condition is
	\bea
	\frac{1}{3!}\eta_{\hat{a}_1[\hat{b}_1|}\eta_{\hat{a}_2|\hat{b}_2|}\eta_{\hat{a}_3|\hat{b}_3]}{F}{}^{\hat{b}_1\hat{b}_2\hat{b}_3}&=& \frac{1}{3!}
	\epsilon{}_{\hat{a}_1\cdots \hat{a}_6}{F}{}^{\hat{a}_4\hat{a}_5\hat{a}_6}~~~,\label{selfdualitya}
	\eea
	while the SL(6) covariant selfduality condition is obtained by multiplying ${E}_{\hat{m}_1}{}^{\hat{a}_1}{E}_{\hat{m}_2}{}^{\hat{a}_2}{E}_{\hat{m}_3}{}^{\hat{a}_3}$ and with \bref{SL6E} 
	\bea
	&\displaystyle\frac{1}{3!}G_{\hat{m}_1\hat{m}_2\hat{m}_3;\hat{m}_4\hat{m}_5\hat{m}_6}
	\don\circ{F}{}^{\hat{n}_1\hat{n}_2\hat{n}_3}~=~ \frac{1}{3!}
	\epsilon{}_{\hat{m}_1\cdots \hat{m}_6}\don\circ{F}{}^{\hat{m}_4\hat{m}_5\hat{m}_6}\label{selfduality}&~~~\\
	&
	G_{\hat{m}_1\hat{m}_2\hat{m}_3;\hat{m}_4\hat{m}_5\hat{m}_6}=
	E_{\hat{m}_1}{}^{\hat{a}_1}E_{\hat{m}_2}{}^{\hat{a}_2}E_{\hat{m}_3}{}^{\hat{a}_3}
	\delta_{\hat{a}_1[\hat{a}_3|}\delta_{\hat{a}_2|\hat{a}_4|}\delta_{\hat{a}_3|\hat{a}_6]}
	E_{\hat{m}_4}{}^{\hat{a}_4}E_{\hat{m}_5}{}^{\hat{a}_5}E_{\hat{m}_6}{}^{\hat{a}_6}.
	&\nn
	\eea
	
	The SL(6) transformation brings back to the world-volume gravitational parameters $g$ and $s_m$ 
	in \bref{Hamiltoniangs}
	similar to \cite{Hatsuda:2012vm} as
	\bea
	F{}^{\hat{a}_1\hat{a}_2\hat{a}_3}&=& \don\circ{F}{}^{\hat{m}_1\hat{m}_2\hat{m}_3}{\cal E}_{\hat{m}_1}{}^{\hat{a}_1}{\cal E}_{\hat{m}_2}{}^{\hat{a}_2}{\cal E}_{\hat{m}_3}{}^{\hat{a}_3}\nn\\
	{\cal E} _{\hat{m}}{}^{\hat{a}}&=&
	\begin{array}{c}
		_0\\_m
	\end{array}
	\left(
	\begin{array}{cc}
		g^{-5/6}&0\\
		-g^{-5/6} s_m&g^{1/6}\delta_m^a
	\end{array}\right)~\in{\rm SL}(6)~,\label{wvSL6}
	\eea
	which are equal to
	\bea
	{\renewcommand{\arraystretch}{1.8}
		\left\{\begin{array}{ccl}
			F^{0ab}&=&\displaystyle\frac{1}{\sqrt{g}}(\don\circ{F}{}^{0mn}-s_l\don\circ{F}{}^{lmn})\delta_m^a\delta_n^b\\
			F^{abc}&=&\displaystyle\sqrt{g}\don\circ{F}{}^{mnl}\delta_m^a\delta_n^b\delta_l^c
		\end{array}\right.}~~~.
	\eea 
	
	\vskip 6mm
	
	We proposed the SL(6) covariant Lagrangian for a ${\cal A}$-theory 5-brane as
	\bea
	L&=&\frac{1}{3!}{\Phi} \don\circ{F}{}^{\hat{m}_1\hat{m}_2 \hat{m}_3}
	G_{\hat{m}_1\hat{m}_2\hat{m}_3;\hat{m}_4\hat{m}_5\hat{m}_6}
	\don\circ{F}{}^{\hat{m}_4\hat{m}_5 \hat{m}_6}
	+\frac{1}{2}\don\circ{\Lambda}_{\hat{n}\hat{l}}
	\don\circ{F}{}^{\hat{n}\hat{m}_1 \hat{m}_2}
	G_{\hat{m}_1\hat{m}_2;\hat{m}_3\hat{m}_4}
	\don\circ{F}{}^{\hat{l}}{}^{\hat{m}_3\hat{m}_4}\nn\\&&
	+\frac{1}{12}\epsilon_{\hat{m}_1\cdots \hat{m}_6}
	\don\circ{\Lambda}_{\hat{n}}{}^{\hat{m}_1}\don\circ{F}{}^{\hat{m}_2 \hat{m}_3 \hat{m}_4}
	\don\circ{F}{}^{\hat{m}_5 \hat{m}_6 \hat{l}} 
	\nn \\
	&=&\frac{1}{3!}{\Phi} F^{\hat{a}_1\hat{a}_2 \hat{a}_3}F_{\hat{a}_1\hat{a}_2 \hat{a}_3}
	+\frac{1}{2}\Lambda_{\hat{a}\hat{b}}
	F^{\hat{a}\hat{c}_1 \hat{c}_2}F^{\hat{b}}{}_{\hat{c}_1\hat{c}_2}+
	\frac{1}{12}\epsilon_{\hat{a}_1\cdots \hat{a}_6}
	\Lambda_{\hat{b}}{}^{\hat{a}_1}F^{\hat{a}_2 \hat{a}_3 \hat{a}_4}
	F^{\hat{a}_5 \hat{a}_6 \hat{b}} \label{SL6Lagrangian} 
	\eea
	with $\Lambda_{\hat{a}\hat{b}}$ is a symmetric traceless matrix
	\bea
	\Lambda_{\hat{a}\hat{b}}=\Lambda_{\hat{a}\hat{b}}-\frac{1}{6}\eta_{\hat{a}\hat{b}} ~\eta^{\hat{c}\hat{d}}\Lambda_{\hat{c}\hat{d}}\nn~~~.
	\eea
	Lagrange multipliers $\Lambda$'s and $\don\circ{\Lambda}$'s are arbitrary, 
	where the variation of the multipliers gives the selfduality condition ${F}_{\overline{\rm SD}}=0$ equivalently \bref{selfdualitya} or \bref{selfduality}.
	
	The Lagrangian \bref{SL6Lagrangian} is written in terms of the SL(5) covariant 
	selfdual and anti-selfdual field strengths as
	\bea
	L&=&
	-\frac{1}{2}{\Phi} F_{\rm SD}{}^{ab}F_{\overline{\rm SD}}{}_{ab}
	-\frac{1}{2}\Lambda_{0}{}^{0}(F_{\overline{\rm SD}}{}^{ab})^2
	-\Lambda_{ab}F_{\overline{\rm SD}}{}^{ac}F_{\overline{\rm SD}}{}^{b}{}_{c}-\frac{1}{4}\epsilon_{{a}_1\cdots {a}_5}
	\Lambda_{0}{}^{{a}_1}F_{\overline{\rm SD}}{}^{a_2a_3}F_{\overline{\rm SD}}{}^{a_4a_5}
	~~~.\nn\\
	\eea
	This coincides with \bref{SL5Lagrangian} by corresponding each coefficient in \bref{SL5philambda} as follows
	\bea
	{\renewcommand{\arraystretch}{1.8}
		\left\{	\begin{array}{ccl}
			{\Phi}&=&-\phi\\
			\Lambda_{0}{}^{0}&=&-\bar{\phi}\\
			\Lambda_{0a}&=&-\lambda{}^a\\
			\Lambda_{ab}&=&-\lambda{}^{ab}
		\end{array}\right.}\label{SL6PhiLambda}
	\eea

	\section{Amplitude}\label{section:7}
	One of the important questions on ${\cal A}$-theory is how to construct its amplitude. The general theory on quantization of ${\cal A}$-theory is very 
	limited so one cannot construct the amplitude directly from the Feynman rules. 
	
	In our approach, we shall focus on using the enlarged gauge symmetry of ${\cal A}$-theory to bootstrap the possible forms of the amplitude, and as 
	we will see, in the D$=$3 case the gauge symmetry completely fixed the kinematic factor of the 4-point amplitude. 
	
	We also use the restriction that the amplitude should be reduced to the ordinary 4-graviton amplitude under section conditions. This will allow us
	to write down the complete ${\cal A}$-symmetric amplitude with the lowest order of $\alpha^{'}$ correction. (i.e. the low energy limit)
	
	Let's review how one can write the usual gauge theory amplitudes in a manifest gauge invariant fashion before we generalized it 
	into ${\cal A}$-theory amplitude.
	
	\subsection{Manifestly gauge invariant 4-point amplitude}\label{section:7-1}
	
	In 4D Yang-Mills theory, the color-ordered 4-point tree amplitude can be written in a form that is manifest gauges invariant with 
	respect to linearized gauge invariance of the external fields.  
	
	\begin{equation}
	A_{4}(s, t) = F_{1}F_{2}F_{3}F_{4}\frac{1}{st}
	\end{equation}
	where $F_{i}$ is the linearized field strength for the $i$th momentum.
	
	\begin{equation}
	(F_{i})_{mn} = k_{i[m}\epsilon_{i\;n]}
	\end{equation}
	
	There are various ways to contract the four field strengths, but the contractions are fixed in a way that all double poles are canceled.
	
	We can greatly simplify the expression if we use spinor indices instead.
	
	\begin{equation}
	F_{\mu\nu} = (\sigma_{\mu\nu})^{\alpha\beta}f_{\alpha\beta}+  (\bar{\sigma}_{\mu\nu})^{\dot{\alpha}\dot{\beta}}
	\bar{f}_{\dot{\alpha}\dot{\beta}},
	\end{equation}
	
	where
	\begin{equation}
	f_{\alpha \beta } = k_{(\alpha }{}^{\gamma }\epsilon_{\beta )\gamma }.
	\end{equation}
	
	is the selfdual part of the YM field strength, while the anti-selfdual part $\bar{f}_{\dot{\alpha }\dot{\beta }}$ can be defined similarly.
	
	The colored ordered amplitudes can be written down more explicitly \cite{Feng:2004tg}
	\begin{equation}
	A_{4}(1^{+}, 2^{+}, 3^{-}, 4^{-}) = (f_{1}f_{2})(\bar{f}_{3}\bar{f}_{4})\frac{1}{st}.
	\end{equation}
	There are six other non-zero amplitudes with two plus and two minus helicities, each the plus helicity external leg corresponds to a 
	$f$ factors, while each minus helicity leg correspond to a $\bar{f}$ factor. 
	
	The 4-point gravity tree level amplitude can be written in a similar form
	\begin{equation}
	\frac{W_{1}W_{2}W_{3}W_{4}}{stu}.
	\end{equation}
	
	$W$ is the linearized Weyl tensor $W_{[ab][cd]}$:
	\begin{equation}
	W^{[mn]}_{\quad [pq]} =k_{[p}k^{[m}h_{q]}^{\; n]}.
	\end{equation}
	Just like the YM 4-point amplitude, it can also be written in its spinor form. The 4-point amplitude in this form is
	\begin{equation}
	A_{4}(1^{+}, 2^{+}, 3^{-}, 4^{-}) =  (w_{1}w_{2}) (\bar{w}_{3} \bar{w}_{4}) \frac{1}{stu}.
	\end{equation}
	The Weyl tensor in spinor indices is
	\begin{equation}
	w_{(\alpha \beta \gamma \delta )}  = k_{(\alpha }^{\epsilon}k_{\beta }^{\zeta}h_{\gamma \delta )\epsilon\zeta}.
	\end{equation}
	
	Inspired by this structure, one should also be able to write the 4-point ${\cal A}$-theory amplitude in a similar fashion. If we are able
	to find the ${\cal A}$-theory equivalent of Weyl tensor ``$W$'', then one should able to write the low energy limit of the ${\cal A}$-theory amplitude 
	in the following form.
	
	\begin{equation}
	A_{4} = W_{\textrm{${\cal A}$-theory}}^{4}\frac{1}{stu}.
	\end{equation}
	
	We demand that the indices are contracted in such a way that after solving the section condition, it will reduce to the 
	ordinary 4-graviton amplitude. If one can find a unique contraction that satisfies this condition, it has to be the correct kinematic 
	factor for the ${\cal A}$-theory amplitudes.

	Our goal is first to find the ${\cal A}$-theory equivalent of the Weyl tensor. ( We will just refer it as the ${\cal A}$-theory field strength from now on)
	We will only consider the D$=$3 case. Given that the index structure on both the metric $h$, the field strength $W$ is known. There are 
	only a few ways that one could write down the contractions $W = \partial^{2}h$. One needs to find the linear combination of them 
	that is gauge invariant. 
	
	Unfortunately, as we shall see, one can in fact not find any $W = \partial^{2}h$ that is gauge invariant. The non-existence of (gauge invariant) 
	field strength for ${\cal A}$-theory is expected, as this is consistent with the results from exceptional supergravity. \cite{Cederwall:2013naa}.

	While we cannot define field strength in a gauge invariant way, we shall show that up to the section conditions, the scalars formed by contractions
	of different field strengths are invariant up to the section conditions. We could use them instead as the basic building blocks of the 
	kinematic factor.
	
	To facilitate our calculations we will also introduce the twistor formalism for ${\cal A}$-theory. The twistors are defined in a way that makes 
	solving the section conditions easy. By writing everything in terms of the twistors, the calculations are greatly simplified.

	\subsection{Twistor formalism}\label{section:7-2}
	To analyze the spectrum of on-shell states, we will also need the little group {\it \^H} of each $H$.  
	For the usual Lorentz groups this can be found by reducing the number of space and time dimensions each by 1; 
	for the covering groups (with the spinor as the defining representation), this is essentially reducing the range of the spinor index
	by half.  For our theories, we just look at the known descriptions of each theory in 2 lower dimensions, and Wick rotates $H$ for 
	these D$-$2 dimensional theories to compact groups (for Abelian factors, phases):
	
	\begin{equation}
	\hbox{\it \^H}_D = H_{D-2,\rm{compact}} .
	\end{equation}
	
	The bi-product of the twistors is by definition the on-shell momentum.
	\begin{equation}
	p_{\alpha\beta} \equiv \lambda_{\alpha}^{a}\lambda_{b\beta}
	\end{equation}
	Of course, the twistors are contracted using some invariant tensors that depend on the dimensions.
	
	The twistors can naturally intertwine between the representations of {\it \^H} and $H$, allowing us to write everything
	on-shell in a manifestly $H$ invariant way. They are objects with one $H$ spinor indices and one {\it \^H} spinor indices.
	We denote it as $\lambda_{\alpha}^{a}$. In general, we will use Greek indices to denote $H$ spinors while using Roman indices to denote
	{\it \^H} spinors, although the exact index struture for D$=3, 4$ will be listed below.
	
	Explicitly in D$=$3, all the relevant groups, as well as their anticommuting coordinates, twistors, and momenta quadratic in twistors 
	(where $(\nobreak\  \nobreak\  )$ means symmetrized), are \cite{Cederwall:2015jfa, Julia:1980gr, Cremmer:1979up, Cremmer:1978ds}
\vskip 20mm
	$${\rm Spinor~indices~of~D=3~theories}$$
	$$
	\vcenter{
		\halign{ $#$ && \quad$#$\hfil \cr
	&  G & H & \hbox{\it \^H} & R & \theta & \lambda & p \cr
	{\cal A} & \textrm{SL}(5) & \textrm{O}(3,2) = \textrm{Sp}(4) & \textrm{SO}(2) & \textrm{I} & \bm{\alpha} & a\bm{\alpha} & (\bm{\alpha\beta}) \cr
	{\cal T} & \textrm{O}(3,3)=\textrm{SL}(4) & \textrm{O}(2,1)^{2} = \textrm{Sp}(2)^{2} & \textrm{I} & \textrm{I} & \alpha,\alpha^{\prime} & \alpha,\alpha^{\prime} & (\alpha\beta),(\alpha^{\prime}\beta') \cr
	{\cal M} & \textrm{GL}(4) & \textrm{O}(3,1) =\textrm{Sp}(2,\mathbb{C}) & \textrm{U}(1) & \textrm{U}(1) & \alpha,\dot{\alpha} & \alpha,\dot{\alpha} & \alpha\beta \cr
	 S & \textrm{GL}(3) & \textrm{O}(2,1) = \textrm{Sp}(2) & \textrm{I} & \textrm{SO}(2) & a\alpha & \alpha & (\alpha\beta) \cr
	}
	}
	$$
	We have written spinors of $H$ (defining representations of the covering group) with Greek indices, 
	and those of {\it \^H} (on $\lambda$) or R-symmetry (on $\theta$) with Latin. The boldface Greek indices 
	mean four-component spinors, while the unbold Greek indices are two-component spinors. 
	
	(U(1) {\it \^H} indices on $\lambda$ are single-valued and not displayed.)
	
	Note that Sp(4) is 4-component real, so the SO(2) index is 2-component real, while Sp$(2,\mathbb{C})$ is 2-component complex, 
	so the U$(1)$ index is implicitly 1-component complex. (However, we can use a complex basis for SO(2), with an off-diagonal metric.)
	Of course, the symmetric SO$(2)$ metric is used to construct $F$'s $p$ from $\lambda$s.
	
	For D$=$4 we have
	(where $\langle\nobreak\  \nobreak\  \rangle$ means antisymmetrized Sp-traceless) \cite{Cederwall:2015jfa, Julia:1980gr, Cremmer:1979up, Cremmer:1978ds}
 	$${\rm Spinor~indices~of~D=4~theories}$$
	$$
	\vcenter{\halign{ $#$ && \nobreak\  \nobreak\  \TextOrMath{\thinspace}{\,}$#$\hfil \cr
	& G & H & \hbox{\it \^H} & R & \theta & \lambda & p \cr
	{\cal A} & \textrm{O}(5,5) & \textrm{O}(5,\mathbb{C}) = \textrm{Sp}(4,\mathbb{C}) & \textrm{O}(3)\textrm{O}(2) = \textrm{U}(2) & \textrm{U}(1) & \bm{\alpha},\bm{\dot{\alpha}} & a\bm{\alpha},a\bm{\dot{\alpha}} & \alpha\beta \cr
	{\cal T} & \textrm{O}(4,4) & \textrm{O}(3,1)^{2} = \textrm{Sp}(2,\mathbb{C})^{2} & \textrm{O}(2)^{2} = \textrm{U}(1)^{2} & \textrm{U}(1)^{2} & \alpha,\dot{\alpha},\alpha^{\prime},\dot{\alpha}{}' & \alpha,\dot{\alpha},\alpha^{\prime},\dot{\alpha}{}' & \alpha\beta,\alpha^{\prime}\beta{}' \cr
	{\cal M} & \textrm{GL}(5) & \textrm{O}(4,1) = \textrm{USp}(2,2) & \textrm{O}(3) = \textrm{USp}(2) & \textrm{USp}(2) & a\bm{\alpha} & a\bm{\alpha} & \langle\bm{\alpha\beta}\rangle \cr
	S & \textrm{GL}(4) & \textrm{O}(3,1) = \textrm{Sp}(2,{\mathbb{C}}) & \textrm{O}(2) = \textrm{U}(1) & \textrm{U}(2) & a\alpha,a\dot{\alpha} & \alpha,\dot{\alpha} & \alpha\beta \cr}}
	$$
	
	Note that the momentums are not only on-shell, they also have to obey the $\mathcal{U}$ constraints. In 3D ${\cal A}$-theory, this section
	condition is 
	\begin{equation}
	p_{1\langle \bm{\alpha }}^{\bm{\gamma }}p_{2\bm{\gamma } \bm{\beta }\rangle} = 0.
	\end{equation}
	
	We shall rewrite this condition in terms of the twistors. For the convenience of calculation, we will use U(1) indices  
	or the little group instead of SO(2), i.e. we will write $\lambda _{\bm{\alpha} } \equiv \lambda _{\bm{\alpha} }^{1}+ i\lambda _{\bm{\alpha} }^{2}$
	, and  $\bar{\lambda }_{ \bm{\alpha} } \equiv \lambda _{\bm{\alpha} }^{1}- i\lambda _{\bm{\alpha} }^{2}$.
	 The momentum is $p_{(\bm{\alpha \beta} )} =\bar{\lambda }_{(\bm{\alpha} }\lambda _{\bm{\beta} )} $, and the section condition is 
	
	\begin{equation}
	\frac{1}{4}\left(\bar{\lambda }_{1\bm{\gamma }}\lambda _{2}^{\bm{\gamma }}\bar{\lambda }_{1\langle\bm{\alpha }}\lambda _{2\bm{\beta }\rangle}+ \bar{\lambda }_{1\bm{\gamma }}\lambda _{2}^{\bm{\gamma }}\lambda _{1\langle\bm{\alpha }}\bar{\lambda }_{2\bm{\beta }\rangle}\right)+\frac{1}{4}\left(\bar{\lambda }_{1\bm{\gamma }}\bar{\lambda }_{2}^{\bm{\gamma }}\lambda _{1\langle\bm{\alpha }}\lambda _{2\bm{\beta }\rangle}+\lambda _{1\bm{\gamma }}\lambda _{2}^{\bm{\gamma }}\bar{\lambda }_{1\langle\bm{\alpha }}\bar{\lambda }_{2\bm{\beta }\rangle} \right) = 0.
	\end{equation}
	
	Consistent constraints on twistors are
	
	\begin{equation}\label{eq:3-2-1}
	\lambda _{1\bm{\gamma }}\bar{\lambda }_{2}^{\bm{\gamma }} = 0, \quad  \lambda _{1\bm{\gamma }}\lambda _{2}^{\bm{\gamma }}\bar{\lambda }_{1\langle\bm{\alpha }}\bar{\lambda }_{2\bm{\beta }\rangle} + \textrm{c.c.}= 0.
	\end{equation}
	
	Observe that by letting $\lambda_{1} = \lambda_{2}$ ,  the constraints above also solve the on-shell condition $p^2 = 0$ 
	automatically.
	
	\subsection{D$=$3 $H=$Sp(4;$\mathbb{R}$) and D=4 $H=$Sp(4;$\mathbb{C}$) covariant amplitudes}\label{section:7-3}
	\subsubsection{Representations of prepotential and field strengths}
	
	Just like in ordinary supergravity, the representations of the $H(x, \theta)$ prepotential and super-field strengths $W(x, \theta)$ can be obtained from solving the 
	torsion constraints of the background. Since we are currently only interested in the bosonic part of the amplitude, the relevant physical 
	quantities are the generalized metric $h$ and generalized Weyl tensor $w$. They can be obtained from the $\theta$ expansion from superfields
	$W$ and $H$.
	
	Here we only give a table listing all the representation and index structures of all the above for D = 3 and D = 4.
	For the D$=$3 case, $H$ is a bi-spinor object while $W$ is a tri-spinor object and $h$ always appeared at the $\theta^{2}$ level,
	while $w$ appeared at the linear $\theta$ level. The representations for these objects for $S, {\cal T, M}$ theories could be obtained by 
	reduction on ${\cal A}$-theory case.
	$${\rm Spinor~indices~of~D=3~ supersymmetric~theories}$$
	$$
	\vcenter{\halign{ $#$ && \quad $#$\hfil \cr
	 & p & h & H & \theta^{2} & w & W \cr
	 {\cal A} & (\bm{\alpha \beta }) & \langle\bm{\alpha \beta }\rangle\langle\bm{\gamma \delta }\rangle & \langle\bm{\alpha \beta }\rangle & \langle\bm{\gamma \delta }\rangle & (\bm{\alpha \beta \gamma \delta }) & (\bm{\alpha \beta \gamma }) \cr
	 {\cal T }& (\alpha \beta ),(\alpha^{'}\beta^{'}) & (\alpha \beta )(\gamma^{'}\delta^{'}) & \alpha \gamma^{'} & \beta \delta^{'} & (\alpha \beta )(\gamma^{'}\delta^{'}) & \alpha \beta ' \cr
	 {\cal M} & \alpha \dot{\beta}& (\alpha \beta )(\dot{\gamma} \dot{\delta}) & \alpha \dot{\gamma}& \beta \dot{\delta}& (\alpha \beta \gamma \delta )\nobreak\ (\oplus\hbox{h.c.}) & (\alpha \beta \gamma ) \cr
	 S & (\alpha \beta ) & (\alpha \beta \gamma \delta ) & (\alpha \beta ) & a(\gamma a\delta ) & 0 & \hbox{scalar} \cr}}
	$$
	
	For the D$=$4 case, the prepotential $H$ is a scalar, and the super-field strength $W$ is a bi-spinor. 
	$h$ always appears at the $\theta^4$ level and $w$ at the linear $\theta$ level.
	$${\rm Spinor~indices~of~D=4~supersymmetric~theories}$$	
	$$
	\vcenter{\halign{ $#$ && \quad $#$\hfil \cr
	 & p & h & w & W  \cr
	 {\cal A} & \bm{\alpha \dot{\beta}}& \langle\bm{\alpha \beta }\rangle\langle\bm{\dot{\gamma} \dot{\delta}}\rangle & (\bm{\alpha \gamma })(\bm{\beta \dot{\delta}}) & \bm{\alpha \beta} \cr
	 {\cal T} & \alpha \dot{\beta},\alpha{}'\dot{\beta}{}' & \alpha \dot{\beta} \gamma{}'\dot{\delta}{}' & (\alpha \gamma )(\dot{\beta}{}'\dot{\delta}{}')\nobreak\ (\oplus\hbox{h.c.}) & \alpha \beta{}' \cr
	 {\cal M} & \langle\bm{\alpha \beta }\rangle & (\bm{\alpha \beta })(\bm{\gamma \delta }) & (\bm{\alpha \beta \gamma \delta }) & (\bm{\alpha \beta })  \cr
	 S & \alpha \dot{\beta}& (\alpha \beta )(\gamma \dot{\delta}) & (\alpha \beta \gamma \delta )\nobreak\ (\oplus\hbox{h.c.}) & (\alpha \beta ) \cr}
	 }
	$$
	
	 \subsubsection{Gauge invariance}
	
	 In 3D ${\cal A}$-theory $w_{ (\bm{\alpha \beta \gamma \delta })}$, written in terms of $h$, can only be a linear combination of the following index structure:
	\begin{equation}
	\begin{split}
		&\partial_{\bm{\alpha }}^{\hspace{5pt}\bm{\epsilon }}\partial_{\bm{\gamma }}^{\hspace{5pt}\bm{\zeta }}h_{<\bm{\beta \epsilon }> <\bm{\delta \zeta }>} + (\textrm{sym. perm. over $\bm{\alpha \beta \gamma \delta }$}), \textrm{  and}\\
		&\partial_{\bm{\alpha \gamma }}\partial^{\bm{\epsilon \zeta }}h_{<\bm{\beta \epsilon }> <\bm{\delta \zeta }>} + (\textrm{sym. perm. over $\bm{\alpha \beta \gamma \delta }$}).
	\end{split}
	\end{equation}
	
	Unfortunately, no linear combination of the above two index structures can achieve gauge invariance. 
	This seems to be consistent with the fact that no gauge invariant tensor in exceptional geometry analogous to the
	Riemann curvature tensor is found. \cite{Coimbra:2012af, Cederwall:2013naa}.
	
	The simplify the expression, we could write the on-shell metric $h$ in terms of twistors. Note that since $h$ is not a gauge
	invariant object, one needs to introduce extra reference twisters $\eta$ and $\xi$. Any gauge invariant object ( Up to section conditions)
	have to be independent of $\eta$ and $\xi$.
	
	The ${\cal A}$-theory metric should contain the same on-shell degrees of freedom as the ${\cal M}$-theory metric. When we reduce the metric from F to ${\cal M}$-theory, only the $h_{(\alpha \beta )(\dot{\gamma }\dot{\delta })}$ component should be non-zero, and equals the expression we just derived. The correct answer is 
	\begin{equation}\label{eq:4-4-1}
	h_{\langle \bm{\alpha \gamma } \rangle \langle \bm{\beta \delta } \rangle} = \left(\lambda_{\langle \bm{\alpha }}\bar{\eta}_{\bm{\gamma }\rangle}\lambda_{\langle \bm{\beta }}\bar{\eta}_{\bm{\delta }\rangle} +  \bar{\lambda}_{\langle\bm{\alpha }}\xi_{\bm{\gamma }\rangle}\bar{\lambda}_{\langle\bm{\beta }}\xi_{\bm{\delta }\rangle}\right) +(\bm{\alpha } \leftrightarrow \bm{\beta })  + (\bm{\gamma } \leftrightarrow\bm{\delta }).
	\end{equation} 
	 The details on the reduction can be found in \cite{Siegel:2020gro}.  These reference twistors have residual gauge transformations $\eta \rightarrow \eta + c \lambda$, and   $\xi \rightarrow \xi + c \lambda$.
	
	Plug the result into the two possible index structures of $w$. For the first index structure, we get 
	\begin{equation}
	\partial_{(\bm{\alpha }}^{\bm{\epsilon}}\partial_{\bm{\beta }}^{\bm{\zeta }} h_{\bm{\gamma \delta } )\bm{\epsilon\zeta }} = \lambda_{(\bm{\alpha }}\lambda_{\bm{\beta }}\lambda_{\bm{\gamma }}\lambda_{\bm{\delta })} + \bar{\lambda}_{(\bm{\alpha }}\bar{\lambda}_{\bm{\beta }}\bar{\lambda}_{\bm{\gamma }}\bar{\lambda}_{\bm{\delta })}.
	\end{equation}
	For the second index structure, we get
	\begin{equation}
	\partial_{(\bm{\alpha \beta }}\partial^{\bm{\epsilon\zeta }}h_{\bm{\gamma \delta }) \bm{\epsilon\zeta }} = 0.
	\end{equation}
	Therefore, the second index structure will not contribute to the Weyl tensor on-shell:
	
	\begin{equation}
	w_{(\bm{\alpha \beta \gamma \delta})} = \partial_{(\bm{\alpha }}^{\bm{\epsilon}}\partial_{\bm{\beta }}^{\bm{\zeta}} h_{\bm{\gamma \delta})\bm{\epsilon\zeta}} = \lambda_{(\bm{\alpha }}\lambda_{\bm{\beta }}\lambda_{\bm{\gamma }}\lambda_{\bm{\delta})} + \bar{\lambda}_{(\bm{\alpha }}\bar{\lambda}_{\bm{\beta }}\bar{\lambda}_{\bm{\gamma }}\bar{\lambda}_{\bm{\delta})}.
	\end{equation}
	The final form of the Weyl tensor does not depend on the reference twistors. 
	
	The gauge transformation of $h$
	
	\begin{equation}
	\delta h_{\langle\bm{\alpha \beta }\rangle \langle \bm{\gamma \delta} \rangle} = \partial_{\langle \bm{\alpha }}^{\langle\bm{\gamma }}\xi_{\bm{\beta }\rangle}^{\bm{\delta} \rangle} 
	\end{equation}  
	The variance of the Weyl tensor $w$ is complicated. One can avoid direct calculation by observing that up to the section condition, one can see that $\delta w$ must be in the form:
	
	\begin{equation}
	\delta w_{(\bm{\alpha \beta \gamma \delta})} = c_{1}\partial_{(\bm{\alpha \beta }}\partial_{\bm{\gamma \delta})}\partial^{\bm{\epsilon\zeta}}\xi_{\bm{\epsilon\zeta}} + c_{2}\partial_{(\bm{\alpha \beta }}\partial_{\bm{\gamma }}^{\bm{\epsilon}}\partial_{\bm{\delta})}^{\bm{\zeta}}\xi_{\bm{\epsilon\zeta}}.
	\end{equation}
	This is because any contraction between the partial derivatives vanishes. 
	Therefore, the above form exhausts all the possible contractions of $\partial^{3}\xi$ that are compatible with the symmetries. 
	
	If we write the partial derivative in terms of twistors, we get 
	\begin{equation}
	\delta w_{(\bm{\alpha \beta \gamma \delta})} = c_{1}\lambda_{(\bm{\alpha }}\lambda_{\bm{\beta }}\bar{\lambda}_{\bm{\gamma }}\bar{\lambda}_{\bm{\delta})}(\lambda_{\bm{\epsilon}}\lambda_{\bm{\zeta}}\xi^{\bm{\epsilon\zeta}}) + c_{2}\left[\lambda_{(\bm{\alpha }}\lambda_{\bm{\beta }}\lambda_{\bm{\gamma }}\bar{\lambda}_{\bm{\delta})}(\bar{\lambda}_{\bm{\epsilon}}\lambda_{\bm{\zeta}}\xi^{\bm{\epsilon\zeta}}) + \textrm{c.c}\right].
	\end{equation}
	Note that both terms in $\delta w$ have at least one uncontracted $\lambda$ and one uncontracted $\bar{\lambda}$. 
	This will be important later for proving gauge invariance.
	
	Now we wish to prove that $w_{1\bm{\alpha \beta \gamma \delta}}w_{2}^{\bm{\alpha \beta \gamma \delta}}$ is gauge invariant.
	First, observe that $\delta (w_{1}w_{2}) =\delta w_{1} w_{2} + w_{1}\delta w_{2}$, and $w = \lambda^{4} + \bar{\lambda}^{4}$. 
	Using the fact that $\delta w$ contains at least one uncontracted $\lambda$ and one uncontracted $\bar{\lambda}$, 
	there must be either a $\lambda_{1}\bar{\lambda_{2}}$ or $\lambda_{2}\bar{\lambda_{1}}$ in every term of $\delta w_{1} w_{2}$. 
	Therefore it vanishes by the section condition. The same argument applies for $ w_{1}\delta w_{2}$. 
	
	This argument can apply to any contractions $w_{1}w_{2}\cdots w_{n}$, as long as all indices are contracted. 
	The argument goes as follows:
	Without lost of generality, we assume that $w_{1}w_{2}\cdots w_{n}$ cannot be factored into smaller pieces, 
	since any index contractions of the form $w_{1}w_{2}\cdots w_{n}$ can be decomposed into products of 
	terms that cannot be decomposed further. 
	
	The gauge transformation is 
	 \begin{equation}
	 \delta(w_{1}w_{2}\cdots w_{n}) = (\delta w_{1}) w_{2}\cdots w_{n} + \cdots\;\;.
	  \end{equation}
	  
	Note that since $w_{2}\cdots w_{n}$ cannot be decomposed into smaller pieces, it can only contain terms 
	with all $\lambda$ or terms with all $\bar{\lambda}$; any terms with both $\lambda$ and $\bar{\lambda}$ will disappear by the section condition. 
	And since $\delta w_{1}$ contains both uncontracted $\lambda$ and $\bar{\lambda}$, every term will vanish.
	 By this argument, we can see that $w_{1}w_{2}\cdots w_{n}$ is indeed gauge invariant. 
	
	 \subsubsection{Construction of the amplitude}
	 In this section, we use boldface $\bm{w}$ to denote the Weyl tensor of ${\cal A}$-theory and non-boldface $w$ to denote the Weyl tensor in ${\cal M}$ theory.
	 
	The last step in constructing the amplitude is to find the correct kinematic factor $\bm{w}^{4}$ that will 
	reduce to the 4D gravity kinematic factor under reduction to ${\cal M}$-theory.
	In general, there are two types of contraction of $\bm{w}^4$. The disconnected type $(\bm{w})^{2}(\bm{w})^{2}$ and the connected type
	$(\bm{w})^{4}$. With the reduction  $\pm{w} \rightarrow w + \bar{w}$. One can see that both types of terms will generate $\lambda^{4}$
	and $\bar{\lambda}^{4}$. These correspond to $R^{4}$ interactions in the Lagrangian, which is unwanted.
	
The 4-point graviton amplitude is 
	\begin{equation}
	A_{4}(1, 2, 3, 4) = \left[ (w_{1}w_{2}) \bar{w}_{3} \bar{w}_{4} + \bar{w}_{1} \bar{w}_{2} w_{3} w_{4} + (2\leftrightarrow 3)  + (2\leftrightarrow 4) \right]\frac{1}{stu}.
	\end{equation}
	
	This one could conjecture candidate ${\cal A}$-theory kinematic factor:
	\begin{equation}
	 (\bm{w}_{1}\bm{w}_{2})(\bm{w}_{3}\bm{w}_{4}) + (2\leftrightarrow 3) + (2 \leftrightarrow 4).
	\end{equation}
	This term will be reduced to the correct 4-graviton kinematic factor and an additional $\bar{\lambda}^{4}$ and $\lambda^{4}$ term, but 
	the  $\bm{w}^{4}$ contractions.
	\begin{equation}
	w_{1\bm{\alpha\beta\gamma\delta}}w^{\bm{\gamma\delta\epsilon\zeta}}_{2}w_{3 \bm{\epsilon\zeta\rho\pi}}w^{\bm{\rho\pi\alpha\beta} }_{4}
	\end{equation}
	will cancel out exactly the unwanted terms. We will arrive at the final amplitude:
	
	\begin{equation}
	\left[ (w_{1\bm{\alpha\beta\gamma\delta}}w_{2}^{\bm{\alpha\beta\gamma\delta}})(w_{3\bm{\epsilon\zeta\rho\pi}}w_{4}^{\bm{\epsilon\zeta\rho\pi}}) - 2w_{1\bm{\alpha\beta\gamma\delta}}w^{\bm{\gamma\delta\epsilon\zeta}}_{2 }w_{3 \bm{\epsilon\zeta\rho\pi}}w^{\bm{\rho\pi\alpha\beta}}_{4} \right]\frac{1}{stu}+  (2\leftrightarrow 3) + (2 \leftrightarrow 4).
	\end{equation}
	\vskip 6mm

\section{Future topics }
${\cal A}$-theory is a theory with manifestly U-duality including all superstring theories.
	We have presented the following elements of the ${\cal A}$-theory formulation;
	duality symmetry groups and their representations,
	gauge symmetries generated by the brane current algebras,
	relation between theories by sectionings,
	U-duality covariant gravity theories,
	quantizable perturbative brane Lagrangians with manifestly U-duality
	and $A$-symmetry covariant amplitude.
	There are interesting topics for future problems as follows.   
\begin{enumerate}
	
    \item Amplitudes: The covariant tree amplitudes for ${\cal A}$-theories in general dimensions are yet to be found. As we can see in the D=3 case, one does not need to explicitly quantize  ${\cal A}$-theories to get the tree amplitude. The $H$-symmetry and gauge covariance is enough constraints to find the amplitude. One could ask if this method could be extended to higher point amplitude or higher dimensional amplitude. This might be possible since we didn't use the full $H$-symmetery in the D=3 case. The full $H$-symmetry also includes the usual supersymmetry and more.  
    
	One possible strategy to do this is to find the Ward identities of the $H$-symmetries for the amplitude, and then attempt to partially or fully solve them. This strategy has been used to find the MHV amplitudes for ${\cal N} = 4$ Yang-Mills \cite{Elvang:2009wd, Elvang:2010xn}.

    \item Covariant quantization: It's currently not known how to quantize ${\cal A}$-theory in a covariant way.
    The biggest challenges are the find the correct BRST formalism. The so-called zeroth-quantized ghost proposed by 
	Siegel might give a hint at how to solve this problem \cite{Siegel:2016dek}.
	Since the ${\cal U}$-constraints will kill the world-volume dimensions down to 2, one could expect that ${\cal A}$-theory has the same oscillating modes as usual string theories. But the fact that it admits non-perturbative supergravity solutions at
	the first-quantized level means that the zero modes are enlarged. The ${\cal U}$-constraints cannot kill zero modes so this is consistent. One way to achieve this is to introduce fermionic world-volume dimensions, which will cancel out the bosonic world-volume dimensions when one performs integration over the vertex operator's positions.  Similar to the ghosts cancel out the gauge redundancy in the path integral. Therefore the name zeroth quantized ghost.
 
     There is a way to utilize the operator product expansion of Virasoro algebra which includes tree and 1-loop graphs leading to the quantum effects. 
	
	\item  Higher rank: The Lie groups for ${\cal A}$-symmetry is infinite dimensional for $D > 5$ and $G$-symmetry is infinite
	dimensional for $D > 7$. The infinite dimensions pose a challenge because many explicit constructions like finding the Lagrangian and finding the constraints are done on a case-by-case basis.
	
	The exceptional supergravity with $E_{9}$ $G$-symmetry was constructed in \cite{Bossard:2018utw, Bossard:2021jix}, it is interesting to see if one can generalize it to ${\cal A}$-theory version with all the brane modes included. 
	
	\item  

	The supersymmetric world-volumes: Current formalism of  ${\cal A}$-theory, the world-voulme only contain bosonic component. It is interesting to see if one could (or should) build ${\cal A}$-theory on world-volume superspace. The zeroth quantized ghost method suggested above suggests that ${\cal A}$-theory with world-volume superspace might be a more natural formalism \cite{Siegel:2020gws}.

	\item  Type I and heterotic dualities \cite{Hatsuda:2021ezo}: Type I and heterotic string theories have 10-dimensional N=1 supersymmetry. These theories with manifestly U-duality symmetry are described by open branes. Quantum consistency such as the Green-Schwarz mechanism is a crucial problem restricting the gauge sector.  Fermions and heterotic fermion section conditions may be interesting.

    \item ${\cal A}$-theory in critical dimensions.
  
    \item String field theory and brane field theory.
\end{enumerate}

\subsection*{Acknowledgments}

We are grateful to Martin Ro\v{c}ek and Yuqi Li for the fruitful discussions. 
We also acknowledge the Simons Center for Geometry and Physics for its hospitality during
``The Simons Summer Workshop in Mathematics and Physics 2022 and 2023" 
where this work has been developed.
W.S. is supported by NSF award PHY-19105093. 
M.H. is supported in part by 
Grant-in-Aid for Scientific Research (C), JSPS KAKENHI
Grant Numbers JP22K03603 and JP20K03604.

\printbibliography

\end{document}